\documentclass[a4paper,11pt]{article}
\pdfoutput=1
\usepackage{jheppub}

\usepackage{amssymb,amsmath}
\usepackage[normalem]{ulem}
\usepackage[utf8x]{inputenc}
\usepackage{slashed}
\usepackage{graphicx}
\usepackage{here}
\usepackage{color}
\usepackage{csquotes} 
\usepackage{comment}
\usepackage{mathrsfs}
\usepackage{float}
\usepackage{ascmac}
\usepackage{multirow}
\usepackage{longtable}
\usepackage{bbm}
\usepackage{booktabs}
\usepackage{mathtools}
\usepackage{longtable}
\usepackage{cancel}

\usepackage[italicdiff]{physics}

\usepackage[hang,small,bf]{caption}
\usepackage[subrefformat=parens]{subcaption}
\usepackage{hyperref}

\makeatletter
\newcommand*\rel@kern[1]{\kern#1\dimexpr\macc@kerna}
\newcommand*\widebar[1]{%
  \begingroup
  \def\mathaccent##1##2{%
    \rel@kern{0.8}%
    \overline{\rel@kern{-0.8}\macc@nucleus\rel@kern{0.2}}%
    \rel@kern{-0.2}%
  }%
  \macc@depth\@ne
  \let\math@bgroup\@empty \let\math@egroup\macc@set@skewchar
  \mathsurround\z@ \frozen@everymath{\mathgroup\macc@group\relax}%
  \macc@set@skewchar\relax
  \let\mathaccentV\macc@nested@a
  \macc@nested@a\relax111{#1}%
  \endgroup
}
\makeatother

\numberwithin{equation}{section}

\preprint{
\begin{minipage}{5cm}
\small
\flushright
KYUSHU-HET-315
\end{minipage}}

\title{Classification of Modular Symmetries in Non-Supersymmetric Heterotic String theories}

\author{Shuta Funakoshi$^{1}$,} 
\author{Yuichi Koga$^{2}$, and} 
\author{Hajime Otsuka$^{1}$} 
\affiliation{
$^1$Department of Physics, Kyushu University, 744 Motooka, Nishi-ku, Fukuoka 819-0395, Japan\\
$^2$Institute for Advanced Study, Kyushu University, 744 Motooka, Nishi-ku, Fukuoka 819-0395, Japan
}
\emailAdd{funakoshi.shuta@phys.kyushu-u.ac.jp}
\emailAdd{koga.yuichi.593@m.kyushu-u.ac.jp}
\emailAdd{otsuka.hajime@phys.kyushu-u.ac.jp}

\abstract{
We study modular symmetries in non-supersymmetric heterotic string theories on toroidal backgrounds with Wilson line modulus, constructed by stringy Scherk-Schwartz compactification. 
In particular, we focus on a subgroup of the T-duality group ${\cal O}(D+16,D,\mathbb{Z})$ with $D=2$ given by an outer automorphism of the Narain lattice, which can be mapped to the Siegel modular group $\mathrm{Sp}(4,\mathbb{Z})$. 
We classify the modular symmetries and a ${\cal CP}$-like symmetry on $T^2$ and its orbifolds with symmetric and asymmetric orbifold twists. 
It turns out that the non-supersymmetric heterotic string theories only 
enjoy a part of modular symmetries in contrast to supersymmetric ones. 
Furthermore, the gauge symmetry is maximally enhanced at fixed points of modular symmetries on $T^2$ on which we analyze the vacuum structure of eight-dimensional tachyon-free vacua.}

\makeatletter
\gdef\@fpheader{}
\makeatother

\begin{document}

\maketitle

\section{Introduction}
\label{sec:intro}

The symmetry of target space in string theory is known as target space dualities or T-dualities \cite{Kikkawa:1984cp,Sakai:1985cs,Giveon:1988tt,Giveon:1994fu}. Since they can not be captured in a theory of point particles, revealing the target space symmetry is of interest in various contexts. 
For instance, a part of T-dualities (including the modular symmetry) will be regarded as a flavor symmetry of particle physics in heterotic string theory 
on toroidal orbifolds \cite{Ferrara:1989qb,Lerche:1989cs,Lauer:1989ax,Lauer:1990tm} and Calabi-Yau threefolds \cite{Ishiguro:2021ccl,Ishiguro:2024xph} as well as type II string theories \cite{Kobayashi:2018rad,Kobayashi:2018bff,Ohki:2020bpo,Kikuchi:2020frp,Kikuchi:2020nxn,
Kikuchi:2021ogn,Almumin:2021fbk,Kikuchi:2023awe}.

When string theory is compactified on a $D$-dimensional torus, the duality symmetry is known as ${\cal O}(d_L, d_R, \mathbb{Z})$ symmetry for $d_L=d_R=D$ in bosonic and type II string theory and $d_L=d_R+16=D+16$ in heterotic string theory \cite{Giveon:1988tt,Giveon:1994fu}. 
So far, T-dualities and their phenomenological applications were well studied in superstring theory. 
For instance, in the case of heterotic string theory with preserving the supersymmetry in the target space, the T-duality group is determined by the outer automorphisms of the Narain lattice. 
When we introduce the Wilson line modulus, two $\mathrm{SL}(2,\mathbb{Z})$ modular symmetries associated with the K\"ahler modulus $T$ and the complex structure modulus $U$ are enhanced to the Siegel modular group $\mathrm{Sp}(4,\mathbb{Z})$. 
Then, the relation between the T-dualities ${\cal O}(D+16, D, \mathbb{Z})$ and the Siegel modular group $\mathrm{Sp}(4,\mathbb{Z})$ was analyzed in a two-dimensional torus \cite{Baur:2020yjl} and its orbifolds with symmetric and asymmetric twists \cite{Nilles:2021glx}. Phenomenological applications of such a Siegel modular group and its congruence subgroup were studied in e.g., \cite{Ding:2020zxw}.

In this paper, we deal with non-supersymmetric heterotic string theories. 
As one of the mechanisms of breaking the supersymmetry, the Scherk-Schwarz mechanism \cite{Scherk:1978ta,Rohm:1983aq,Dixon:1986iz,Kounnas:1989dk} 
was proposed in non-supersymmetric string theories in ten dimensions \cite{Dixon:1986jc,Alvarez-Gaume:1986ghj} and general dimension \cite{Nair:1986zn,Ginsparg:1986wr}. Furthermore, T-dualities were explored in \cite{Itoyama:2021itj} for $D=1,2$ in type 0 string theory and $D=1$ in non-supersymmetric heterotic string theory. 
However, it is quite important to reveal T-dualities in a more generic case to 
explore the vacuum structure of non-supersymmetric string theories and the phenomenological application of T-dualities. 
For concreteness, we focus on a two-dimensional torus and its orbifolds with symmetric and asymmetric orbifold twists. 
Using the technique of \cite{Baur:2020yjl,Nilles:2021glx}, we map the outer automorphism of the Narain lattice in non-supersymmetric heterotic string theories to the Siegel modular group $\mathrm{Sp}(4,\mathbb{Z})$. 
Since the T-duality group is partially broken by the Scherk-Schwarz twist, only the subgroup of $\mathrm{Sp}(4,\mathbb{Z})$ 
remains in the system. 
We classify the remaining symmetries on two-dimensional toroidal backgrounds with Wilson line modulus, constructed by stringy Scherk-Schwartz compactification. 
As concrete examples, we consider non-supersymmetric heterotic stings on $T^2$ which are directly connected to the $SO(16)\times SO(16)$ heterotic string. We then find that the fixed points of the remaining modular symmetries correspond to maximal enhancements of gauge groups. We finally obtain five 8d tachyon-free vacua. One of them has a negative cosmological constant and is at an unstable knife edge, which implies a boundary of a tachyonic region \cite{Ginsparg:1986wr}, while the others have positive cosmological constants. We analyze the Hessian at these four vacua and find out that one of them is a local maximum, while the other three are saddle points.

This paper is organized as follows. 
After reviewing $\mathrm{Sp}(4,\mathbb{Z})$ symmetry in section \ref{sec:Narain&Sp4Z}, we discuss the relation between ${\mathrm{Sp}(4,\mathbb{Z}})$ modular symmetry and the subgroup of the T-duality group in the context of the supersymmetric heterotic string theory on $T^2$ following \cite{Baur:2020yjl}. 
Then, we move to the T-duality group of non-supersymmetric string theories in section \ref{sec:T-duality}. 
In section \ref{sec:classification}, we classify the modular symmetry in non-supersymmetric string theories on $T^2$ by mapping the T-duality group to the Siegel modular group $\mathrm{Sp}(4,\mathbb{Z})$. Gauge symmetry enhancements and cosmological constants are also discussed.
This analysis is extended to various toroidal orbifolds with symmetric and asymmetric orbifold twists in section \ref{sec:T2ZN}. 
Finally, section \ref{sec:conclusion} is devoted to the conclusion. 
In Appendix \ref{app:T2ZN}, we present all the possible configurations of $\hat{Z}$ on $T^2$ orbifolds.

\section{Narain formulation and $\mathrm{Sp}(4,\mathbb{Z})$ modular symmetry}
\label{sec:Narain&Sp4Z}
In this section, we review the Narain formulation and the relation between the subgroup of the T-duality group and ${\mathrm{Sp}(4,\mathbb{Z}})$ modular symmetry in 2d toroidal compactifications of the supersymmetric heterotic string theory \cite{Baur:2020yjl}.

\subsection{Siegel modular group $\mathrm{Sp}(4,\mathbb{Z})$}
\label{sec:Sp4Z}
The symplectic group $\mathrm{Sp}(2g,\mathbb{Z})$, which is also called the Siegel modular group of genus $g$, is defined as the set of linear transformations of a $2g$-dimensional vector space over the integers which preserve a skew-symmetric bilinear form $J$:
\begin{align}
    \mathrm{Sp}(2g,\mathbb{Z}):=\left\lbrace M\in \mathrm{GL}(2g,\mathbb{Z})\left|~M^{\mathrm{T}}JM=J \right.\right\rbrace,
\end{align}
where $J$ is the $2g\times 2g$-dimensional matrix given by
\begin{align}
    J:=\left(
    \begin{array}{cc}
        0 & \mathbbm{1}_{g}\\
        -\mathbbm{1}_{g} & 0
    \end{array}
    \right).
\end{align}
The Siegel upper half plane $\mathbb{H}_{g}$ is defined as
\begin{align}
    \mathbb{H}_{g}:=\left\lbrace \Omega \in \mathrm{GL}(g,\mathbb{C})\left|~\Omega^{\mathrm{T}}=\Omega,~\mathrm{Im}~\Omega>0 \right.\right\rbrace,
\end{align}
and the element $\Omega$ of $\mathbb{H}_{g}$ has $g(g+1)/2$ complex numbers corresponding to moduli. The Siegel modular group $\mathrm{Sp}(2g,\mathbb{Z})$ acts on $\mathbb{H}_{g}$ as
\begin{align}\label{transformation_Omega}
    \Omega\stackrel{M}{\longrightarrow}(A\Omega+B)(C\Omega+D)^{-1},
\end{align}
where $M\in \mathrm{Sp}(2g,\mathbb{Z})$ is expressed with $g\times g$-dimensional blocks $A,B,C$ and $D$ as
\begin{align}
    M=\left(
    \begin{array}{cc}
        A & B\\
        C & D
    \end{array}
    \right).
\end{align}

From now on, we focus on the $\mathrm{Sp}(4,\mathbb{Z})$ group. In the case of $g=2$, the element $\Omega$ of $\mathbb{H}_{g}$ is a $2\times2$ matrix with $3$ moduli $U,T$ and $Z$ expressed as
\begin{align}
     \Omega=\left(
    \begin{array}{cc}
        U & Z\\
        Z & T
    \end{array}
    \right).
\end{align}
It is known that the Siegel modular group $\mathrm{Sp}(4,\mathbb{Z})$ include the two $\mathrm{SL}(2,\mathbb{Z})$ factors, which associate two moduli $T$ and $U$. The corresponding embedded subgroup of $\mathrm{Sp}(4,\mathbb{Z})$ is represented by
\begin{align}
    M_{(\gamma_T,\gamma_U)}:=\left(
    \begin{array}{cccc}
        a_U & 0 & b_U & 0\\
        0 & a_T & 0 & b_T\\
        c_U & 0 & d_U & 0\\
        0 & c_T & 0 & d_T\\
    \end{array}
    \right),
\end{align}
where $a_Td_T-b_Tc_T=a_Ud_U-b_Uc_U=1$ and
\begin{align}
    \gamma_i:=\left(
    \begin{array}{cc}
        a_i & b_i\\
        c_i & d_i
    \end{array}
    \right)\in \mathrm{SL}(2,\mathbb{Z})_i,~~~i=T,U.
\end{align}
The group $\mathrm{SL}(2,\mathbb{Z})$ is generated by the $\mathrm{T}$- and $\mathrm{S}$-transformations defined by
\begin{align}\label{STmatrix}
    \mathrm{T}:=\left(
    \begin{array}{cc}
        1 & 1\\
        0 & 1
    \end{array}
    \right),~~~
    \mathrm{S}:=\left(
    \begin{array}{cc}
        0 & 1\\
        -1 & 0
    \end{array}
    \right).
\end{align}
Using \eqref{transformation_Omega}, we obtain the transformation law of the moduli induced by $M_{(\mathbbm{1}_2,\gamma_U)}$ as
\begin{subequations}\label{complex_SP}
    \begin{align}
        T&\xrightarrow{M_{(\mathbbm{1}_2,\gamma_U)}}
        T-\frac{c_U Z^2}{c_U U+d_U},\\
        U&\xrightarrow{M_{(\mathbbm{1}_2,\gamma_U)}}
        \frac{a_U U+b_U}{c_U U+d_U},\\
        Z&\xrightarrow{M_{(\mathbbm{1}_2,\gamma_U)}}
        \frac{Z}{c_U U+d_U},
    \end{align}
\end{subequations}
and similarly by $M_{(\gamma_T,\mathbbm{1}_2)}$ as
\begin{subequations}\label{Kahler_SP}
    \begin{align}
        T&\xrightarrow{M_{(\gamma_T,\mathbbm{1}_2)}}
        \frac{a_T T+b_T}{c_T T+d_T},\\
        U&\xrightarrow{M_{(\gamma_T,\mathbbm{1}_2)}}
        U-\frac{c_T Z^2}{c_T T+d_T},\\
        Z&\xrightarrow{M_{(\gamma_T,\mathbbm{1}_2)}}
        \frac{Z}{c_T T+d_T}.
    \end{align}
\end{subequations}
The Siegel modular group $\mathrm{Sp}(4,\mathbb{Z})$ includes a $\mathbb{Z}_2$ mirror transformation
\begin{align}
    M_{\times}:=\left(
    \begin{array}{cccc}
        0 & 1 & 0 & 0\\
        1 & 0 & 0 & 0\\
        0 & 0 & 0 & 1\\
        0 & 0 & 1 & 0\\
    \end{array}
    \right),
\end{align}
and one can check $M_{\times}^2=\mathbbm{1}_4$. From \eqref{transformation_Omega}, $M_{\times}$ interchanges $T$ and $U$:
\begin{subequations}\label{mirror_SP}
    \begin{align}
        T&\xrightarrow{M_{\times}}U,\\
        U&\xrightarrow{M_{\times}}T,\\
        Z&\xrightarrow{M_{\times}}Z.
    \end{align}
\end{subequations}
The final elements of $\mathrm{Sp}(4,\mathbb{Z})$ we focus on are
\begin{align}
    M\begin{psmallmatrix}
        \ell\\
        m
    \end{psmallmatrix}:=\left(
    \begin{array}{cccc}
        1 & 0 & 0 & -\ell\\
        m & 1 & -\ell & 0\\
        0 & 0 & 1 & -m\\
        0 & 0 & 0 & 1\\
    \end{array}
    \right),~~~\ell,m\in\mathbb{Z}.
\end{align}
These elements lead to
\begin{subequations}\label{WLshift_SP}
    \begin{align}
        T&\xrightarrow{M\begin{psmallmatrix}
        \ell\\
        m
    \end{psmallmatrix}}T+m\left( mU+2Z-\ell \right),\\
        U&\xrightarrow{M\begin{psmallmatrix}
        \ell\\
        m
    \end{psmallmatrix}}U,\\
        Z&\xrightarrow{M\begin{psmallmatrix}
        \ell\\
        m
    \end{psmallmatrix}}Z+mU-\ell.
    \end{align}
\end{subequations}

Finally, we comment on a $\cal CP$-like transformation. 
Let us introduce 
\begin{align}
    M_\ast :=
\begin{pmatrix}
    -1 & 0 & 0 & 0\\
    0 & -1 & 0 & 0\\
    0 & 0 & 1 & 0\\
    0 & 0 & 0 & 1\\
\end{pmatrix}
\notin \mathrm{Sp}(4,\mathbb{Z}),
\end{align}
satisfying $M_\ast^{\mathrm{T}} J M_\ast = -J$. 
Then, a $\mathrm{CP}$-like transformation is 
defined as
\begin{align}
    M \xrightarrow{M_\ast} M^\prime := M^{-1}_\ast M M_\ast,
\end{align}
for all $M\in \mathrm{Sp}(4,\mathbb{Z})$, 
which belongs to $\mathrm{Sp}(4,\mathbb{Z})$, i.e., $M^\prime \in \mathrm{Sp}(4,\mathbb{Z})$ for all $M \in \mathrm{Sp}(4,\mathbb{Z})$; thereby such a $\mathrm{CP}$-like transformation defines an outer automorphism due to $M_\ast \notin \mathrm{Sp}(4,\mathbb{Z})$, and it enhances $\mathrm{Sp}(4,\mathbb{Z})$ to \cite{Ishiguro:2020nuf,Ding:2021iqp}
\begin{align}
    \mathrm{GSp}(4,\mathbb{Z}):=
    \left\lbrace M \in \mathbb{Z}^{4\times 4}\left|~M^{\mathrm{T}}JM = \pm J \right.\right\rbrace,
\end{align}
Under $M_\ast$, the moduli fields transform as
\begin{align}
    T \xrightarrow{M_\ast}-\bar{T},
    \nonumber\\
    U \xrightarrow{M_\ast}-\bar{U},
    \nonumber\\
    Z \xrightarrow{M_\ast}-\bar{Z},
\end{align}
which is an extension of the outer automorphism of $\mathrm{Sp}(2,\mathbb{Z})\cong \mathrm{SL}(2,\mathbb{Z})$ \cite{Baur:2019kwi,Novichkov:2019sqv}.

\subsection{Narain formulation on the torus}
\label{sec:Narainformulation}
The toroidal compactification of supersymmetric heterotic strings can be described by the Narain formulation \cite{Narain:1985jj,Narain:1986am,Narain:1986qm}. For the the $D$-dimensional torus, the $16+D$ left- and $D$ right-moving string modes $(y_{L},y_{R})$ satisfy the following identification:
\begin{align}
    \begin{pmatrix}
        y_L\\
        y_R
    \end{pmatrix}
    \sim 
    \begin{pmatrix}
        y_L\\
        y_R
    \end{pmatrix}
    +\mathcal{E}\hat{N},
\end{align}
where $\mathcal{E}\hat{N}$ is used for defining the Narain lattice, discussed in detail below. Then the heterotic string theory on the $D$-dimensional torus has the pairing of states
\begin{align}\label{SUSY_pairing}
    \left(\Gamma_{16+D,D};\bar{v},\bar{s}\right),
\end{align}
where $\Gamma_{16+D,D}$ is called the Narain lattice and $v$ and $s$ represent the vector and spinor conjugacy classes of $SO(8)$, respectively. The one-loop partition function is expressed as
\begin{align}
    Z_{\mathrm{SUSY}}^{(10-D)}(\tau)=\frac{1}{\tau_2^{\frac{8-D}{2}}\eta^{24}\bar{\eta}^{12}}(\bar V_8-\bar S_8)\sum_{P\in\Gamma_{16+D,D}}q^{\frac{1}{2}P_L^2}\bar{q}^{\frac{1}{2}P_R^2},
\end{align}
where $\tau=\tau_1+i\tau_2$ is the modulus of the world-sheet torus and $\eta(\tau)$ is the Dedekind eta function given by
\begin{align}
    \eta(\tau)=q^{\frac{1}{24}}\prod_{n=1}^{\infty}(1-q^n),~~~q=e^{2\pi i\tau}.
\end{align}
The $SO(2n)$ characters $O_{2n},V_{2n},S_{2n},C_{2n}$ are defined as
\begin{align}\label{SO(2n)character}
	O_{2n}
	&=\frac{1}{2\eta^{n}}\left( \vartheta^{n}
	\begin{bmatrix} 
		0\\ 
		0\\ 
	\end{bmatrix}(0,\tau)+ \vartheta^{n}
	\begin{bmatrix} 
		0\\ 
		1/2\\ 
	\end{bmatrix}(0,\tau)
	\right),\\
	V_{2n}
	&=\frac{1}{2\eta^{n}}\left( \vartheta^{n}
	\begin{bmatrix} 
		0\\ 
		0\\ 
	\end{bmatrix}(0,\tau)- \vartheta^{n}
	\begin{bmatrix} 
		0\\ 
		1/2\\ 
	\end{bmatrix}(0,\tau)
	\right),\\
	S_{2n}
	&=\frac{1}{2\eta^{n}}\left( \vartheta^{n}
	\begin{bmatrix} 
		1/2\\ 
		0\\ 
	\end{bmatrix}(0,\tau)+ \vartheta^{n}
	\begin{bmatrix} 
		1/2\\ 
		1/2\\ 
	\end{bmatrix}(0,\tau)
	\right),\\
	C_{2n}
	&=\frac{1}{2\eta^{n}}\left( \vartheta^{n}
	\begin{bmatrix} 
		1/2\\ 
		0\\ 
	\end{bmatrix}(0,\tau)-\vartheta^{n}
	\begin{bmatrix} 
		1/2\\ 
		1/2\\ 
	\end{bmatrix}(0,\tau)
	\right),
\end{align}
with the theta functions with characteristics defined as
\begin{align}
	\vartheta
	\begin{bmatrix} 
		\alpha\\ 
		\beta\\ 
	\end{bmatrix}(z,\tau)&=\sum_{n=-\infty}^{\infty}\exp\left( \pi i (n+\alpha)^2 \tau +2\pi i (n+\alpha)(z+\beta) \right). 
\end{align}
From the modular invariance of the one-loop partition function, $\Gamma_{16+D,D}$ must be an even self-dual lattice with Lorentzian signature $(16+D,D)$ and can be spanned by the Narain vielbein $\mathcal{E}$ expressed as a $(16+2D)\times(16+2D)$ matrix
\begin{align}
    \Gamma_{16+D,D}=\left\lbrace  P=\mathcal{E}\hat{N}~\left.\right|~ \hat{N}=\begin{pmatrix}q\\ w\\ n\end{pmatrix} \in \mathbb{Z}^{16+2D}\right\rbrace,
\end{align}
where $q\in\mathbb{Z}^{16}$ is the gauge quantum numbers, $w\in\mathbb{Z}^{D}$ the winding numbers and $n\in\mathbb{Z}^{D}$ the Kaluza–Klein numbers. The Narain vielbein $\mathcal{E}$ satisfies
\begin{align}
    \hat{\eta}:=\mathcal{E}^{\mathrm{T}}~\eta~ \mathcal{E}=\left(
    \begin{array}{ccc}
        g & 0 & 0\\
        0 & 0 & \mathbbm{1}_{D}\\
        0 & \mathbbm{1}_{D} & 0
    \end{array}
\right),~~~~g:=\alpha_{g}^{\mathrm{T}}\alpha_{g}, 
\end{align}
where $\eta=\mathrm{diag}(\mathbbm{1}_{16+D},-\mathbbm{1}_{D})$ and $\alpha_{g}$ denotes a set of the basis of a 16-dimensional even self-dual Euclidean lattice $\Gamma_{16}$, corresponding to the $E_8\times E_8$ or $\mathrm{Spin}(32)/\mathbb{Z}_2$ lattice. For example, we take $\alpha_{g}$ for the $E_8\times E_8$ lattice as
\begin{align}
    \alpha_g:=\left(
    \begin{array}{cc}
        \alpha(E_8) & 0 \\
        0 & \alpha(E_8)
    \end{array}
    \right),
\end{align}
where $\alpha(E_8)$ can be taken as
\begin{align}\label{alpha_E8}
    \alpha(E_8)=\left(
    \begin{array}{cccccccc}
        1 & 0 & 0 & 0 & 0 & 0 & -\frac{1}{2} & 0 \\
        -1 & 1 & 0 & 0 & 0 & 0 & -\frac{1}{2} & 0 \\
        0 & -1 & 1 & 0 & 0 & 0 & -\frac{1}{2} & 0 \\
        0 & 0 & -1 & 1  & 0 & 0 & -\frac{1}{2} & 0 \\
        0 & 0 & 0 & -1 & 1 & 0 & -\frac{1}{2} & 0 \\
        0 & 0 & 0 & 0 & -1 & 1 & -\frac{1}{2} & 1 \\
        0 & 0 & 0 & 0 & 0 & 1 & -\frac{1}{2} & -1 \\
        0 & 0 & 0 & 0 & 0 & 0 & -\frac{1}{2} & 0
    \end{array}
    \right).
\end{align}
Then $g=\alpha_{g}^{\mathrm{T}}\alpha_{g}$ is the Cartan matrix of $E_8\times E_8$ written as
\begin{align}
    g:=\left(
    \begin{array}{cc}
        g(E_8) & 0 \\
        0 & g(E_8)
    \end{array}
    \right)
    =\left(
    \begin{array}{cc}
        \alpha(E_8)^{\mathrm{T}}\alpha(E_8) & 0 \\
        0 & \alpha(E_8)^{\mathrm{T}}\alpha(E_8)
    \end{array}
    \right),
\end{align}
where $g(E_8)$ can be taken as
\begin{align}
    g(E_8)=\left(
    \begin{array}{cccccccc}
        2 & -1 & 0 & 0 & 0 & 0 & 0 & 0 \\
        -1 & 2 & -1 & 0 & 0 & 0 & 0 & 0 \\
        0 & -1 & 2 & -1 & 0 & 0 & 0 & 0 \\
        0 & 0 & -1 & 2  & -1 & 0 & 0 & 0 \\
        0 & 0 & 0 & -1 & 2 & -1 & 0 & -1 \\
        0 & 0 & 0 & 0 & -1 & 2 & -1 & 0 \\
        0 & 0 & 0 & 0 & 0 & -1 & 2 & 0 \\
        0 & 0 & 0 & 0 & -1 & 0 & 0 & 2
    \end{array}
    \right).
\end{align}
The product of two vectors in the Narain lattice is defined as
\begin{align}\label{inner_product}
    P_{1}^{\mathrm{T}}\eta P_{2}
    =\hat{N}_{1}^{\mathrm{T}}\hat{\eta}\hat{N}_{2}
    =\Pi_{1}^{\mathrm{T}}\Pi_{2}
    +w_{1}^{\mathrm{T}}n_{2}
    +n_{1}^{\mathrm{T}}w_{2}\in\mathbb{Z},
\end{align}
where $P_{i}=\mathcal{E}\hat{N}_{i}\in\Gamma_{16+D,D}$ and we define $\Pi_{i}=\alpha_{g}q_{i}\in\Gamma_{16}$ for $i=1,2$. In particular, we obtain
\begin{align}\label{length_p}
    P^{\mathrm{T}}\eta P
    =\hat{N}^{\mathrm{T}}\hat{\eta}\hat{N}
    =\Pi^{\mathrm{T}}\Pi
    +2w^{\mathrm{T}}n
    \in2\mathbb{Z},~~~\Pi:=\alpha_{g}q.
\end{align}

The generalized metric $\mathcal{H}$ of the Narain lattice is expressed by the metric $G$, the anti-symmetric tensor field $B$ and the Wilson lines $A$ as follows:
\begin{align}
    \mathcal{H}:=\mathcal{E}^{\mathrm{T}}\mathcal{E}=
    \begin{pmatrix}
        g+ \alpha'\alpha_{g}^{\mathrm{T}}A G^{-1} A^\mathrm{T}  \alpha_\mathrm{g} & 
        \alpha_\mathrm{g}^\mathrm{T} A (\mathbbm{1}_2 + G^{-1} C) & - \alpha'\alpha_\mathrm{g}^\mathrm{T} A G^{-1}\\
        (\mathbbm{1}_2 + C^\mathrm{T} G^{-1}) A^\mathrm{T} \alpha_\mathrm{g} & \frac{1}{\alpha'}\left(G + A^\mathrm{T} A + C^\mathrm{T} G^{-1}C\right) & -C^\mathrm{T} G^{-1}\\
        -\alpha'G^{-1} A^\mathrm{T} \alpha_\mathrm{g} & -G^{-1} C & \alpha'G^{-1}
    \end{pmatrix}.
\end{align}
where $C:=B+\frac{\alpha'}{2}A^{\mathrm{T}}A$.

The outer automorphism of the Narain lattice, which is known as the T-duality group of toroidally compactified heterotic strings, is given by
\begin{align}\label{T-duality_group}
    O(16+D,D,\mathbb{Z}):=\left\lbrace \hat{\Sigma}\in\mathrm{GL}(16+2D,\mathbb{Z})\left| ~\hat{\Sigma}^{\mathrm{T}}\hat{\eta}\hat{\Sigma}=\hat{\eta} \right. \right\rbrace.
\end{align}
In order for the product of the Narain vectors \eqref{inner_product} to be invariant under the outer automorphisms, $\hat{\Sigma}\in O(16+D,D,\mathbb{Z})$ should act on the Narain vielbein $\mathcal{E}$ as 
\begin{align}\label{E_transformation}
    \mathcal{E} \stackrel{\hat{\Sigma}}{\longrightarrow} \mathcal{E}~\hat{\Sigma}^{-1}.
\end{align}
Here we have used $\hat{\Sigma}^{-1}$ as in \cite{Baur:2019iai,Baur:2020yjl,Nilles:2021glx} for the latter convenience.\footnote{One can prove $\hat{\Sigma}^{-1}\in O(16+D,D,\mathbb{Z})$ if $\hat{\Sigma}\in O(16+D,D,\mathbb{Z})$.}

We now focus on the $D=2$ case and the following configuration of Wilson lines $A$
\begin{align}\label{WL}
    A= \begin{pmatrix}a_1 & a_2\\-a_1&-a_2\\0&0\\ \vdots & \vdots\\ 0 & 0 \end{pmatrix},
\end{align}
which corresponds to the direction of the simple root $(1,-1,0,\ldots,0)^{\mathrm{T}}$ of $E_8\times E_8$ or $SO(32)$. We can then define the following moduli $(T,U,Z)$:
\begin{subequations}
    \begin{align}
        T:&=\frac{1}{\alpha'}\left(B_{12}+i\sqrt{\mathrm{det}G}\right) +a_1(-a_2+Ua_1),\\
        U:&=\frac{1}{G_{11}}\left(G_{12}+i\sqrt{\mathrm{det}G}\right),\\
        Z:&=-a_2+Ua_1.
    \end{align}
\end{subequations}
We define the transformation of the moduli $(T,U,Z)$ under the modular transformation from $O(16+2,2,\mathbb{Z})$ through the generalized metric $\mathcal{H}$ as \cite{Baur:2019iai,Baur:2020yjl,Nilles:2021glx}
\begin{align}\label{transformation_moduli}
    \mathcal{H}(T,U,Z)\stackrel{\hat{\Sigma}}{\longrightarrow}\hat{\Sigma}^{-\mathrm{T}}\mathcal{H}(T,U,Z)\hat{\Sigma}^{-1}:=\mathcal{H}(T',U',Z').
\end{align}
Using \eqref{transformation_moduli}, one can identify the correspondence between the Siegel modular group $\mathrm{Sp}(4,\mathbb{Z})$ and the T-duality group $SO(16+D,D,\mathbb{Z})$. The elements we focus on are listed in \cite{Baur:2020yjl}.
\begin{itemize}
    \item Mirror transformation:
	\begin{align}\label{mirror_t-duality}
            \hat{M}=
            \begin{pmatrix}
            \mathbbm{1}_{16}&0&0&0&0\\
            0&0&0&-1&0\\0&0&1&0&0\\0&-1&0&0&0\\0&0&0&0&1
            \end{pmatrix}.
	\end{align}
        From \eqref{transformation_moduli}, we obtain
        \begin{align}
            T\leftrightarrow U,~~~Z\leftrightarrow Z.
        \end{align}
        These transformations correspond to the mirror transformation \eqref{mirror_SP} of $\mathrm{Sp}(4,\mathbb{Z})$.
    
    \item Modular group of the complex structure modulus:
        \begin{align}\label{complex_t-duality}
            \hat{C}_{\mathrm{T}}=
            \begin{pmatrix}
            \mathbbm{1}_{16}&0&0&0&0\\
            0&1&-1&0&0\\0&0&1&0&0\\0&0&0&1&0\\0&0&0&1&1
            \end{pmatrix},~~~
            \hat{C}_{\mathrm{S}}=
            \begin{pmatrix}
            \mathbbm{1}_{16}&0&0&0&0\\
            0&0&-1&0&0\\0&1&0&0&0\\0&0&0&0&-1\\0&0&0&1&0
            \end{pmatrix}.
	\end{align}
        Using \eqref{transformation_moduli}, we get
        \begin{subequations}
            \begin{align}
                T&\xrightarrow{\hat{C}_{\mathrm{T}}}T,\\
                U&\xrightarrow{\hat{C}_{\mathrm{T}}}U + 1,\\
                Z&\xrightarrow{\hat{C}_{\mathrm{T}}}Z,
            \end{align}
        \end{subequations}
        and
        \begin{subequations}
            \begin{align}
                T&\xrightarrow{\hat{C}_{\mathrm{S}}}T-\frac{Z^2}{U},\\
                U&\xrightarrow{\hat{C}_{\mathrm{S}}}
                -\frac{1}{U},\\
                Z&\xrightarrow{\hat{C}_{\mathrm{S}}}-\frac{Z}{U},
            \end{align}
        \end{subequations}
         which correspond to the $\mathrm{T}$- and $\mathrm{S}$-transformations for the complex structure moduli of $\mathrm{Sp}(4,\mathbb{Z})$ given in \eqref{complex_SP}. 
	
    \item Modular group of the K\"ahler modulus:
	\begin{align}\label{Kahler_t-duality}
            \hat{K}_{\mathrm{T}}  :=  \hat{M}\, \hat{C}_{\mathrm{T}}\,\hat{M}^{-1} = \begin{pmatrix}
            \mathbbm{1}_{16}&0&0&0&0\\
            0&1&0&0&0\\0&0&1&0&0\\0&0&1&1&0\\0&-1&0&0&1
            \end{pmatrix},~~
            \hat{K}_{\mathrm{S}}  :=  \hat{M}\, \hat{C}_{\mathrm{S}}\,\hat{M}^{-1} = \begin{pmatrix}
            \mathbbm{1}_{16}&0&0&0&0\\
            0&0&0&0&1\\0&0&0&-1&0\\0&0&1&0&0\\0&-1&0&0&0
            \end{pmatrix}.
	\end{align}
        Similarly by \eqref{transformation_moduli}, it is verified that these induce the $\mathrm{T}$- and $\mathrm{S}$-transformations for the K\"ahler structure moduli of $\mathrm{Sp}(4,\mathbb{Z})$ given in \eqref{Kahler_SP}. 
        
    \item Wilson line shifts:
	\begin{align}\label{WLshift_t-duality}
            \hat{W}(\Delta A)  := 
            \left(\begin{array}{ccc}
            \mathbbm{1}_{16}&-\Delta A& 0\\
            0 & \mathbbm{1}_{2}& 0\\
            \Delta A^{\mathrm{T}}g & -\frac{1}{2}\Delta A^{\mathrm{T}} g \Delta A & \mathbbm{1}_{2}
            \end{array}\right).
	\end{align}
        We focus on the following configuration of $\Delta A$ as in \cite{Baur:2020yjl}:
        \begin{align}\label{WL_lm}
            \Delta A= \begin{pmatrix}m & \ell\\0&0\\ \vdots & \vdots\\ 0 & 0 \end{pmatrix}~~~(m,\ell\in \mathbb{Z}),
        \end{align}
        and we define $\hat{W}\begin{psmallmatrix}
        \ell\\
        m
    \end{psmallmatrix}$ as $\hat{W}(\Delta A)$ with \eqref{WL_lm}. From \eqref{transformation_moduli}, we get
        \begin{subequations}
            \begin{align}
                a_1&\xrightarrow{\hat{W}(m,\ell)}a_1+m,\\
                a_2&\xrightarrow{\hat{W}(m,\ell)}a_2+\ell,\\
                B_{12}&\xrightarrow{\hat{W}(m,\ell)}B_{12}+a_1\ell-a_2 m,
            \end{align}
        \end{subequations}
        which reproduces the $\mathrm{Sp}(4,\mathbb{Z})$ transformation given in \eqref{WLshift_SP}.

    \item $\mathcal{CP}$-like transformation:
        \begin{align}\label{CP_t-duality}
            \hat{\Sigma}_{*} := 
            \begin{pmatrix}-\mathbbm{1}_{16}&0&0&0&0\\0&-1&0&0&0\\0&0&1&0&0\\0&0&0&-1&0\\0&0&0&0&1
            \end{pmatrix}.
        \end{align}
        This leads to the following transformation by \eqref{transformation_moduli}:
        \begin{subequations}
            \begin{align}
                T&\xrightarrow{\hat{\Sigma}_{*}}-\bar{T},\\
                U&\xrightarrow{\hat{\Sigma}_{*}}-\bar{U},\\
                Z&\xrightarrow{\hat{\Sigma}_{*}}-\bar{Z}.
            \end{align}
        \end{subequations}
\end{itemize}
In summary, we obtain the correspondence:
\begin{subequations}
    \begin{align}
        &M_{(\mathbbm{1}_2,\mathrm{T})}
        \leftrightarrow\hat{C}_{\mathrm{T}},\qquad
        M_{(\mathbbm{1}_2,\mathrm{S})}
        \leftrightarrow\hat{C}_{\mathrm{S}},\qquad
        M_{\times}\leftrightarrow\hat{M},~\\
        &M_{(\mathrm{T},\mathbbm{1}_2)}
        \leftrightarrow\hat{K}_{\mathrm{T}},\qquad
        M_{(\mathrm{S},\mathbbm{1}_2)}
        \leftrightarrow\hat{K}_{\mathrm{S}},\qquad
        M\begin{psmallmatrix}
        \ell\\
        m
    \end{psmallmatrix}\leftrightarrow\hat{W}\begin{psmallmatrix}
        \ell\\
        m
    \end{psmallmatrix}.
\end{align}
\end{subequations}

\section{T-duality of non-supersymmetric strings}
\label{sec:T-duality}

In this section, we briefly review the construction of non-supersymmetric string theories, which is so-called the stringy Scherk-Schwarz mechanism, and their T-dualities.

\subsection{Construction of non-supersymmetric heterotic string theories}
\label{sec:construction}
The non-supersymmetric string theory can be constructed from the parent supersymmetric string theory by $\mathbb{Z}_2$ freely acting orbifold. The $\mathbb{Z}_2$ shift action we focus on in this paper is expressed as an eigenvalue $(-1)^{F}\exp(2\pi iP^{\mathrm{T}}\eta\delta)$ for a state with internal momenta $P$.\footnote{One can introduce another $\mathbb{Z}_2$ action which induces the rank reductions of the gauge symmetry \cite{Nakajima:2023zsh,DeFreitas:2024ztt,Hamada:2024cdd}.} Here $F$ is the spacetime fermion number and $\delta$ is called the shift vector in the Narain lattice satisfying $2\delta\in\Gamma_{16+D,D}$. In order to construct the non-supersymmetric models with this twist, it is convenient to split the Narain lattice $\Gamma_{16+D,D}$ into the following two subsets $\Gamma_{16+D,D}^{+}$ and $\Gamma_{16+D,D}^{-}$:
\begin{subequations}\label{Gamma_splitting}
    \begin{align}
        &\Gamma_{16+D,D}^{+}(\delta)=\left\lbrace P\in\Gamma_{16+D,D}~\left|~P^{\mathrm{T}}\eta\delta\in\mathbb{Z} \right.\right\rbrace,\\
        &\Gamma_{16+D,D}^{-}(\delta)=\left\lbrace P\in\Gamma_{16+D,D}~\left|~P^{\mathrm{T}}\eta\delta\in\mathbb{Z}+\frac{1}{2} \right.\right\rbrace.
    \end{align}
\end{subequations}
After modding out the supersymmetric heterotic strings \eqref{SUSY_pairing} by the $\mathbb{Z}_2$ shift action, we get the untwisted sectors in which the pairing of states is expressed as \cite{Ginsparg:1986wr}
\begin{align}
    \left(\Gamma_{16+D,D}^{+};\bar{v}\right),
    ~\left(\Gamma_{16+D,D}^{-};\bar{s}\right).
\end{align}
The modular invariance of the partition functions of the non-supersymmetric string theories imposes the condition $\delta^2\in\mathbb{Z}$ and requires the twisted sectors to be added. For the case with $\delta^2\in2\mathbb{Z}+1$, the pairing of states in the twisted sectors is expressed as
\begin{align}
    \left(\Gamma_{16+D,D}^{+}+\delta;\bar{o}\right),
    ~\left(\Gamma_{16+D,D}^{-}+\delta;\bar{c}\right),
\end{align}
where $o$ and $c$ represent the scalar and co-spinor conjugacy classes of $SO(8)$ respectively, and for the case with $\delta^2\in2\mathbb{Z}$,
\begin{align}
    \left(\Gamma_{16+D,D}^{-}+\delta;\bar{o}\right),
    ~\left(\Gamma_{16+D,D}^{+}+\delta;\bar{c}\right).
\end{align}
The one-loop partition function of the non-supersymmetric strings can be written as
\begin{align}\label{partition function}
    &Z_{\cancel{\mathrm{SUSY}}}^{(10-D)}(\tau)=\frac{1}{\tau_2^{\frac{8-D}{2}}\eta^{24}\bar{\eta}^{12}}
    \left\lbrace
    \bar{V}_{8} \sum_{P\in\Gamma_{16+D,D}^{+}}q^{\frac{1}{2}P_L^2}\bar{q}^{\frac{1}{2}P_R^2}
    -\bar{S}_{8} \sum_{P\in\Gamma_{16+D,D}^{-}}q^{\frac{1}{2}P_L^2}\bar{q}^{\frac{1}{2}P_R^2}\right.\nonumber\\
    &\left.\hspace{100pt}
    +\bar{O}_{8} \sum_{P\in\Gamma_{16+D,D}^{\pm}+\delta}q^{\frac{1}{2}P_L^2}\bar{q}^{\frac{1}{2}P_R^2}
    -\bar{C}_{8}  \sum_{P\in\Gamma_{16+D,D}^{\mp}+\delta}q^{\frac{1}{2}P_L^2}\bar{q}^{\frac{1}{2}P_R^2}
    \right\rbrace,
\end{align} 
where the upper (lower) signs in the twisted sectors apply for $\delta^2$ odd (even).

Since $2\delta$ is an element of $\Gamma_{16+D,D}$, the shift vector can be expressed as
\begin{align}\label{shift_vector}
    \delta_{(\hat{Z})}=\frac{1}{2}\mathcal{E}\hat{Z},~~\hat{Z}=\begin{pmatrix}\hat{q}\\ \hat{w}\\ \hat{n}\end{pmatrix} \in \mathbb{Z}^{16+2D}.
\end{align}
From the definitions of $\Gamma_{16+D,D}^{\pm}$, the two shift vectors labeled by $\hat{Z}$ and $\hat{Z}'$ give the same splitting of the Narain lattice if $\hat{V}\in\mathbb{Z}^{16+2D}$ exists such that $\hat{Z}'=\hat{Z}+2\hat{V}$. Thus we can only focus on the $\hat{Z}$ whose components take either 0 or 1, except for $\hat{Z}=(0^{16+2D})$. In addition, from the condition $\delta^{2}\in\mathbb{Z}$, $\hat{Z}$ must satisfy
\begin{align}
\label{Z_condition}
    \hat{Z}^{\mathrm{T}}\hat{\eta}\hat{Z}=0~(\text{mod 4}).
\end{align}
Equivalently, we can rewrite this condition as
\begin{align}
\label{Z_condition2}
    \hat{\Pi}^{\mathrm{T}}\hat{\Pi}
    +2\hat{w}^{\mathrm{T}}\hat{n}=0~(\text{mod 4}),
\end{align}
where we define $\hat{\Pi}=\alpha_{g}\hat{q}\in\Gamma_{16}$. For the simplest case of $D=0$, the condition \eqref{Z_condition2} leads to $\delta_{16}^2\in\mathbb{Z}$ where $\delta_{16}:=\hat{\Pi}/2$. In this case, the vectors $\hat{Z}$ are equivalent not only up to the shift by $2\hat{V}$ but also up to permutations of the components. Thus we obtain the three or four allowed shift vectors for $\hat{\Pi}$ living in the $E_8\times E_8$ or $Spin(32)/\mathbb{Z}$ lattice, which correspond to the well-known ten-dimensional non-supersymmetric heterotic strings \cite{Dixon:1986iz}. For the case of $D\geq1$, the string models with $\hat{w}=\hat{n}=0$ can be understood as the ones obtained from the ten-dimensional non-supersymmetric heterotic strings by toroidal compactifications while the other models in general are so-called interpolating models (See e.g. \cite{Itoyama:1986ei,Itoyama:1987rc,Itoyama:2019yst,Itoyama:2020ifw,Itoyama:2021fwc,Koga:2022qch}). 

\subsection{T-duality and congruence subgroup}
\label{sec:congruence}
Since the non-supersymmetric string theory has the more complicated partitioning of the states than the supersymmetric string, the T-duality group of the non-supersymmetric string is obtained by constraining \eqref{T-duality_group}. To be more precise, in the non-supersymmetric strings, the duality transformation \eqref{E_transformation} must maintain the partitioning of the states:
\begin{align}\label{partitioning_condition}
  P^{\mathrm{T}}\eta\delta=P'^{\mathrm{T}}\eta\delta~~~(\mathrm{mod}~1)~~~~\text{for any } P,  
\end{align}
where $P'$ is defined as
\begin{align}
    P=\mathcal{E}\hat{N}\stackrel{\hat{\Sigma}}{\longrightarrow} P'=\mathcal{E}~\hat{\Sigma}^{-1}\hat{N}.
\end{align}
Substituting these and \eqref{shift_vector} into \eqref{partitioning_condition}, we get the condition for $\hat{Z}$:\footnote{The condition \eqref{condition_T-duality} can be also obtained when the Narain vielbein $E$ transforms as $E\to E~\hat{\Sigma}$.}
\begin{align}\label{condition_T-duality}
    \hat{\Sigma}\hat{Z}=\hat{Z}~~(\mathrm{mod}~2).
\end{align}
In conclusion, the T-duality group of the non-supersymmetric strings constructed by using the shift-vector $\delta_{(\hat{Z})}$ is given as the congruence subgroup of T-duality group \cite{Itoyama:2021itj}
\begin{align}\label{nonsusy_duality}
D_{(\hat{Z})}\left(16+D,D \right) =\left\lbrace \hat{\Sigma}\in O(16+D,D,\mathbb{Z}) \left| ~\hat{\Sigma}\hat{Z}=\hat{Z}~(\text{mod 2})\right. \right\rbrace.
\end{align}

\section{Classification of modular symmetries on $T^2$}
\label{sec:classification}
Here we focus on the two-dimensional toroidal compactification in non-supersymmetric heterotic strings, i.e., the case with $d_{L}=16+2,~d_{R}=2$. In the following, using \eqref{nonsusy_duality}, we specify the elements of $\mathrm{O}_{\hat\eta}(16+2,2,\mathbb{Z})$ listed in Subsection \ref{sec:Narainformulation} which remain symmetries in the non-supersymmetric strings.

\subsection{Remaining symmetries depending on $\hat{Z}$}
\label{sec:remaining_elements}
In the following, we derive the conditions of $\hat{Z}$ for each element listed in Subsection \ref{sec:Narainformulation} to remain a symmetry:
\begin{itemize}
    \item \uline{Mirror transformation \eqref{mirror_t-duality}:}
    
        From the condition \eqref{nonsusy_duality}, $\hat{M}$ is a symmetry in the models with $\hat{w}_{1}=\hat{n}_{1}$.
    
    \item \uline{Modular group of the complex structure modulus \eqref{complex_t-duality}:}
    
        One can find $\hat{C}_{\mathrm{T}}$ is a symmetry in the models with $\hat{w}_{2}=0,~\hat{n}_{1}=0$ while $\hat{C}_{\mathrm{S}}$ is with $\hat{w}_{1}=\hat{w}_{2},~\hat{n}_{1}=\hat{n}_{2}$.
	
    \item \uline{Modular group of the K\"ahler modulus \eqref{Kahler_t-duality}:}
    
        In the same way as in the complex structure modulus, $\hat{K}_{\mathrm{T}}$ is a symmetry in the models with $\hat{w}_{1}=0,~\hat{w}_{2}=0$ while $\hat{K}_{\mathrm{S}}$ is with $\hat{w}_{1}=\hat{n}_{2},~\hat{n}_{1}=\hat{w}_{2}$. These conditions can be obtained by the replacement $\hat{w}_{1}\leftrightarrow\hat{n}_{1}$ in the cases of $\hat{C}_{\mathrm{T}},~\hat{C}_{\mathrm{S}}$, respectively.
        
    \item \uline{Wilson line shifts \eqref{WLshift_t-duality}:}
    
        The action $\hat{W}(\Delta A)$ can be an element of $D_{(\hat{Z})}(16+2,2)$ if it satisfies both of the following conditions
        \begin{align}\label{condition for WL shift}
            \Delta A \hat{w}=0~(\text{mod~}2),~~\Delta A^{\mathrm{T}}g\left(\hat{q}-\frac{1}{2}\Delta A\hat{w}\right)=0~(\text{mod~}2).
        \end{align}
        In the case of \eqref{WL_lm}, the condition \eqref{condition for WL shift} leads to
        \begin{subequations}\label{condition for Delta A}
            \begin{align}
                &m\hat{w}_{1}+\ell\hat{w}_{2}=0~(\text{mod~}2),\\
                &2m\hat{q}_{1}-m\hat{q}_{2}-m^{2}\hat{w}_{1}-m\ell\hat{w}_{2}=0~(\text{mod~}2),\\
                &2\ell\hat{q}_{1}-\ell\hat{q}_{2}-m\ell\hat{w}_{1}-\ell^{2}\hat{w}_{2}=0~(\text{mod~}2),
            \end{align}
        \end{subequations}
        where we have used $g_{11}=2,~g_{12}=-1,~g_{1i}=0~(i\neq 1,2)$ since $g$ is the Cartan matrix of the $E_{8}\times E_{8}$ or $SO(32)$ gauge symmetry.
        From these conditions, it can be found that any $m$ and $\ell$ are allowed in the models with $(\hat{q}_{1},\hat{q}_{2};\hat{w}_{1},\hat{w}_{2})=(0,0;0,0),(1,0;0,0)$. In the other cases, $m$ and $\ell$ must be constrained. For example, the model constructed by $(\hat{q}_{1},\hat{q}_{2};\hat{w}_{1},\hat{w}_{2})=(1,0;1,0)$ or $(1,0;0,1)$ has the shift symmetry when $m$ or $\ell$ is even, respectively.

    \item \uline{$\mathcal{CP}$-like transformation \eqref{CP_t-duality}:}
    
        This cannot impose any constraints on $\hat{Z}$, that is, this is always a symmetry in the non-supersymmetric strings constructed in Section \ref{sec:construction}.
\end{itemize}
For the modular groups of the complex structure and K\"ahler moduli, we have only focused on the $\mathrm{T}$- and $\mathrm{S}$-transformations induced by $\hat{C}_{\mathrm{T}}, \hat{C}_{\mathrm{S}}$ and $\hat{K}_{\mathrm{T}},\hat{K}_{\mathrm{S}}$. However, considering the full $SL(2,\mathbb{Z})$ transformations in the non-supersymmetric strings, we can see the symmetry associated with these moduli as the congruence subgroups of the parent $SL(2,\mathbb{Z})$ groups. To do so, let us consider the following elements of $O(18,2,\mathbb{Z})$:
\begin{align}
    \hat{C}_{\gamma}:=
    \begin{pmatrix}
        \mathbbm{1}_{16}&0&0\\
        0&\gamma^{-1}&0\\0&0&\gamma^{\mathrm{T}}
    \end{pmatrix},
    ~~~
    \hat{K}_{\gamma}:=\hat{M}\, \hat{C}_{\gamma}\,\hat{M}^{-1},~~~\gamma\in SL(2,\mathbb{Z}).
\end{align}
We get the transformations \eqref{complex_SP} and \eqref{Kahler_SP} from these elements, which implies that there is a correspondence $M_{(\mathbbm{1}_2,\gamma_U)}\leftrightarrow\hat{C}_{\gamma}$ and $M_{(\gamma_T,\mathbbm{1}_2)}\leftrightarrow\hat{K}_{\gamma}$. One can easily check $\hat{C}_{\gamma=\mathrm{T}}=\hat{C}_{\mathrm{T}},~\hat{C}_{\gamma=\mathrm{S}}=\hat{C}_{\mathrm{S}}$ and $\hat{K}_{\gamma=\mathrm{T}}=\hat{K}_{\mathrm{T}},~\hat{K}_{\gamma=\mathrm{S}}=\hat{K}_{\mathrm{S}}$ by \eqref{STmatrix}. We obtain the condition from \eqref{nonsusy_duality} for $\hat{C}_{\gamma}$
\begin{align}\label{gamma_condition}
    \gamma \hat{w}=\hat{w}~(\text{mod~}2),~~~\gamma^{-\mathrm{T}} \hat{n}=\hat{n}~(\text{mod~}2),
\end{align}
and for $\hat{K}_{\gamma}$ obtained by interchanging $\hat{w}_1\leftrightarrow-\hat{n}_1$ from \eqref{gamma_condition}
\begin{align}
    \gamma \begin{pmatrix} -\hat{n}_1 \\ \hat{w}_2\end{pmatrix} =\begin{pmatrix} -\hat{n}_1 \\ \hat{w}_2\end{pmatrix}~(\text{mod~}2),~~~\gamma^{-\mathrm{T}} \begin{pmatrix} -\hat{w}_1 \\ \hat{n}_2\end{pmatrix}=\begin{pmatrix} -\hat{w}_1 \\ \hat{n}_2\end{pmatrix}~(\text{mod~}2).
\end{align}

\begin{center}
\begin{longtable}{|c||c||c|c|c|c|c|}
\caption{The elements of $D_{(\hat{Z})}(18,2)$ which depend on the choice of $\hat{Z}$.}
\label{tab:torus} \\

\hline $\#$ & $\left(\hat{q}_{1}\;\mathrm{mod}\;1,\hat{q}_{2};\hat{w}_{1},\hat{w}_{2};\hat{n}_{1},\hat{n}_{2} \right)$  & $\hat{C}_\gamma$ &$\hat{K}_\gamma$& $\hat{W}\begin{psmallmatrix}
        -1\\
        0
    \end{psmallmatrix}$&$\hat{M}$& $\hat{\mathcal{CP}}$\\ \hline 
\endfirsthead

\multicolumn{7}{c}%
{{\bfseries \tablename\ \thetable{} -- continued from previous page}} \\
\hline $\#$ & $\left(\hat{q}_{1}\;\mathrm{mod}\;1,\hat{q}_{2};\hat{w}_{1},\hat{w}_{2};\hat{n}_{1},\hat{n}_{2} \right)$ & $\hat{C}_\gamma$ &$\hat{K}_\gamma$& $\hat{W}\begin{psmallmatrix}
        -1\\
        0
    \end{psmallmatrix}$ &$\hat{M}$&$\hat{\mathcal{CP}}$\\ \hline 
\endhead

\hline \multicolumn{7}{|r|}{{Continued on next page}} \\ \hline
\endfoot

\hline 
\endlastfoot

    $1$ & $\left(0,0;0,0;0,0\right)$& $\mathrm{SL}(2,\mathbb{Z})$& $\mathrm{SL}(2,\mathbb{Z})$ & $m,\ell\in\mathbb{Z}$&\checkmark & \checkmark \\
    $2$ & $\left(0,0;0,0;0,1\right)$& $\Gamma_1(2)$ &$\Gamma_1(2)$&  $m,\ell\in\mathbb{Z}$&\checkmark & \checkmark \\
    $3$ & $\left(0,0;0,0;1,0\right)$& $\Gamma^1(2)$ &$\Gamma_1(2)$& $m,\ell\in\mathbb{Z}$&--- & \checkmark \\
    $4$ & $\left(0,0;0,0;1,1\right)$& $\Gamma_\vartheta$ & $\Gamma_1(2)$&$m,\ell\in\mathbb{Z}$&--- & \checkmark \\
    $5$ & $\left(0,0;0,1;0,0\right)$& $\Gamma^1(2)$ &$\Gamma^1(2)$&$\ell\in2\mathbb{Z}$&\checkmark & \checkmark \\
    $6$ & $\left(0,0;0,1;0,1\right)$& $\Gamma(2)$ &$\Gamma(2)$&  $\ell\in2\mathbb{Z}$&\checkmark & \checkmark \\
    $7$ & $\left(0,0;0,1;1,0\right)$& $\Gamma^1(2)$ &$\Gamma_\vartheta$& $\ell\in2\mathbb{Z}$&--- & \checkmark \\
    $8$ & $\left(0,0;0,1;1,1\right)$& $\Gamma(2)$ &$\Gamma(2)$& $\ell\in2\mathbb{Z}$&--- & \checkmark \\
    $9$ & $\left(0,0;1,0;0,0\right)$& $\Gamma_1(2)$ &$\Gamma^1(2)$& $m\in2\mathbb{Z}$&--- & \checkmark \\
    $10$ & $\left(0,0;1,0;0,1\right)$& $\Gamma_1(2)$ &$\Gamma_\vartheta$& $m\in2\mathbb{Z}$&--- & \checkmark \\
    $11$ & $\left(0,0;1,0;1,0\right)$& $\Gamma(2)$ &$\Gamma(2)$& $m\in2\mathbb{Z}$& \checkmark & \checkmark \\
    $12$ & $\left(0,0;1,0;1,1\right)$& $\Gamma(2)$ &$\Gamma(2)$&  $m\in2\mathbb{Z}$& \checkmark & \checkmark \\
    $13$ & $\left(0,0;1,1;0,0\right)$& $\Gamma_\vartheta$ & $\Gamma^1(2)$&$m+\ell\in2\mathbb{Z}$&--- & \checkmark \\
    $14$ & $\left(0,0;1,1;0,1\right)$& $\Gamma(2)$ &$\Gamma(2)$& $m+\ell\in2\mathbb{Z}$&--- & \checkmark \\
    $15$ & $\left(0,0;1,1;1,0\right)$& $\Gamma(2)$ &$\Gamma(2)$&  $m+\ell\in2\mathbb{Z}$& \checkmark & \checkmark \\
    $16$ & $\left(0,0;1,1;1,1\right)$& $\Gamma_\vartheta$ &$\Gamma_\vartheta$& $m+\ell\in2\mathbb{Z}$& \checkmark & \checkmark \\
    $17$ & $\left(0,1;0,0;0,0\right)$& $\mathrm{SL}(2,\mathbb{Z})$ & $\mathrm{SL}(2,\mathbb{Z})$&$m,\ell\in2\mathbb{Z}$& \checkmark & \checkmark \\
    $18$ & $\left(0,1;0,0;0,1\right)$& $\Gamma_1(2)$ &$\Gamma_1(2)$& $m,\ell\in2\mathbb{Z}$& \checkmark & \checkmark \\
    $19$ & $\left(0,1;0,0;1,0\right)$& $\Gamma^1(2)$ & $\Gamma_1(2)$& $m,\ell\in2\mathbb{Z}$&--- & \checkmark \\
    $20$ & $\left(0,1;0,0;1,1\right)$& $\Gamma_\vartheta$ & $\Gamma_1(2)$&$m,\ell\in2\mathbb{Z}$ &---& \checkmark \\
    $21$ & $\left(0,1;0,1;0,0\right)$& $\Gamma^1(2)$&$\Gamma^1(2)$& $m,\ell\in2\mathbb{Z}$& \checkmark & \checkmark \\
    $22$ & $\left(0,1;0,1;0,1\right)$& $\Gamma(2)$ &$\Gamma(2)$&  $m,\ell\in2\mathbb{Z}$& \checkmark & \checkmark \\
    $23$ & $\left(0,1;0,1;1,0\right)$& $\Gamma^1(2)$ &$\Gamma_\vartheta$&  $m,\ell\in2\mathbb{Z}$&--- & \checkmark \\
    $24$ & $\left(0,1;0,1;1,1\right)$& $\Gamma(2)$ &$\Gamma(2)$&  $m,\ell\in2\mathbb{Z}$&--- & \checkmark \\
    $25$ & $\left(0,1;1,0;0,0\right)$& $\Gamma_1(2)$ &$\Gamma^1(2)$&  $m,\ell\in2\mathbb{Z}$&--- & \checkmark \\
    $26$ & $\left(0,1;1,0;0,1\right)$& $\Gamma_1(2)$ &$\Gamma_\vartheta$&  $m,\ell\in2\mathbb{Z}$&--- & \checkmark \\
    $27$ & $\left(0,1;1,0;1,0\right)$& $\Gamma(2)$ &$\Gamma(2)$&  $m,\ell\in2\mathbb{Z}$& \checkmark & \checkmark \\
    $28$ & $\left(0,1;1,0;1,1\right)$& $\Gamma(2)$ &$\Gamma(2)$&  $m,\ell\in2\mathbb{Z}$& \checkmark & \checkmark \\
    $29$ & $\left(0,1;1,1;0,0\right)$& $\Gamma_\vartheta$ &$\Gamma^1(2)$&  $m,\ell\in2\mathbb{Z}$&--- & \checkmark \\
    $30$ & $\left(0,1;1,1;0,1\right)$& $\Gamma(2)$ &$\Gamma(2)$& $m,\ell\in2\mathbb{Z}$&--- & \checkmark \\
    $31$ & $\left(0,1;1,1;1,0\right)$& $\Gamma(2)$ &$\Gamma(2)$& $m,\ell\in2\mathbb{Z}$ & \checkmark& \checkmark \\
    $32$ & $\left(0,1;1,1;1,1\right)$& $\Gamma_\vartheta$ &$\Gamma_\vartheta$&  $m,\ell\in2\mathbb{Z}$& \checkmark & \checkmark \\

\end{longtable}
\end{center}
Using these conditions, we identify the elements of \eqref{nonsusy_duality} characterized by the congruence subgroup of the modular $SL(2,\mathbb{Z})$ group. We summarize the results including the other symmetries such as the Mirror transformation, Wilson line shift and ${\cal CP}$-like symmetry in Table \ref{tab:torus}. Conditions of $\ell,m$ are additional ones to $\ell,m\in\mathbb{Z}$. A check mark implies that the transformation is the symmetry in the non-supersymmetric heterotic strings labeled by $\left(\hat{q}_{1},\hat{q}_{2};\hat{w}_{1},\hat{w}_{2};\hat{n}_{1},\hat{n}_{2} \right)$.
Here $\Gamma(n)$ is the principal congruence subgroup of the modular group
\begin{align}
    \Gamma(n):=\left\lbrace \left. 
    \left(\begin{array}{cc}
    a & b \\
    c & d
    \end{array}
    \right) \in PSL(2,\mathbb{Z})~\right| ~a,d=1~(\text{mod}~n),~b,c=0~(\text{mod}~n) \right\rbrace,
\end{align}
and $\Gamma_1(n)$ and $\Gamma^1(n)$ are the Hecke congruence subgroup
\begin{align}
    \Gamma_{1}(n)&:=\left\lbrace \left. 
    \left(\begin{array}{cc}
    a & b \\
    c & d
    \end{array}
    \right) \in PSL(2,\mathbb{Z})~\right| ~a,d=1~(\text{mod}~n),~c=0~(\text{mod}~n) \right\rbrace,\\
    \Gamma^{1}(n)&:=\left\lbrace \left. \left(\begin{array}{cc}
    a & b \\
    c & d
    \end{array}
    \right) \in PSL(2,\mathbb{Z})~\right| ~a,d=1~(\text{mod}~n),~b=0~(\text{mod}~n) \right\rbrace,
\end{align}
and $\Gamma_{\vartheta}$ is the theta subgroup
\begin{align}
    \Gamma_{\vartheta}&:=\left\lbrace \left. \left(\begin{array}{cc}
    a & b \\
    c & d
    \end{array}
    \right) \in PSL(2,\mathbb{Z})~\right| ~ac=0~(\text{mod}~2),~bd=0~(\text{mod}~2) \right\rbrace.
\end{align}

Note that $\hat{q}_i~(i=3,\ldots16)$ are irrelevant to identifying the symmetry in non-supersymmetric models with the Wilson lines \eqref{WL_lm}. However, if one needs to specify a non-supersymmetric model, $\hat{q}\in\mathbb{Z}^{16}$ and \eqref{Z_condition} must be considered. In the forthcoming subsection, we treat concrete examples related to the $SO(16)\times SO(16)$ non-supersymmetric heterotic string, which is tachyon-free in ten dimensions.

\subsection{Gauge symmetry enhancement}
\label{sec:enhancement}
In this subsection, we see the relation between the modular symmetry and the gauge symmetry enhancement in the $E_8\times E_8$ and $SO(16)\times SO(16)$ heterotic strings on $T^2$. 

\subsubsection{The $E_8\times E_8$ string}
\label{sec:E8E8}
As a preliminary step, let us identify the gauge group of the $E_8\times E_8$ supersymmetric heterotic string on $T^2$ with the Wilson lines \eqref{WL}. The mass formula and level-matching condition are given by
\begin{subequations}
    \begin{align}
        m^2&=P_L^2+p_R^2+2\left( N_L + N_R -
        \begin{cases}
            1 & \text{R sector} \\
            \frac{3}{2} & \text{NS sector} 
        \end{cases}
        \right),\\
        0&=P_L^2-p_R^2+2\left( N_L - N_R -
        \begin{cases}
            1 & \text{R sector} \\
            \frac{1}{2} & \text{NS sector} 
        \end{cases}
        \right),
    \end{align}
\end{subequations}
where $N_L$ and $N_R$ are left- and right-moving oscillator numbers and the internal momenta $P=\mathcal{E}\hat{N}$ are given by
\begin{align}
    P=\binom{P_{\mathrm{L}}}{p_{\mathrm{R}}}=
    \left(\begin{array}{c}
    \alpha_{\mathrm{g}} q+A w\\
    \sqrt{\frac{\alpha'}{2}} e^{-\mathrm{T}}\left(-A^{\mathrm{T}} \alpha_{\mathrm{g}} q+\frac{1}{\alpha'}\left(2G-E\right) w+n\right) \\
    \sqrt{\frac{\alpha'}{2}} e^{-\mathrm{T}}\left(-A^{\mathrm{T}} \alpha_{\mathrm{g}} q- \frac{1}{\alpha'}Ew+n\right) 
    \end{array}\right),~~~E:=G+C.
\end{align}
The lowest lying states have $N_R=\frac{1}{2}$ or $N_R=0$ in the NS or R sector. These can be massless if
\begin{align}
    P_L^2+2(N_L-1)=0,~~~p_R=0.
\end{align}
The latter condition leads to
\begin{align}
    n=A^{\mathrm{T}}\Pi+\frac{1}{\alpha'}Ew\in\mathbb{Z}^{2}.
\end{align}
In addition, using $P^{\mathrm{T}}\eta P=P_L^2-p_R^2$ and \eqref{length_p}, we obtain 
\begin{align}
    P_L^2=\Pi^{\mathrm{T}}\Pi
    +2w^{\mathrm{T}}n.
\end{align}
Whatever the value of the moduli takes, there exist massless states satisfying $\Pi=0,w=0$ and $n=0$ with $N_L=1$, which correspond to the gravity multiplet plus gauge multiplets of $U(1)^{16+2}$. Moreover, in the case with the specific values of the moduli, the additional massless states appear with $P_L^2=2$ and $N_L=0$. These $P_L$ correspond to roots of a product of ADE types of Lie groups $G_r$ whose total rank $r$ satisfies $r\leq16+2$. Then the gauge group is enhanced from $U(1)^{16+2}$ to $G_r\times U(1)^{16+2-r}$. If the gauge group has no $U(1)$ factors, it is called the maximally enhanced gauge group.

The conditions $P_L^2=2$ and $p_R=0$ with the Wilson lines \eqref{WL} lead to
\begin{subequations}\label{massless_condition}
    \begin{align}
        &\Pi^{\mathrm{T}}\Pi
        +2\left(w_1n_1+w_2n_2\right)=2,\\
        &n_1=a_1(\Pi_1-\Pi_2)+\frac{1}{\alpha'}\left(E_{11}w_1+E_{12}w_2\right)\in\mathbb{Z},\\
        &n_2=a_2(\Pi_1-\Pi_2)+\frac{1}{\alpha'}\left(E_{21}w_1+E_{22}w_2\right)\in\mathbb{Z},
    \end{align}
\end{subequations}
with $E_{ij}=G_{ij}+C_{ij}~(i,j=1,2)$.
Note that from \eqref{alpha_E8}, $\Pi_1-\Pi_2=2q_1-q_2\in\mathbb{Z}$. Several cases can be considered depending on the moduli $a_1,a_2$ and $E$. Let us first consider the case with general values of $a_1,a_2$ and $E$. The massless conditions \eqref{massless_condition} give
\begin{align}\label{general_massless_solution}
    \Pi^2=2,~\Pi_1=\Pi_2,~w_1=w_2=n_1=n_2=0.
\end{align}
These imply that the gauge group is $E_8\times E_7\times U(1)^3$.
In the following, we consider the specific configurations of the moduli leading to the fixed points of the $\mathrm{Sp}(4,\mathbb{Z})$ group, which are expected to give the maximally enhanced gauge groups. It is known that there are six fixed points (dimension 0) of $\mathrm{Sp}(4,\mathbb{Z})$, but one can easily see that two of them do not lead to maximal enhancements, since the moduli $a_1,a_2$ and $E$ in those cases do not take rational numbers. We then focus on the four cases with rational moduli. For convenience, we define 
\begin{align}
    E_1:=\begin{pmatrix}
            1 & 0\\
            0 & 1
        \end{pmatrix},
    ~~~E_2:=\begin{pmatrix}
            1 & -1\\
            0 & 1
        \end{pmatrix}.
\end{align}

\begin{enumerate}
    \item $a_1=0,~a_2=0$ and $E=\alpha'E_1$:
    
        These moduli correspond to
        \begin{align}
            \begin{pmatrix}
                U & Z\\
                Z & T
            \end{pmatrix}
            =
            \begin{pmatrix}
                i & 0\\
                0 & i
            \end{pmatrix}.
        \end{align}
        The massless conditions \eqref{massless_condition} lead to 
        \begin{subequations}
            \begin{align}
                &\Pi^2=2,~w_1=w_2=n_1=n_2=0,\\
                &\Pi=0,~w_1=n_1=\pm1,~w_2=n_2=0,
                \label{example_not_solution}\\
                &\Pi=0,~w_1=n_1=0,~w_2=n_2=\pm1.
            \end{align}
        \end{subequations}
        Then the gauge group is enhanced to $E_8\times E_8\times SU(2)^2$.
        
    \item $a_1=0,~a_2=0$ and $E=\alpha'E_2$:

       These moduli correspond to
        \begin{align}
            \begin{pmatrix}
                U & Z\\
                Z & T
            \end{pmatrix}
            =
            \begin{pmatrix}
                \omega & 0\\
                0 & \omega
            \end{pmatrix},~~~\omega := e^{2\pi i/3}.
        \end{align}
       From \eqref{massless_condition}, we get
       \begin{subequations}
            \begin{align}
                &\Pi^2=2,~w_1=w_2=n_1=n_2=0,\\
                &\Pi=0,~w_1=n_1=\pm1,~w_2=n_2=0,\\
                &\Pi=0,~w_1=\pm1,~n_1=0,~w_2=n_2=\pm1,\\
                &\Pi=0,~w_1=0,~n_1=\pm1,~w_2=n_2=\mp1.
            \end{align}
        \end{subequations}
        These solutions give the gauge group $E_8\times E_8\times SU(3)$.

    \item $a_1=\frac{1}{2},~a_2=0$ and $E=\alpha'E_1$:

       These moduli correspond to
        \begin{align}
            \begin{pmatrix}
                U & Z\\
                Z & T
            \end{pmatrix}
            =
            \frac{i}{\sqrt{3}}
            \begin{pmatrix}
                 2 & 1\\
                 1 & 2
            \end{pmatrix}.
        \end{align}
        Using \eqref{massless_condition}, we obtain
        \begin{subequations}
            \begin{align}
                &\Pi^2=2,~\Pi_1=\Pi_2,~w_1=w_2=n_1=n_2=0,\\
                &\Pi^2=2,~\Pi_1-\Pi_2=\pm2,~w_1=0,~n_1=\pm1,~w_2=n_2=0,\\
                &\Pi^2=2,~\Pi_1-\Pi_2=\pm2,~w_1=\mp1,~n_1=0,~w_2=n_2=0,\\
                &\Pi=0,~w_1=n_1=\pm1,~w_2=n_2=0,\\
                &\Pi=0,~w_1=n_1=0,~w_2=n_2=\pm1.
            \end{align}
        \end{subequations}
        These lead to the gauge group $E_8\times E_7\times SU(3)\times SU(2)$.

    \item $a_1=\frac{1}{2},~a_2=0$ and $E=\alpha'E_2$:\footnote{More precisely, this configuration is dual to the following fixed point of $\mathrm{Sp}(4,\mathbb{Z})$
    \begin{align}
        \begin{pmatrix}
                U & Z\\
                Z & T
            \end{pmatrix}
            =
            \begin{pmatrix}
                \tilde{\eta} & \frac{1}{2}(\tilde{\eta}-1)\\
                \frac{1}{2}(\tilde{\eta}-1)& \tilde{\eta}
            \end{pmatrix},
    \end{align}
     which also leads to the gauge group $E_8\times E_7\times SU(4)$, as mentioned in \cite{Font:2020rsk}.}

       These moduli correspond to
        \begin{align}
            \begin{pmatrix}
                U & Z\\
                Z & T
            \end{pmatrix}
            =
            \begin{pmatrix}
                \tilde{\eta}-1 & \frac{1}{2}(\tilde{\eta}-1)\\
                \frac{1}{2}(\tilde{\eta}-1)& \tilde{\eta}-1
            \end{pmatrix},~~~\tilde{\eta} := \frac{1}{3}(1 + 2\sqrt{2} i).
        \end{align}
        The conditions \eqref{massless_condition} give
       \begin{subequations}
            \begin{align}
                &\Pi^2=2,~\Pi_1=\Pi_2,~w_1=w_2=n_1=n_2=0,\\
                &\Pi^2=2,~\Pi_1-\Pi_2=\pm2,~w_1=0,~n_1=\pm1,~w_2=n_2=0,\\
                &\Pi^2=2,~\Pi_1-\Pi_2=\pm2,~w_1=\mp1,~n_1=0,~w_2=n_2=0,\\
                &\Pi^2=2,~\Pi_1-\Pi_2=\pm2,~w_1=\mp1,~n_1=\pm1,~w_2=n_2=\mp1,\\
                &\Pi=0,~w_1=n_1=\pm1,~w_2=n_2=0,\\
                &\Pi=0,~w_1=0,~n_1=\pm1,~w_2=n_2=\mp1,\\
                &\Pi=0,~w_1=\pm1,~n_1=0,~w_2=n_2=\pm1.
            \end{align}
        \end{subequations}
        We can read off the gauge group as $E_8\times E_7\times SU(4)$.
\end{enumerate}
We summarize the realized gauge symmetries depending on the moduli in Table \ref{tab:gauge_sym_SUSY}. The maximally enhanced gauge groups in the supersymmetric heterotic strings on $T^2$ were completely studied in \cite{Font:2020rsk}, which shows there are no maximal enhancements other than these four with the Wilson lines \eqref{WL}.

\begin{table}[H]
\begin{center}
\caption{Maximally enhanced gauge symmetries in $E_8\times E_8$ string.}
\begin{tabular}{|c||c|c|c||c|} \hline
    & $a_1$ & $a_2$ & $E$ & Gauge symmetry \\ \hline\hline
    $(1)$ & $0$ & $0$ & $\alpha'E_1$ & $E_8\times E_8\times SU(2)^2$ \\
    $(2)$ & $0$ & $0$ & $\alpha'E_2$ & $E_8\times E_8\times SU(3)$ \\
    $(3)$ & $\frac{1}{2}$ & $0$ & $\alpha'E_1$ & $E_8\times E_7\times SU(3)\times SU(2)$ \\
    $(4)$ & $\frac{1}{2}$ & $0$ & $\alpha'E_2$ & $E_8\times E_7\times SU(4)$ \\
    \hline
\end{tabular}
\label{tab:gauge_sym_SUSY}
\end{center}
\end{table}

\subsubsection{The $SO(16)\times SO(16)$ string}
\label{sec:SO(16)SO(16)}
We now consider the non-supersymmetric heterotic strings. In particular, we focus on the models with $\hat{q}_*^{\mathrm{T}}:=(0, 0, 0, 0, 0, 1, 0, 1; 0, 0, 0, 0, 0, 1, 0, 1)$, which leads to the $SO(16)\times SO(16)$ non-supersymmetric heterotic string theory in ten dimensions. Since this $\hat{q}$ leads to $\hat{\Pi}^{\mathrm{T}}=(0, 0, 0, 0, 0, 2, 0, 0; 0, 0, 0, 0, 0, 2, 0, 0)$, the condition for $\hat{Z}$ \eqref{Z_condition2} allows $\hat{w}$ and $\hat{n}$ satisfying $\hat{w}^{\mathrm{T}}\hat{n}=0~(\text{mod 2})$. Thus these $\hat{Z}$ can lead to the model $\#1,2,3,4,5,7,9,10,13$ or $\#16$ with $\hat{q}_*$ in Table \ref{tab:torus}.

The mass formula and level-matching condition are the same as in the supersymmetric cases, but we must consider the additional constraint coming from the Scherk-Schwarz twist. In order to identify the gauge group, we only focus on the massless vectors in the non-supersymmetric strings, paired with $\Gamma_{16+2,2}^{+}$. Using \eqref{Gamma_splitting}, the additional constraint is written as
\begin{align}\label{SS_condition}
    2\left(\Pi_6+\Pi_{14}\right)+w^{\mathrm{T}}\hat{n}
    +n^{\mathrm{T}}\hat{w}\in2\mathbb{Z}.
\end{align}
For any non-supersymmetric strings with $\hat{q}_*$, the condition \eqref{general_massless_solution} can be solutions of massless conditions if $\Pi_6+\Pi_{14}\in\mathbb{Z}$. Then we find that the gauge group is $SO(16)\times SO(12)\times SU(2)\times U(1)^3$ in any model with $\hat{q}_*$ with general values of moduli $a_1, a_2$ and $E$. Note that the constraint \eqref{SS_condition} excludes the non-zero roots of $E_8\times E_8$ satisfying $\Pi^2=2$ such as
\begin{align}
    \Pi=\frac{1}{2}\left(\underline{\pm,\pm,\pm,\pm,\pm,\pm,\pm,\pm}_{+};0^{8}\right),~~\frac{1}{2}\left(0^{8};\underline{\pm,\pm,\pm,\pm,\pm,\pm,\pm,\pm}_{+}\right),
\end{align}
where underlines with the subscript $+$ mean permutations with the number of $+$ even, and
breaks $E_8$ and $E_7$ in the supersymmetric case to $SO(16)$ and $SO(12)\times SU(2)$, respectively. 

We first consider the model $\#1$ with $\hat{q}_*$, which can be simply realized by $T^2$ compactification of the non-supersymmetric $SO(16)\times SO(16)$ string. From \eqref{SS_condition}, we obtain $\Pi_6+\Pi_{14}\in\mathbb{Z}$. Taking this condition into account, we find that the gauge symmetries in the model $\#1$ with $\hat{q}_*$ can be maximally enhanced with the four sets of moduli $(1),\ldots,(4)$ as listed in Table \ref{tab:gauge_sym_1}. As shown in Table \ref{tab:torus}, the model $1$ has the same modular symmetries as the supersymmetric heterotic strings. Therefore, the four fixed points of $\mathrm{Sp}(4,\mathbb{Z})$ we consider in Section \ref{sec:E8E8} also lead to the maximal enhancements in the model $\#1$ with $\hat{q}_*$. 
\begin{table}[H]
\begin{center}
\caption{Realized gauge symmetries in the model $\#1$ with $\hat{q}_*$.}
\begin{tabular}{|c||c|c|c||c|} \hline
    & $a_1$ & $a_2$ & $E$ & Gauge symmetry \\ \hline\hline
    $(1)$ & $0$ & $0$ & $\alpha'E_1$ & $SO(16)\times SO(16)\times SU(2)^2$ \\
    $(2)$ & $0$ & $0$ & $\alpha'E_2$ & $SO(16)\times SO(16)\times SU(3)$ \\
    $(3)$ & $\frac{1}{2}$ & $0$ & $\alpha'E_1$ & $SO(16)\times SO(12)\times SU(3)\times SU(2)^2$ \\
    $(4)$ & $\frac{1}{2}$ & $0$ & $\alpha'E_2$ & $SO(16)\times SO(12)\times SU(4)\times SU(2)$ \\
    \hline
\end{tabular}
\label{tab:gauge_sym_1}
\end{center}
\end{table}

As a more interesting case, we next consider the model $\#9$ with $\hat{q}_*$, which is called an interpolating model in the sense that $R_1\to\infty$ gives a 9d supersymmetric string while $R_1\to0$ leads to a 9d non-supersymmetric string.\footnote{Decompactification limits of the 8d non-supersymmetric heterotic strings were studied in \cite{Koga:2022qch}.} The gauge symmetry enhancement patterns in the model $\#9$ are different from those in the model $\#1$, since the model $\#9$ is labeled by $\hat{w}_1=1$ while the model $\#1$ by $\hat{w}_1=0$. For example, in the case of moduli $(1)$, the conditions \eqref{example_not_solution} cannot be solutions of the massless condition because of \eqref{SS_condition}. We list the realized gauge symmetry in the model $\#9$ with $\hat{q}_*$ depending on the four sets of moduli $(1),\ldots,(4)$ in Table \ref{tab:gauge_sym_9}. We can find the gauge symmetries in the model $\#9$ with $\hat{q}_*$ cannot be maximally enhanced with the four sets of moduli $(1),\ldots,(4)$. 
\begin{table}[H]
\begin{center}
\caption{Realized gauge symmetries in the model $\#9$ with $\hat{q}_*$.}
\begin{tabular}{|c||c|c|c||c|} \hline
    & $a_1$ & $a_2$ & $E$ & Gauge symmetry \\ \hline\hline
    $(1)$ & $0$ & $0$ & $\alpha'E_1$ & $SO(16)\times SO(16)\times SU(2)\times U(1)$ \\
    $(2)$ & $0$ & $0$ & $\alpha'E_2$ & $SO(16)\times SO(16)\times SU(2)\times U(1)$ \\
    $(3)$ & $\frac{1}{2}$ & $0$ & $\alpha'E_1$ & $SO(16)\times SO(12)\times SU(2)^3\times U(1)$ \\
    $(4)$ & $\frac{1}{2}$ & $0$ & $\alpha'E_2$ & $SO(16)\times SO(12)\times SU(2)^3\times U(1)$ \\
    \hline
\end{tabular}
\label{tab:gauge_sym_9}
\end{center}
\end{table}	
The reason why the four sets of moduli $(1),\ldots,(4)$ do not lead to the maximal enhancements in the model $\#9$ can be understood as follows: the model $\#9$ has the broken modular symmetry rather than $\mathrm{Sp}(4,\mathbb{Z})$ and its fixed points are different from the ones of the $\mathrm{Sp}(4,\mathbb{Z})$ group. Table \ref{tab:torus}
shows that the elements $\hat{C}_{\gamma}$ and $\hat{K}_{\gamma}$ in the model $\#9$ belong to the duality group of the non-supersymmetric string if $\gamma\in\Gamma_1(2)$ for $\hat{C}_{\gamma}$ and $\gamma\in\Gamma^1(2)$ for $\hat{K}_{\gamma}$. Thus we should consider the fixed points of $\Gamma_1(2)$ and $\Gamma^1(2)$ for the complex structure moduli $U$ and K\"ahler moduli $T$, respectively:\footnote{Note that the fixed point of $\Gamma_1(2)$ is given by
\begin{align}
    \begin{pmatrix}
        -1 & -1\\
        2 & 1
    \end{pmatrix}
    \tau = \tau
    \qquad
    \mathrm{with}
        \begin{pmatrix}
        -1 & -1\\
        2 & 1
    \end{pmatrix}
    \in \Gamma_1(2).
\end{align} 
The fixed point of $\Gamma^1(2)$ can be understood by the fact that $\Gamma^1(2)$ is conjugated to $\Gamma_1(2)$ under the $S$ transformation.}
\begin{align}
    \begin{pmatrix}
        U & Z\\
        Z & T
    \end{pmatrix}
    =
    \begin{pmatrix}
        -\frac{1}{2}+\frac{i}{2} & 0\\
        0 & -1+i
    \end{pmatrix},
\end{align}
where we take $a_1=a_2=0$ for simplicity. The corresponding moduli $E$ is written as
\begin{align}
    E=\alpha'
    \begin{pmatrix}
        2 & -2\\
        0 & 1
    \end{pmatrix}.
\end{align}
The massless solutions satisfying the constraint \eqref{SS_condition} are
\begin{subequations}
    \begin{align}
        &\Pi^2=2,~\Pi_6+\Pi_{14}\in\mathbb{Z},
        ~w_1=w_2=n_1=n_2=0,\\
        &\Pi=0,~w_1=n_1=\pm2,~w_2=n_2=\mp1,\\
        &\Pi=0,~w_1=\pm1,~n_1=0,~w_2=n_2=\pm1.
    \end{align}
\end{subequations}
Therefore the gauge group is maximally enhanced to $SO(16)\times SO(16)\times SU(2)^2$.

The gauge groups of the other models with $\hat{q}_*$ are also maximally enhanced at the fixed points of remaining modular symmetries. We summarize the maximal enhancements of the non-supersymmetric heterotic strings with $\hat{q}_*$ in Table \ref{tab:gauge_sym_all}. Here we have used the fact that the fixed points of the subgroups $\Gamma_1(2),~\Gamma^1(2)$ and $\Gamma_{\vartheta}$ are given by $-\frac{1}{2}+\frac{i}{2},~-1+i$ and $i$, respectively.\footnote{Note that the fixed point of $\Gamma_{\vartheta}$ is given by the $S$ transformation. $\Gamma(2)$ has no fixed points in the upper-half plane.}

Using the data of fixed moduli and $q,\bar{q}$-expansions of the partition functions \eqref{partition function}, we find that all the partition functions of the models with maximal enhancements except for the model $\#1$ are completely equivalent, which means they all are dual to each other. However, the case with $T=U=i$ and $Z=0$ in the model $\#1$, which also leads to the same gauge group $SO(16)\times SO(16)\times SU(2)^2$, is not equivalent to them. We leave for future work a detailed investigation of the differences in the moduli spaces between model $\#1$ and the other models.

\begin{center}
\begin{longtable}{|c|c|c|c|}
    \caption{Maximally enhanced gauge symmetries and 8d cosmological constants $\Lambda^{(8)}$ in the models with $\hat{q}_*$. All of the maximally enhanced models except for the model $\#1$ are dual to each other and correspond to knife edges.}
    \label{tab:gauge_sym_all}\\
    
\hline $\#$ & Fixed moduli $(U,T,Z)$ & Gauge symmetry & $\Lambda^{(8)}$\\ \hline\hline 
\endfirsthead

\multicolumn{4}{c}
{{\bfseries \tablename\ \thetable{} -- continued from previous page}} \\
\hline $\#$ & Fixed moduli $(U,T,Z)$ & Gauge symmetry & $\Lambda^{(8)}$\\ \hline\hline 
\endhead

\hline \multicolumn{4}{|r|}{{Continued on next page}} \\ \hline
\endfoot

\hline 
\endlastfoot

    \multirow{7}{*}{$1$} & $\begin{pmatrix}
                i & 0\\
                0 & i
            \end{pmatrix}$ & $SO(16)\times SO(16)\times SU(2)^2$ & $293.6$\\
    & $\begin{pmatrix}
                \omega & 0\\
                0 & \omega
            \end{pmatrix}$ & $SO(16)\times SO(16)\times SU(3)$ & $274.5$\\
    & $\frac{i}{\sqrt{3}}
            \begin{pmatrix}
                 2 & 1\\
                 1 & 2
            \end{pmatrix}$ & $SO(16)\times SO(12)\times SU(3)\times SU(2)^2$ & $254.3$\\
    & $\begin{pmatrix}
                \tilde{\eta}-1 & \frac{1}{2}(\tilde{\eta}-1)\\
                \frac{1}{2}(\tilde{\eta}-1)& \tilde{\eta}-1
            \end{pmatrix}$ & $SO(16)\times SO(12)\times SU(4)\times SU(2)$ & $221.8$\\ \hline
    \multirow{1}{*}{$2$} & $\begin{pmatrix}
                -\frac{1}{2}+\frac{i}{2} & 0\\
                0 & -\frac{1}{2}+\frac{i}{2}
            \end{pmatrix}$ & $SO(16)\times SO(16)\times SU(2)^2$ & $-74.29$\\ \hline
    \multirow{1}{*}{$3$} & $\begin{pmatrix}
                -1+i & 0\\
                0 & -\frac{1}{2}+\frac{i}{2}
            \end{pmatrix}$ & $SO(16)\times SO(16)\times SU(2)^2$ & $-74.29$\\ \hline
    \multirow{1}{*}{$4$} & $\begin{pmatrix}
                i & 0\\
                0 & -\frac{1}{2}+\frac{i}{2}
            \end{pmatrix}$ & $SO(16)\times SO(16)\times SU(2)^2$ & $-74.29$\\ \hline
    \multirow{1}{*}{$5$} & $\begin{pmatrix}
                -1+i & 0\\
                0 & -1+i
            \end{pmatrix}$ & $SO(16)\times SO(16)\times SU(2)^2$ & $-74.29$\\ \hline
    \multirow{1}{*}{$7$} & $\begin{pmatrix}
                -1+i & 0\\
                0 & i
            \end{pmatrix}$ & $SO(16)\times SO(16)\times SU(2)^2$ & $-74.29$\\ \hline
    \multirow{1}{*}{$9$} & $\begin{pmatrix}
                -\frac{1}{2}+\frac{i}{2} & 0\\
                0 & -1+i
            \end{pmatrix}$ & $SO(16)\times SO(16)\times SU(2)^2$ & $-74.29$\\ \hline
    \multirow{1}{*}{$10$} & $\begin{pmatrix}
                -\frac{1}{2}+\frac{i}{2} & 0\\
                0 & i
            \end{pmatrix}$ & $SO(16)\times SO(16)\times SU(2)^2$ & $-74.29$\\ \hline
    \multirow{1}{*}{$13$} & $\begin{pmatrix}
                i & 0\\
                0 & -1+i
            \end{pmatrix}$ & $SO(16)\times SO(16)\times SU(2)^2$ & $-74.29$\\ \hline
    \multirow{1}{*}{$16$} & $\begin{pmatrix}
                i & 0\\
                0 & i
            \end{pmatrix}$ & $SO(16)\times SO(16)\times SU(2)^2$ & $-74.29$\\
\end{longtable}
\end{center}

\subsection{Tachyons, knife edges and cosmological constants}
\label{sec:CC}
The $SO(16)\times SO(16)$ string in ten dimensions is a well-known tachyon-free theory. However, if we compactify it on $T^d$, physical (level-matched) tachyons may appear. We should thus consider whether theories with maximally enhanced gauge groups in Table \ref{tab:gauge_sym_all} have tachyons or not. Tachyons would be in the scalar conjugacy classes of $SO(8)$, paired with $\Gamma_{18,2}^{+}+\delta$ in the model $16$ while $\Gamma_{18,2}^{-}+\delta$ in the other possible models. They may appear if there exist $(\Pi,w,n)$ satisfying
\begin{align}\label{cond_tachyon}
    P_L^2-p_R^2=1,~~~p_R^2<1.
\end{align} 
However, we cannot find the solutions of the conditions \eqref{cond_tachyon} in each model with maximal enhancements. Therefore, we find that all the theories in Table \ref{tab:gauge_sym_all} are tachyon-free.

In addition to tachyons, massless scalars are also involved in the stability of the theory. They can appear if there exist $(\Pi,w,n)$ satisfying
\begin{align}\label{cond_scalar}
    P_L^2-p_R^2=1,~~~p_R^2=1.
\end{align}
If a massless scalar exists in a point of the moduli space, then a certain infinitesimal deformation of the moduli corresponding to that point gives a tachyonic point since the conditions \eqref{cond_scalar} give a boundary of the tachyonic region obtained by the conditions \eqref{cond_tachyon}. Such an unstable point is called a knife edge \cite{Ginsparg:1986wr}. From \eqref{cond_scalar}, we find that four maximal enhancements in the model $\#1$ do not lead to massless scalars, while in the other cases, there are $1024$ massless scalars. The maximal enhancements except for four cases in the model $\#1$ correspond to knife edges.

If tachyons do not appear, the one-loop cosmological constant takes a finite value. It is defined by integrating the partition function over the fundamental domain $\mathcal{F}$ of the $SL(2,\mathbb{Z})$ transformation
\begin{align}
    \Lambda^{(10-D)}=-\frac{1}{2}(4\pi^2\alpha')^{-\frac{10-D}{2}}\int_{\mathcal{F}}\frac{d^2\tau}{\tau_2^2}Z_{\cancel{\mathrm{SUSY}}}^{(10-D)}(\tau).
\end{align}
We numerically calculate the 8d cosmological constants $\Lambda^{(8)}$,  included in Table \ref{tab:gauge_sym_all}. All of the maximal enhancement points except for the model $\#1$ have the same value of the cosmological constant as expected since they are equivalent theories. Moreover, we can see that four maximal enhancements without tachyons and massless scalars have positive cosmological constants while the others do negative ones.

As proved in \cite{Ginsparg:1986wr}, maximal enhancements in the non-supersymmetric heterotic strings on $T^D$ extremize the cosmological constant. That is, the maximally enhanced theories correspond to the extrema of the effective potential. A natural question is whether the maximal enhancement points can be local minima, which is verified by checking that the Hessian of the cosmological constant is positive definite. We evaluate the integral of the Hessian of the partition function:
\begin{align}
    H_{ij}=\nabla_i\nabla_j\Lambda^{(8)}=-\frac{1}{2}(4\pi^2\alpha')^{-4}\int_{\mathcal{F}}\frac{d^2\tau}{\tau_2^2} \nabla_i\nabla_j Z_{\cancel{\mathrm{SUSY}}}^{(8)}(\tau),
\end{align}
where $\nabla$ is defined as
\begin{align}
    \nabla=\left(\frac{\partial}{\partial a_{1}},\frac{\partial}{\partial a_{2}},\frac{\partial}{\partial E_{11}},\frac{\partial}{\partial E_{12}},\frac{\partial}{\partial E_{21}},\frac{\partial}{\partial E_{22}}\right).
\end{align}
By numerical calculations, we obtain the eigenvalues of
the Hessian at each of the four maximal enhancements in the model $\#1$ as listed in Table \ref{tab:Hessian}. It leads us to the conclusion that one of them is a local maximum while the others are saddle points. The situation in that we cannot find local minima is the same as in the $SO(16)\times SO(16)$ string on $S^1$ studied in \cite{Fraiman:2023cpa}. We leave for future work a study of maximal enhancements given by more general moduli and their stabilities.

\begin{table}[H]
\begin{center}
\caption{Eigenvalues of
the Hessian at four maximal enhancements in the model $\#1$.}
\scalebox{0.85}{
\begin{tabular}{|c||c|c|c||cccccc|} \hline
    & $a_1$ & $a_2$ & $E$ & \multicolumn{6}{|c|}{$H_{ij}$} \\ \hline\hline
    $(1)$ & $0$ & $0$ & $\alpha'E_1$ & $-67.46$ & $-47.31$ & $-4.486\times 10^5$ & $-4.486\times 10^5$ & $-4.485\times 10^5$ & $-4.484\times 10^5$ \\
    $(2)$ & $0$ & $0$ & $\alpha'E_2$ & $-761.1$ & $172.8$ & $-8083$ & $-5650$ & $-1883$ & $-1711$ \\
    $(3)$ & $\frac{1}{2}$ & $0$ & $\alpha'E_1$ & $-62.31$ & $54.89$ & $-3825$ & $-1670$ & $507.9$ & $-186.7$ \\
    $(4)$ & $\frac{1}{2}$ & $0$ & $\alpha'E_2$ & $127.4$ & $-55.04$ & $-2982$ & $1172$ & $478.1$ & $-355.4$ \\
    \hline
\end{tabular}
}
\label{tab:Hessian}
\end{center}
\end{table}

\section{Classification of modular symmetries on toroidal orbifolds}
\label{sec:T2ZN}

In this section, we move to the toroidal orbifolds in the context of non-supersymmetric string theories.

\subsection{Narain formulation on toroidal orbifolds}
\label{sec:Narain_orbifolds}

In contrast to $T^D$, the Narain lattice is modified by the existence of a Narain twist $\Theta$:
\begin{align}
    \begin{pmatrix}
        y_L\\
        y_R
    \end{pmatrix}
    \sim 
    \Theta
    \begin{pmatrix}
        y_L\\
        y_R
    \end{pmatrix}
    +\mathcal{E}\hat{N},
\end{align}
with $\hat{N}\in \mathbb{Z}^{2D+16}$. 
The set of Narain twists
\begin{align}
    \Theta
    := 
    \begin{pmatrix}
        \theta_L & 0\\
         0 & \theta_R 
    \end{pmatrix}
    ,
\end{align}
obeying $\Theta^T \Theta= \mathbb{I}_{2D+16}$ and $\Theta^T \eta \Theta = \eta$, generates the Narain point group $P_{\mathrm{Narain}}$, and transformations with $(\Theta, \mathcal{E}\hat{N})$ generate the Narain space group $S_{\mathrm{Narain}}$. 
As discussed in \cite{Nilles:2021glx}, let us move to the so-called Narain lattice basis such that the Narain space group $\hat{S}_{\mathrm{Narain}}$ is constructed by $(\hat{\Theta}, \hat{N})$ with 
\begin{align}
    \hat{\Theta} := \mathcal{E}^{-1} \Theta \mathcal{E}
\end{align}
obeying 
\begin{align}
    \hat{\Theta}^T {\cal H} \hat{\Theta} = {\cal H},
    \qquad 
    \hat{\Theta}^T \hat{\eta} \hat{\Theta} = \hat{\eta}.
\label{eq:H_cond}
\end{align}
The Narain twist $\hat{\Theta}$ in this basis generates the Narain point group $\hat{P}_{\mathrm{Narain}}$.

In the following analysis, we explore the modular group in the context of non-supersymmetric heterotic string theories:
\begin{align}
    G_{\mathrm{modular}}:= \left\lbrace \hat{\Sigma}\in O(16+D,D,\mathbb{Z}) \left|~\hat{\Sigma}^{\mathrm{T}}\hat{\eta}\hat{\Sigma}=\hat{\eta},
    \,~\hat{\Sigma}\hat{Z}=\hat{Z}~(\text{mod 2})\right.,\,~ \hat{\Sigma}^{-1}\hat{\Theta}\hat{\Sigma}\in \hat{P}_{\mathrm{Narain}} \right\rbrace,
\end{align}
for all $\hat{\Theta}\in \hat{P}_{\mathrm{Narain}}$, 
where we study the Narain point group as shown in Table \ref{tab:orbifolds}. 
Note that it is required to satisfy \eqref{Z_condition} for the whole $\hat{q}\in\mathbb{Z}^{16}$, 
but our discussion is relevant for only $\hat{q}_1$ and $\hat{q}_2$. 
Hence, we focus on the modular group on each orbifold. 
In particular, following the analysis of supersymmetric heterotic string theory \cite{Baur:2020yjl}, we study the modular symmetries on toroidal compactifications with symmetric and asymmetric orbifold twists, as shown in Table \ref{tab:orbifolds} in which we introduce the stabilizer group of $\mathrm{Sp}(4,\mathbb{Z})$. 

To define the stabilizer group, let us rewrite the $\mathrm{Sp}(2g,\mathbb{Z})$ modular group as $\Gamma_g=\mathrm{Sp}(2g,\mathbb{Z})$. 
Then, we consider points $\tau$ in a region $\Omega$ which are individually invariant by some element $h$ of $\Gamma_g$:
\begin{align}
    h\tau=\tau,\quad \mathrm{with}\,\, \tau\in\Omega,
    \label{eq:fixedpoints}
\end{align}
Hence, one can define subgroup $H$ of $\Gamma_g$ whose element satisfy \eqref{eq:fixedpoints}, 
i.e., $H\tau = \tau$. 
Since the transformations of $-\gamma$ and $\gamma$ are the same for $\tau$, it leads to the group  $\bar{H}=H/\{\pm\mathbb{I}_{2g}\}$. 
In this paper, we call $\bar{H}$ the stabilizer group. 

Furthermore, one can define the so-called normalizer group $N(H)$ 
which leaves the region $\Omega$ invariant: 
\begin{align}
    \gamma\tau=\tau^\prime,\qquad \tau,\tau^\prime\in\Omega
\end{align}
satisfying
\begin{align}
    H\tau=\tau,\qquad H\tau^\prime=\tau^\prime.
\end{align}
We arrive at  
\begin{align}
    \gamma^{-1}H\gamma=H,
\end{align}
indicating that $N(H)$ is the normalizer of $H$. 
Note that the stabilizer group $H$ is in general a subgroup of the normalizer of $H$, $N(H)$, 
but the normalizer group is the same as the stabilizer group for the fixed points \eqref{eq:fixedpoints} with dimension zero.

\small
\begin{table}[H]
    \centering
    \caption{Toroidal compactifications with symmetric and asymmetric orbifold twists. We show the Narain point groups for each value of moduli $(U,T, Z)$. Here, we define $\omega := e^{2\pi i/3}$, $\zeta := e^{2\pi i/5}$, and $\tilde{\eta} := \frac{1}{3}(1 + 2\sqrt{2} i)$.}
    \label{tab:orbifolds}
    \scalebox{0.9}{
    \begin{tabular}{c|ccc}
        \toprule
        \textbf{Complex dimension} &
        \textbf{Fixed moduli} & \textbf{Narain Point Group} & \textbf{Type of} \\
        \textbf{of the moduli space} &
        \textbf{$(U, T, Z)$} & \textbf{(Stabilizer group of Sp$(4,\mathbb{Z})$)} & \textbf{ orbifolds} \\    
        \hline
        \midrule
        \multirow{5}{*}{2} &
        $\begin{pmatrix}
            U & 0\\
            0 & T
        \end{pmatrix}$ & $\mathbb{Z}_2$ & Symmetric\\
         & $\begin{pmatrix}
            T & Z\\
            Z & T
        \end{pmatrix}$ & $\mathbb{Z}_2$ & Asymmetric\\
        & $\begin{pmatrix}
            U & 1/2\\
            1/2 & T
        \end{pmatrix}$ & $\mathbb{Z}_2$ & Symmetric\\
        \hline\hline
        \multirow{10}{*}{1}        
        & $\begin{pmatrix}
            i & 0\\
            0 & T
        \end{pmatrix}$ & $\mathbb{Z}_4$ & Symmetric\\
        & $\begin{pmatrix}
            \omega & 0\\
            0 & T
        \end{pmatrix}$ & $\mathbb{Z}_3$,\,\,$\mathbb{Z}_6$ & Symmetric\\
        & $\begin{pmatrix}
            T & 0\\
            0 & T
        \end{pmatrix}$ & $\mathbb{Z}_2 \times \mathbb{Z}_2$ & Asymmetric\\
        & $\begin{pmatrix}
            T & 1/2\\
            1/2 & T
        \end{pmatrix}$ & $\mathbb{Z}_2 \times \mathbb{Z}_2$ & Asymmetric\\
        & $\begin{pmatrix}
            T & T/2\\
            T/2 & T
        \end{pmatrix}$ & $S_3$ & Asymmetric\\
        \hline \hline
        \multirow{12}{*}{0} &
        $\begin{pmatrix}
            \zeta & \zeta + \zeta^{-2}\\
            \zeta + \zeta^{-2} & -\zeta^{-1}
        \end{pmatrix}$ & $\mathbb{Z}_5$ & Asymmetric\\
        & $\begin{pmatrix}
            \tilde{\eta} & \frac{1}{2}(\tilde{\eta} - 1)\\
            \frac{1}{2}(\tilde{\eta} - 1) & \tilde{\eta}
        \end{pmatrix}$ & $S_4$ & Asymmetric\\
       & $\begin{pmatrix}
            i & 0\\
            0 & i
        \end{pmatrix}$ & $(\mathbb{Z}_4 \times \mathbb{Z}_2) \rtimes \mathbb{Z}_2$ & Asymmetric\\
        & $\begin{pmatrix}
            \omega & 0\\
            0 & \omega
        \end{pmatrix}$ & $S_3 \times \mathbb{Z}_6$ & Asymmetric\\
        & $i\frac{1}{\sqrt{3}}
        \begin{pmatrix}
            2 & 1\\
            1 & 2
        \end{pmatrix}$ & $S_3 \times \mathbb{Z}_2 \cong D_{12}$ & Asymmetric\\
        & $\begin{pmatrix}
            \omega & 0\\
            0 & i
        \end{pmatrix}$ & $\mathbb{Z}_{12}$ & Asymmetric\\
        \hline        
         \bottomrule
    \end{tabular}
    }
\end{table}
\normalsize

\subsection{Sp$(4,\mathbb{Z})$ fixed points with dimension 2}
\label{sec:dim2}

In this section, we discuss the $\mathrm{Sp}(4,\mathbb{Z})$ fixed points where the complex dimension of the moduli space is two. 
There are three possibilities: (i) symmetric $\mathbb{Z}_2$ orbifold without Wilson line in Sec. \ref{sec:Z21}, (ii) asymmetric $\mathbb{Z}_2$ orbifold in Sec. \ref{sec:Z22}, and (iii) symmetric $\mathbb{Z}_2$ orbifold with discrete Wilson line in Sec. \ref{sec:Z23}.

\subsubsection{Symmetric $\mathbb{Z}_2$ orbifold without Wilson line}
\label{sec:Z21}

We start with the $\mathbb{Z}_2$ Narain point group $\hat{P}_{\mathrm{Narain}}$ of the $\mathbb{Z}_2$ orbifold, generated by
\begin{align}
    \hat{\Theta} := \left(\hat{K}_S\right)^2=\left(\hat{C}_S\right)^2.
\end{align}
Under this $\mathbb{Z}_2$ twist, the metric ${\cal H}$ is constrained by \eqref{eq:H_cond}, which leads to the vanishing Wilson line modulus $Z=0$. 
We arrive at the following configuration of the moduli in the Siegel upper half-plane:
\begin{align}
 \tau=\begin{pmatrix}
        U&0\\
        0&T\\
    \end{pmatrix}
    .
    \label{eq:tauZ21}
\end{align}
In the language of $\mathrm{Sp}(4,\mathbb{Z})$, the above moduli matrix corresponds to the fixed point of $\mathrm{Sp}(4,\mathbb{Z})$ generated by the following stabilizer $\bar{H}\cong\mathbb{Z}_2$:
\begin{align}
    h=\begin{pmatrix}
        1&0&0&0\\
        0&-1&0&0\\
        0&0&1&0\\
        0&0&0&-1\\
    \end{pmatrix}
    .
\end{align}
Furthermore, the normalizer $N(H)$ on \eqref{eq:tauZ21}
\begin{align}
    \langle M_{(\mathrm{S},\,\mathbb{I}_2)},\, M_{(\mathrm{T},\,\mathbb{I}_2)},\, M_{(\mathbb{I}_2,\,\mathrm{S})},\, M_{(\mathbb{I}_2,\,\mathrm{T})},\, M_\times\rangle
\end{align}
respectively corresponds to an outer automorphism of $\hat{P}_{\mathrm{Narain}}$, namely the modular group of the symmetric $\mathbb{Z}_2$ orbifold in the case of supersymmetric heterotic string theory:
\begin{align}
    \langle \hat{K}_S, \hat{K}_T, \hat{C}_S, \hat{C}_T, \hat{M}\rangle\cong \frac{\mathrm{SL}(2,\mathbb{Z})_T\times\mathrm{SL}(2,\mathbb{Z})_U}{\mathbb{Z}_2}\rtimes \mathbb{Z}_2^{\hat{M}},
\label{eq:modularZ21}
\end{align}
where $\mathbb{Z}_2$ originates from $\left(\hat{K}_S\right)^2=\left(\hat{C}_S\right)^2$.

In the case of non-supersymmetric heterotic strings, we have to search for $\hat{Z}$ satisfying Eqs. \eqref{condition_T-duality} and \eqref{Z_condition}. 
We list all the possible configurations of $\hat{Z}$ in Appendix \ref{app:Z21}, where the moduli fields 
transform as 
\begin{align}
\begin{array}{ll}
    T\xrightarrow{\hat{K}_S}-\frac{1}{T},& U\xrightarrow{\hat{K}_S}U,\\
    T\xrightarrow{\hat{K}_T}T+1,& U\xrightarrow{\hat{K}_T}U,\\
    T\xrightarrow{\hat{C}_S}T,& U\xrightarrow{\hat{C}_S}-\frac{1}{U},\\
    T\xrightarrow{\hat{C}_T}T,& U\xrightarrow{\hat{C}_T}U+1,\\
    T\xrightarrow{\hat{M}}U,& U\xrightarrow{\hat{M}}T.
\end{array}
\end{align}
It turns out that one can realize the same congruence subgroups as shown in Table \ref{tab:torus} without the Wilson line, depending on the choice of $\hat{Z}$. 
Hence, only a subgroup of \eqref{eq:modularZ21} remains in non-supersymmetric heterotic string theories. 
Note that the $\mathcal{CP}$-like transformation, which is an outer automorphism of the Narain space group, 
is defined as
\begin{align}
    \hat{\mathcal{CP}} = \hat{\Sigma}_\ast,
\end{align}
with $\hat{\mathcal{CP}} \hat{\Theta} \hat{\mathcal{CP}}^{-1}=\hat{\Theta}^{-1}$.

\subsubsection{Asymmetric $\mathbb{Z}_2$ orbifold}
\label{sec:Z22}

In this section, we discuss the following $\mathbb{Z}_2$ Narain point group $\hat{P}_{\mathrm{Narain}}$ of the $\mathbb{Z}_2$ orbifold:
\begin{align}
    \hat{\Theta} := \hat{M}= 
    \hat{W}
    \begin{psmallmatrix}
        0\\
        1
    \end{psmallmatrix}
    \hat{K}_S
    \hat{W}
    \begin{psmallmatrix}
        1\\
        0
    \end{psmallmatrix}
    \hat{K}_S
    \hat{W}
    \begin{psmallmatrix}
        0\\
        -1
    \end{psmallmatrix}
.
\label{eq:twist_Z22}
\end{align}
Under this $\mathbb{Z}_2$ twist, the metric ${\cal H}$ is constrained by \eqref{eq:H_cond}, which leads to the following configuration of the moduli in the Siegel upper half-plane:
\begin{align}
 \tau=\begin{pmatrix}
        T&Z\\
        Z&T\\
    \end{pmatrix}
    ,
    \label{eq:tauZ22}
\end{align}
with
\begin{align}
    T &= U = \frac{1}{\alpha^\prime (1-a_1^2)}\left( G_{12}+i\sqrt{-G_{12}^2 + \alpha^\prime G_{22}(1-a_1^2)}\right),
\end{align}
where
\begin{align}
    G_{11} &= \alpha^\prime (1-a_1^2),\qquad
    B_{12}=\alpha^\prime a_1a_2 +G_{12}.
\end{align}
Hence, the Wilson line $Z=-a_2 +T a_1$ is constrained as $a_1^2 <1$ due to $G_{11}>0$. 
In the language of $\mathrm{Sp}(4,\mathbb{Z})$, the above moduli matrix corresponds to the fixed point of $\mathrm{Sp}(4,\mathbb{Z})$ generated by the following stabilizer $\bar{H}\cong\mathbb{Z}_2$:
\begin{align}
    h=\begin{pmatrix}
        0&1&0&0\\
        1&0&0&0\\
        0&0&0&1\\
        0&0&1&0\\
    \end{pmatrix}.
\end{align}
Furthermore, the normalizer $N(H)$ on \eqref{eq:tauZ22}
\begin{align}
    \langle M_{(\mathrm{S},\,\mathrm{S})}, \,M_{(\mathrm{T},\,\mathrm{T})},\, M_{(\mathrm{S}^2,\,\mathbb{I}_2)},\, M\begin{psmallmatrix}
        -1\\
        0
    \end{psmallmatrix}\rangle
\label{eq:normalizer_Z22}
\end{align}
respectively corresponds to an outer automorphism of $\hat{P}_{\mathrm{Narain}}$, namely the modular group of the symmetric $\mathbb{Z}_2$ orbifold in the case of supersymmetric heterotic string theory:
\begin{align}
    \langle \hat{M}_1\,,\hat{M}_2\,,\hat{M}_3\,,\hat{M}_4\rangle,
\label{eq:modularZ22}
\end{align}
with 
\begin{align}
    \hat{M}_1:=\hat{K}_S\hat{C}_S,
    \quad
    \hat{M}_2:=\hat{K}_T\hat{C}_T,
    \quad
    \hat{M}_3:=\left(\hat{K}_S\right)^2,
    \quad
    \hat{M}_4:=\hat{W}\begin{psmallmatrix}
        -1\\
        0
\end{psmallmatrix}
,
\end{align}
obeying
\begin{align}
    &(\hat{M}_1)^2=(\hat{M}_1\hat{M}_2)^3=(\hat{M}_1\hat{M}_4)^6=(\hat{M}_3)^2= \mathbb{I}_{20},
    \nonumber\\
    &[\hat{M}_1, \hat{M}_3]=[\hat{M}_2, \hat{M}_3]=[\hat{M}_2, \hat{M}_4]=[\hat{M}_i, (\hat{M}_1\hat{M}_4)^3]=[\hat{M}_2, \hat{M}_1\hat{M}_2]=[\hat{M}_3, \hat{M}_1\hat{M}_2]=0,
    \nonumber\\
    &\hat{M}_3\hat{M}_4\hat{M}_3^{-1}=\hat{M}_4^{-1},
\label{eq:alb_Z22}
\end{align}
with $i=1,2,3,4$.

In the case of non-supersymmetric heterotic strings, we have to search for $\hat{Z}$ satisfying Eqs. \eqref{condition_T-duality} and \eqref{Z_condition}. 
We list all the possible configurations of $\hat{Z}$ in Appendix \ref{app:Z22}, where the moduli fields 
transform as 
\begin{align}
    \begin{array}{ll}
    T\xrightarrow{\hat{M}_1}-\frac{T}{T^2-Z^2},& Z\xrightarrow{\hat{M}_1}\frac{Z}{T^2-Z^2},\\
    T\xrightarrow{\hat{M}_2}T+1,& Z\xrightarrow{\hat{M}_2}Z,\\
    T\xrightarrow{\hat{M}_3}T,& Z\xrightarrow{\hat{M}_3}-Z,\\
    T\xrightarrow{\hat{M}_4}T,& Z\xrightarrow{\hat{M}_4}Z+1.
    \end{array}
\end{align}
Table \ref{tab:Z22} indicates that the non-supersymmetric string theories have the same modular symmetry as the supersymmetric one for a specific $\hat{Z}$, but the modular symmetries are in general broken down. The remaining symmetries are summarized in Appendix \ref{app:Z22}. 
It turns out that depending on the choice of $\hat{Z}$, only a subgroup of \eqref{eq:modularZ22} remains in non-supersymmetric heterotic string theories. 
Note that the $\mathcal{CP}$-like transformation, which is an outer automorphism of the Narain space group, 
is defined as
\begin{align}
    \hat{\mathcal{CP}} = \hat{\Sigma}_\ast,
\end{align}
with $\hat{\mathcal{CP}} \hat{\Theta} \hat{\mathcal{CP}}^{-1}=\hat{\Theta}^{-1}$.

\subsubsection{Symmetric $\mathbb{Z}_2$ orbifold with discrete Wilson line}
\label{sec:Z23}

As a last example with the moduli space of complex dimension two, we discuss the $\mathbb{Z}_2$ Narain point group $\hat{P}_{\mathrm{Narain}}$, generated by
\begin{align}
    \hat{\Theta}^\prime := \left(\hat{K}_S\right)^2
    \hat{W}
    \begin{psmallmatrix}
        1\\
        0
\end{psmallmatrix}
=\left(\hat{C}_S\right)^2
    \hat{W}
    \begin{psmallmatrix}
        1\\
        0
\end{psmallmatrix}
\label{eq:twist_Z23}
.
\end{align}
Under this $\mathbb{Z}_2$ twist, the metric ${\cal H}$ is constrained by \eqref{eq:H_cond}, which leads to the vanishing Wilson line modulus $Z=1/2$ with $a_1=0$ and $a_2=-1/2$. 
We arrive at the following configuration of the moduli in the Siegel upper half-plane:
\begin{align}
 \tau=\begin{pmatrix}
        U&1/2\\
        1/2&T\\
    \end{pmatrix}
    .
    \label{eq:tauZ23}
\end{align}
Note that the the fixed point \eqref{eq:tauZ22} in Sec. \ref{sec:Z22} is related to \eqref{eq:tauZ23} under the following basis change:
\begin{align}
    b:=\begin{pmatrix}
        1&0&0&0\\
        1&0&0&-1\\
        0&-1&1&0\\
        0&1&0&0\\
    \end{pmatrix},
\end{align}
i.e.,
\begin{align}
    b \begin{pmatrix}
        U^\prime&1/2\\
        1/2&T^\prime\\
    \end{pmatrix}
    =
    \begin{pmatrix}
        T&Z\\
        Z&T\\
    \end{pmatrix}
    ,
\end{align}
with $U^\prime = (T + Z)/2$ and $T^\prime = (2(Z - T))^{-1}$. 
Hence, the Narain twist \eqref{eq:twist_Z23} is related to \eqref{eq:twist_Z22} in Sec. \ref{sec:Z22}:
\begin{align}
    \hat{\Theta}^\prime = \hat{B}^{-1}\hat{\Theta}\hat{B},
\end{align}
with
\begin{align}
    \hat{B}:= \left(\hat{K}_T\right)^{-1}
        \hat{W}
    \begin{psmallmatrix}
        0\\
        1
\end{psmallmatrix}
\hat{K}_T \left(\hat{K}_S\right)^3
\in {\cal O}_{\hat{\eta}}(18, 2, \mathbb{Z}).
\end{align}

In the language of $\mathrm{Sp}(4,\mathbb{Z})$, the above moduli matrix corresponds to the fixed point of $\mathrm{Sp}(4,\mathbb{Z})$ generated by the following stabilizer $\bar{H}\cong\mathbb{Z}_2$:
\begin{align}
    h=\begin{pmatrix}
        1&0&0&-1\\
        0&-1&1&0\\
        0&0&1&0\\
        0&0&0&-1\\
    \end{pmatrix}.
\end{align}
Furthermore, the normalizer $N(H)$ on \eqref{eq:tauZ23} 
can be obtained by the basis change of \eqref{eq:normalizer_Z22}, i.e.,
\begin{align}
    \langle b^{-1}M_{(\mathrm{S},\,\mathrm{S})}b, \,b^{-1}M_{(\mathrm{T},\,\mathrm{T})}b,\, b^{-1}M_{(\mathrm{S}^2,\,\mathbb{I}_2)}b,\, b^{-1}M\begin{psmallmatrix}
        -1\\
        0
    \end{psmallmatrix}
    b\rangle
\end{align}
respectively corresponds to an outer automorphism of $\hat{P}_{\mathrm{Narain}}$, namely the modular group of the symmetric $\mathbb{Z}_2$ orbifold with discrete Wilson line in the case of supersymmetric heterotic string theory:
\begin{align}
    \langle \hat{M}_1, \hat{M}_2, \hat{M}_3, \hat{M}_4\rangle,
\label{eq:modularZ23}
\end{align}
with
\begin{align}
    \hat{M}_1:= \hat{B}^{-1}\hat{K}_S \hat{C}_S \hat{B},\quad
    \hat{M}_2:= \hat{B}^{-1}\hat{K}_T \hat{C}_T \hat{B},\quad
    \hat{M}_3:= \hat{B}^{-1}\left(\hat{K}_S\right)^2\hat{B},\quad
    \hat{M}_4:= \hat{B}^{-1}\hat{W}
    \begin{psmallmatrix}
        -1\\
        0
    \end{psmallmatrix} \hat{B},
\end{align}
obeying the same algebra \eqref{eq:alb_Z22}.

In the case of non-supersymmetric heterotic strings, we have to search for $\hat{Z}$ satisfying Eqs. \eqref{condition_T-duality} and \eqref{Z_condition}. 
We list all the possible configurations of $\hat{Z}$ in Appendix \ref{app:Z23}, where the moduli fields 
transform as
\begin{align}
    \begin{array}{ll}
    T\xrightarrow{\hat{M}_1}-\frac{1}{4T},& U\xrightarrow{\hat{M}_1}-\frac{1}{4U},\\
    T\xrightarrow{\hat{M}_2}-\frac{T}{2T-1},& U\xrightarrow{\hat{M}_2}U+\frac{1}{2},\\
    T\xrightarrow{\hat{M}_3}-\frac{1}{4U},& U\xrightarrow{\hat{M}_3}-\frac{1}{4T},\\
    T\xrightarrow{\hat{M}_4}\frac{T}{2T+1},& U\xrightarrow{\hat{M}_4}U+\frac{1}{2}.
    \end{array}
\end{align}
Since the modular group is given by the basis change of \eqref{eq:normalizer_Z22}, we arrive at the same remaining modular symmetries of Sec. \ref{sec:Z22}. 
The transformation of $T$ and $U$ under $\hat{M}_1$ 
is embedded into $\mathrm{SL}(2,\mathbb{R})$ not on $\mathrm{SL}(2,\mathbb{Z})$, i.e.,
\begin{align}
\begin{pmatrix}
    0 & -\frac{1}{\sqrt{4}}\\
    \sqrt{4} & 0
\end{pmatrix}
\in \mathrm{SL}(2,\mathbb{R}),
\end{align}
which is the so-called ``Scaling duality'' discussed in the context of type IIB flux compactifications on toroidal orbifolds \cite{Ishiguro:2023flux,Ishiguro:2025pwa}. 
Note that this transformation can be embedded into $\mathrm{Sp}(4,\mathbb{Z})$. 
The $\mathcal{CP}$-like transformation, which is an outer automorphism of the Narain space group, 
is defined as
\begin{align}
    \hat{\mathcal{CP}}^\prime = \hat{B}^{-1}\hat{\Sigma}_\ast\hat{B},
\end{align}
with $\hat{\mathcal{CP}}^\prime \hat{\Theta} (\hat{\mathcal{CP}}^{\prime})^{-1}=\hat{\Theta}^{-1}$.

\subsection{$\mathrm{Sp}(4,\mathbb{Z})$ fixed points with dimension 1}
\label{sec:dim1}

In this section, we discuss the $\mathrm{Sp}(4,\mathbb{Z})$ fixed points where the complex dimension of the moduli space is one. 
There are five possibilities: (i) symmetric $\mathbb{Z}_4$ orbifold in Sec. \ref{sec:Z4}, (ii) symmetric $\mathbb{Z}_3$ and $\mathbb{Z}_6$ orbifolds in Sec. \ref{sec:Z6}, (iii) asymmetric $\mathbb{Z}_2\times \mathbb{Z}_2$ orbifold without Wilson line in Sec. \ref{sec:Z2Z21}, (iv) asymmetric $\mathbb{Z}_2\times \mathbb{Z}_2$ orbifold with discrete Wilson line in Sec. \ref{sec:Z2Z22}, and (v) asymmetric $S_3$ orbifold in Sec. \ref{sec:S3}.

\subsubsection{Symmetric $\mathbb{Z}_4$ orbifold without Wilson line}
\label{sec:Z4}

For the moduli space of complex dimension one, we start with the Narain point group $\hat{P}_{\mathrm{Narain}}$ of the $\mathbb{Z}_4$ orbifold, generated by
\begin{align}
    \hat{\Theta} := \hat{C}_S.
\end{align}
Under this $\mathbb{Z}_4$ twist, the metric ${\cal H}$ is constrained by \eqref{eq:H_cond}, which leads to the vanishing Wilson line modulus $Z=0$ and fixed complex structure modulus $U=i$ due to $G_{11}=G_{22}$ and $G_{12}=0$. On the other hand, $T$ is still unconstrained. 
Then, we arrive at the following configuration of the moduli in the Siegel upper half-plane:
\begin{align}
 \tau=\begin{pmatrix}
        i& 0\\
        0& T\\
    \end{pmatrix}
    .
    \label{eq:tauZ4}
\end{align}
In the language of $\mathrm{Sp}(4,\mathbb{Z})$, the above moduli matrix corresponds to the fixed point of $\mathrm{Sp}(4,\mathbb{Z})$ generated by the following stabilizer $\bar{H}\cong \mathbb{Z}_4$:
\begin{align}
    h=\begin{pmatrix}
        0&0&1&0\\
        0& 1&0&0\\
        -1&0&0&0\\
        0&0&0&1\\
    \end{pmatrix}
    .
\end{align}
Furthermore, the normalizer $N(H)$ on \eqref{eq:tauZ4}
\begin{align}
    \langle M_{(\mathrm{S},\,\mathbb{I}_2)}, \,M_{(\mathrm{T},\,\mathbb{I}_2)},\, M_{(\mathbb{I}_2,\,\mathrm{S})}\rangle
\end{align}
respectively corresponds to an outer automorphism of $\hat{P}_{\mathrm{Narain}}$, namely the modular group of the symmetric $\mathbb{Z}_4$ orbifold in the case of supersymmetric heterotic string theory:
\begin{align}
    \langle \hat{K}_S, \hat{K}_T, \hat{C}_S\rangle \cong 
    \frac{\mathrm{SL}(2,\mathbb{Z})_{\mathrm{T}}\times \mathbb{Z}_4}{\mathbb{Z}_2},
\label{eq:modularZ4}
\end{align}
where $\mathbb{Z}_2$ originates from $\left(\hat{K}_S\right)^2=\left(\hat{C}_S\right)^2$.

In the case of non-supersymmetric heterotic strings, we have to search for $\hat{Z}$ satisfying Eqs. \eqref{condition_T-duality} and \eqref{Z_condition}. 
We list all the possible configurations of $\hat{Z}$ in Appendix \ref{app:Z4}, where the moduli fields 
transform as under the generator of the modular group:
\begin{align}
    T\xrightarrow{\hat{K}_S}-\frac{1}{T},\qquad T\xrightarrow{\hat{K}_T}T+1,\qquad T\xrightarrow{\hat{C}_S}T.
\label{eq:modulartrfZ4}
\end{align}
It turns out that depending on the choice of $\hat{Z}$, only a subgroup of \eqref{eq:modularZ4} remains in non-supersymmetric heterotic string theories. 
Since the modular transformation \eqref{eq:modulartrfZ4} is already discussed in Sec. \ref{sec:classification}, the remaining modular symmetries are the same ones as summarized in Table \ref{tab:modularZ4}:
\begin{table}[H]
\begin{center}
\caption{The modular group in the case of symmetric $\mathbb{Z}_4$ orbifold without Wilson line.}
\begin{tabular}{|c||c||c|} \hline
    $\#$ & $\left(\hat{q}_{1}\;\mathrm{mod}\;1,\hat{q}_{2}\;\mathrm{mod}\;1; \hat{w}_{1},\hat{w}_{2};\hat{n}_{1},\hat{n}_{2} \right)$ & $G_{\mathrm{modular}}$   \\\hline
    $1$ & $\left(0,0;0,0;0,0\right)$ & $(\mathrm{SL}(2,\mathbb{Z})\times \mathbb{Z}_4)/\mathbb{Z}_2$ \\
    $2$ & $\left(0,0;0,0;0,1\right)$ & $\Gamma_1(2)\times\mathbb{Z}_2$ \\
    $3$ & $\left(0,0;0,0;1,0\right)$ & $\Gamma_1(2)\times\mathbb{Z}_2$ \\
    $4$ & $\left(0,0;0,0;1,1\right)$ & $\Gamma_1(2)\times \mathbb{Z}_4$ \\
    $5$ & $\left(0,0;0,1;0,0\right)$ & $\Gamma^1(2)\times\mathbb{Z}_2$ \\
    $6$ & $\left(0,0;0,1;0,1\right)$ & $\Gamma(2)\times\mathbb{Z}_2$ \\
    $7$ & $\left(0,0;0,1;1,0\right)$ & $\Gamma_{\vartheta}\times\mathbb{Z}_2$ \\
    $8$ & $\left(0,0;0,1;1,1\right)$ & $\Gamma(2)\times\mathbb{Z}_2$ \\
    $9$ & $\left(0,0;1,0;0,0\right)$ & $\Gamma^1(2)\times\mathbb{Z}_2$ \\
    $10$ & $\left(0,0;1,0;0,1\right)$ & $\Gamma_{\vartheta}\times\mathbb{Z}_2$ \\
    $11$ & $\left(0,0;1,0;1,0\right)$ & $\Gamma(2)\times\mathbb{Z}_2$ \\
    $12$ & $\left(0,0;1,0;1,1\right)$ & $\Gamma(2)\times\mathbb{Z}_2$  \\
    $13$ & $\left(0,0;1,1;0,0\right)$ & $\Gamma^1(2)\times \mathbb{Z}_4$ \\
    $14$ & $\left(0,0;1,1;0,1\right)$ & $\Gamma(2)\times\mathbb{Z}_2$ \\
    $15$ & $\left(0,0;1,1;1,0\right)$ & $\Gamma(2)\times\mathbb{Z}_2$ \\
    $16$ & $\left(0,0;1,1;1,1\right)$ & $\Gamma_\vartheta\times \mathbb{Z}_4$ \\
    \hline
\end{tabular}
\label{tab:modularZ4}
\end{center}
\end{table}	
Note that the $\mathcal{CP}$-like transformation, which is an outer automorphism of the Narain space group, 
is defined as
\begin{align}
    \hat{\mathcal{CP}} = \hat{\Sigma}_\ast,
\end{align}
with $\hat{\mathcal{CP}} \hat{\Theta} \hat{\mathcal{CP}}^{-1}=\hat{\Theta}^{-1}$. 
Such a $\mathcal{CP}$-like transformation always exists in this class of $\mathbb{Z}_4$ orbifold.

\subsubsection{Symmetric $\mathbb{Z}_3$ and $\mathbb{Z}_6$ orbifolds without Wilson line}
\label{sec:Z6}

Let us deal with the Narain point group $\hat{P}_{\mathrm{Narain}}$ of the $\mathbb{Z}_6$ orbifold, generated by
\begin{align}
    \hat{\Theta} := \left(\hat{C}_S\right)^3 \hat{C}_T\hat{C}_S\hat{C}_T,
\end{align}
as an element of ${\cal O}_{\hat{\eta}}(18,2,\mathbb{Z})$. 
Under this $\mathbb{Z}_6$ twist, the metric ${\cal H}$ is constrained by \eqref{eq:H_cond}, which leads to the vanishing Wilson line modulus $Z=0$ and fixed complex structure modulus $U=\omega$ with $\omega =e^{\frac{2\pi i}{3}}$ due to $G_{11}=G_{22}=-2G_{12}$ and $a_1=a_2=0$. On the other hand, $T$ is still unconstrained. 
Then, we arrive at the following configuration of the moduli in the Siegel upper half-plane:
\begin{align}
 \tau=\begin{pmatrix}
        \omega& 0\\
        0& T\\
    \end{pmatrix}
    .
    \label{eq:tauZ6}
\end{align}
In the language of $\mathrm{Sp}(4,\mathbb{Z})$, the above moduli matrix corresponds to the fixed point of $\mathrm{Sp}(4,\mathbb{Z})$ generated by the following stabilizer $\bar{H}\cong \mathbb{Z}_{6}$:
\begin{align}
    h=\begin{pmatrix}
        1&0&1&0\\
        0& 1&0&0\\
        -1&0&0&0\\
        0&0&0&1\\
    \end{pmatrix}.
\end{align}
Furthermore, the normalizer $N(H)$ on \eqref{eq:tauZ6}
\begin{align}
    \langle M_{(\mathrm{S},\,\mathbb{I}_2)}, \,M_{(\mathrm{T},\,\mathbb{I}_2)},\, M_{(\mathbb{I}_2,\,\mathrm{S}^3\mathrm{T})}\rangle
\end{align}
respectively corresponds to an outer automorphism of $\hat{P}_{\mathrm{Narain}}$, namely the modular group of the symmetric $\mathbb{Z}_6$ orbifold in the case of supersymmetric heterotic string theory:
\begin{align}
    \langle \hat{K}_S, \hat{K}_T, \left(\hat{C}_S\right)^3\hat{C}_T\rangle
    \cong \frac{\left(\mathrm{SL}(2,\mathbb{Z})_{\mathrm{T}} \times \mathbb{Z}_6\right)}{\mathbb{Z}_2},
\label{eq:modularZ6}
\end{align}
with $\left(\hat{C}_S\right)^3\hat{C}_T=\hat{\Theta}^{-1}$ being a $\mathbb{Z}_6$ rotation. 
For a symmetric $\mathbb{Z}_3$ orbifold, the Narain twist is defined as $\Theta^2$ as a subgroup of $\mathbb{Z}_6$, and one can obtain the same modular group.

In the case of non-supersymmetric heterotic strings, we have to search for $\hat{Z}$ satisfying Eqs. \eqref{condition_T-duality} and \eqref{Z_condition}. 
We list all the possible configurations of $\hat{Z}$ in Appendix \ref{app:Z6}, where the moduli fields 
transform as under the generator of the modular group:
\begin{align}
    T\xrightarrow{\hat{K}_S}-\frac{1}{T},\qquad T\xrightarrow{\hat{K}_T}T+1,\qquad T\xrightarrow{(\hat{C}_S)^3\hat{C}_T}T.
\end{align}
It turns out that depending on the choice of $\hat{Z}$, only a subgroup of \eqref{eq:modularZ6} remains in non-supersymmetric heterotic string theories. 
In the case of $\mathbb{Z}_6$ orbifold, we list the remaining symmetries in Table \ref{tab:modularZ6}:\footnote{One can obtain a similar structure for the $\mathbb{Z}_3$ orbifold.}
\begin{table}[H]
\begin{center}
\caption{The modular group in the case of symmetric $\mathbb{Z}_6$ orbifold without Wilson line.}
\begin{tabular}{|c||c||c|} \hline
    $\#$ & $\left(\hat{q}_{1}\;\mathrm{mod}\;1,\hat{q}_{2}\;\mathrm{mod}\;1;\hat{w}_{1},\hat{w}_{2};\hat{n}_{1},\hat{n}_{2} \right)$ & $G_{\mathrm{modular}}$   \\\hline
    $1$ & $\left(0,0;0,0;0,0\right)$ & $(\mathrm{SL}(2,\mathbb{Z})\times \mathbb{Z}_6)/\mathbb{Z}_2$ \\
    $2$ & $\left(0,0;0,0;0,1\right)$ & $\Gamma_1(2)\times\mathbb{Z}_2$ \\
    $3$ & $\left(0,0;0,0;1,0\right)$ & $\Gamma_1(2)\times\mathbb{Z}_2$ \\
    $4$ & $\left(0,0;0,0;1,1\right)$ & $\Gamma_1(2)\times \mathbb{Z}_2$ \\
    $5$ & $\left(0,0;0,1;0,0\right)$ & $\Gamma^1(2)\times\mathbb{Z}_2$ \\
    $6$ & $\left(0,0;0,1;0,1\right)$ & $\Gamma(2)\times\mathbb{Z}_2$ \\
    $7$ & $\left(0,0;0,1;1,0\right)$ & $\Gamma_{\vartheta}\times\mathbb{Z}_2$ \\
    $8$ & $\left(0,0;0,1;1,1\right)$ & $\Gamma(2)\times\mathbb{Z}_2$ \\
    $9$ & $\left(0,0;1,0;0,0\right)$ & $\Gamma^1(2)\times\mathbb{Z}_2$ \\
    $10$ & $\left(0,0;1,0;0,1\right)$ & $\Gamma_{\vartheta}\times\mathbb{Z}_2$ \\
    $11$ & $\left(0,0;1,0;1,0\right)$ & $\Gamma(2)\times\mathbb{Z}_2$ \\
    $12$ & $\left(0,0;1,0;1,1\right)$ & $\Gamma(2)\times\mathbb{Z}_2$  \\
    $13$ & $\left(0,0;1,1;0,0\right)$ & $\Gamma^1(2)\times\mathbb{Z}_2$ \\
    $14$ & $\left(0,0;1,1;0,1\right)$ & $\Gamma(2)\times\mathbb{Z}_2$ \\
    $15$ & $\left(0,0;1,1;1,0\right)$ & $\Gamma(2)\times\mathbb{Z}_2$ \\
    $16$ & $\left(0,0;1,1;1,1\right)$ & $\Gamma_\vartheta\times\mathbb{Z}_2$ \\
    \hline
\end{tabular}
\label{tab:modularZ6}
\end{center}
\end{table}	
Note that the $\mathcal{CP}$-like transformation, which is an outer automorphism of the Narain space group, 
is defined as
\begin{align}
    \hat{\mathcal{CP}} = \hat{W}\begin{psmallmatrix}
        -1\\
        0
    \end{psmallmatrix}\hat{\Sigma}_\ast,
\end{align}
with $\hat{\mathcal{CP}} \hat{\Theta} \hat{\mathcal{CP}}^{-1}=\hat{\Theta}^{-1}$. \\

\subsubsection{Asymmetric $\mathbb{Z}_2\times \mathbb{Z}_2$ orbifold without Wilson line}
\label{sec:Z2Z21}

In this section, we discuss the Narain point group $\hat{P}_{\mathrm{Narain}}$ of the $\mathbb{Z}_2\times \mathbb{Z}_2$ orbifold, each which is generated by
\begin{align}
    \hat{\Theta}_1 := \hat{M},\qquad \hat{\Theta}_2 := \hat{M}\left(\hat{K}_S\right)^2,
\end{align}
as an element of ${\cal O}_{\hat{\eta}}(18,2,\mathbb{Z})$. 
Under this $\mathbb{Z}_2\times \mathbb{Z}_2$ twist, the metric ${\cal H}$ is constrained by \eqref{eq:H_cond}, which leads to the vanishing Wilson line modulus $Z=0$ and fixed moduli $U=T$ due to $G_{11}=\alpha^\prime$, $G_{12}=B_{12}$ and $a_1=a_2=0$. Then, we arrive at the following configuration of the moduli in the Siegel upper half-plane:
\begin{align}
 \tau=\begin{pmatrix}
        T& 0\\
        0& T\\
    \end{pmatrix}
    .
    \label{eq:tauZ2Z21}
\end{align}
In the language of $\mathrm{Sp}(4,\mathbb{Z})$, the above moduli matrix corresponds to the fixed point of $\mathrm{Sp}(4,\mathbb{Z})$ generated by the following stabilizer $\bar{H}\cong D_8$:
\begin{align}
    h_1=\begin{pmatrix}
        0&1&0&0\\
        1&0&0&0\\
        0&0&0&1\\
        0&0&1&0\\
    \end{pmatrix},\quad
    h_2=\begin{pmatrix}
        0&-1&0&0\\
        1&0&0&0\\
        0&0&0&-1\\
        0&0&1&0\\
    \end{pmatrix}.
\end{align}
Furthermore, the normalizer $N(H)$ on \eqref{eq:tauZ2Z21}
\begin{align}
    \langle M_{(\mathrm{S},\,\mathrm{S})}, \,M_{(\mathrm{T},\,\mathrm{T})},\, M_{(\mathrm{S}^2,\,\mathbb{I}_2)},\, M_\times\rangle
\end{align}
respectively corresponds to an outer automorphism of $\hat{P}_{\mathrm{Narain}}$, namely the modular group of the symmetric $\mathbb{Z}_2\times\mathbb{Z}_2$ orbifold in the case of supersymmetric heterotic string theory:
\begin{align}
    \langle \hat{K}_S\hat{C}_S, \hat{K}_T\hat{C}_T, \left(\hat{K}_S\right)^2 , \hat{M}\rangle
    \cong \mathrm{PSL}(2,\mathbb{Z})\times \mathbb{Z}_2 \times \mathbb{Z}_2^{\hat{M}},
\label{eq:modularZ2Z21}
\end{align}
with $\left(\hat{K}_S\hat{C}_S\right)^2=\mathbb{I}$.

In the case of non-supersymmetric heterotic strings, we have to search for $\hat{Z}$ satisfying Eqs. \eqref{condition_T-duality} and \eqref{Z_condition}. 
We list all the possible configurations of $\hat{Z}$ in Appendix \ref{app:Z2Z21}, where the moduli fields 
transform as under the generator of the modular group:
\begin{align}
    &T\xrightarrow{\hat{K}_S\hat{C}_S}-\frac{1}{T},\qquad T\xrightarrow{\hat{K}_T\hat{C}_T}T+1,\qquad T\xrightarrow{(\hat{K}_S)^2}T,\qquad T\xrightarrow{\hat{M}}T.
\end{align}
It turns out that depending on the choice of $\hat{Z}$, only a subgroup of \eqref{eq:modularZ2Z21} remains in non-supersymmetric heterotic string theories. 
We list the remaining symmetries in Table \ref{tab:modularZ2Z21}:
\begin{table}[H]
\begin{center}
\caption{The modular group in the case of asymmetric $\mathbb{Z}_2\times \mathbb{Z}_2$ orbifold without Wilson line.}
\begin{tabular}{|c||c||c|} \hline
    $\#$ & $\left(\hat{q}_{1}\;\mathrm{mod}\;1,\hat{q}_{2}\;\mathrm{mod}\;1;\hat{w}_{1},\hat{w}_{2};\hat{n}_{1},\hat{n}_{2} \right)$ & $G_{\mathrm{modular}}$   \\\hline
    $1$ & $\left(0,0;0,0;0,0\right)$ & $\mathrm{PSL}(2,\mathbb{Z})\times \mathbb{Z}_2 \times \mathbb{Z}_2^{\hat{M}}$ \\
    $2$ & $\left(0,0;0,0;0,1\right)$ & $\Gamma_1(2)\times \mathbb{Z}_2 \times \mathbb{Z}_2^{\hat{M}}$ \\
    $3$ & $\left(0,0;0,0;1,0\right)$ & $\Gamma(2)\times \mathbb{Z}_2 \times \mathbb{Z}_2^{\hat{M}}$ \\
    $4$ & $\left(0,0;0,0;1,1\right)$ & $\Gamma(2) \times \mathbb{Z}_2 \times \mathbb{Z}_2^{\hat{M}}$ \\
    $5$ & $\left(0,0;0,1;0,0\right)$ & $\Gamma^1(2)\times \mathbb{Z}_2 \times \mathbb{Z}_2^{\hat{M}}$ \\
    $6$ & $\left(0,0;0,1;0,1\right)$ & $\Gamma_{\vartheta}\times \mathbb{Z}_2 \times \mathbb{Z}_2^{\hat{M}}$ \\
    $7$ & $\left(0,0;0,1;1,0\right)$ & $\Gamma(2)\times \mathbb{Z}_2 \times \mathbb{Z}_2^{\hat{M}}$ \\
    $8$ & $\left(0,0;0,1;1,1\right)$ & $\mathbb{Z}_3$ \\
    $9$ & $\left(0,0;1,0;0,0\right)$ & $\Gamma(2)\times \mathbb{Z}_2 \times \mathbb{Z}_2^{\hat{M}}$ \\
    $10$ & $\left(0,0;1,0;0,1\right)$ & $\Gamma(2) \times \mathbb{Z}_2 \times \mathbb{Z}_2^{\hat{M}}$ \\
    $11$ & $\left(0,0;1,0;1,0\right)$ & $\mathrm{PSL}(2,\mathbb{Z})\times \mathbb{Z}_2 \times \mathbb{Z}_2^{\hat{M}}$ \\
    $12$ & $\left(0,0;1,0;1,1\right)$ & $\Gamma_1(2)\times \mathbb{Z}_2 \times \mathbb{Z}_2^{\hat{M}}$  \\
    $13$ & $\left(0,0;1,1;0,0\right)$ & $\Gamma(2)\times \mathbb{Z}_2 \times \mathbb{Z}_2^{\hat{M}}$ \\
    $14$ & $\left(0,0;1,1;0,1\right)$ & Residue class of $\Gamma(2)$ \\
    $15$ & $\left(0,0;1,1;1,0\right)$ & $\Gamma^1(2)\times \mathbb{Z}_2 \times \mathbb{Z}_2^{\hat{M}}$ \\
    $16$ & $\left(0,0;1,1;1,1\right)$ & $\Gamma_\vartheta \times \mathbb{Z}_2 \times \mathbb{Z}_2^{\hat{M}}$ \\
    \hline
\end{tabular}
\label{tab:modularZ2Z21}
\end{center}
\end{table}	
Note that the $\mathcal{CP}$-like transformation, which is an outer automorphism of the Narain space group, 
is defined as
\begin{align}
    \hat{\mathcal{CP}} = \hat{\Sigma}_\ast
\end{align}
with $\hat{\mathcal{CP}} \hat{\Theta}_i \hat{\mathcal{CP}}^{-1}=\hat{\Theta}_i^{-1}$ ($i=1,2$).\\

\subsubsection{Asymmetric $\mathbb{Z}_2\times \mathbb{Z}_2$ orbifold with discrete Wilson line}
\label{sec:Z2Z22}

In this section, we focus on a different asymmetric $\mathbb{Z}_2\times \mathbb{Z}_2$ orbifold, each which is generated by
\begin{align}
    \hat{\Theta}_1 := \hat{M},\qquad \hat{\Theta}_2 := 
    \hat{W}\begin{psmallmatrix}
        -1\\
        0
    \end{psmallmatrix}\hat{M}\left(\hat{C}_S\right)^2,
\end{align}
as an element of ${\cal O}_{\hat{\eta}}(18,2,\mathbb{Z})$. 
Under this $\mathbb{Z}_2\times \mathbb{Z}_2$ twist, the metric ${\cal H}$ is constrained by \eqref{eq:H_cond}, which leads to the Wilson line modulus $Z=1/2$ and fixed moduli $U=T$ due to $G_{11}=\alpha^\prime$, $G_{12}=B_{12}$, $a_1=0$ and $a_2=-1/2$. Then, we arrive at the following configuration of the moduli in the Siegel upper half-plane:
\begin{align}
 \tau=\begin{pmatrix}
        T& 1/2\\
        1/2& T\\
    \end{pmatrix}
    .
    \label{eq:tauZ2Z22}
\end{align}
In the language of $\mathrm{Sp}(4,\mathbb{Z})$, the above moduli matrix corresponds to the fixed point of $\mathrm{Sp}(4,\mathbb{Z})$ generated by the following stabilizer $\bar{H}\cong D_8/\{\pm\mathbb{I}_4\}\cong D_4$:
\begin{align}
    h_1=\begin{pmatrix}
        0&1&0&0\\
        1&0&0&0\\
        0&0&0&1\\
        0&0&1&0\\
    \end{pmatrix},\quad
    h_2=\begin{pmatrix}
        0&1&-1&0\\
        -1&0&0&1\\
        0&0&0&1\\
        0&0&-1&0\\
    \end{pmatrix}.
\end{align}
These generators satisfy $h_1^2=h_2^4=(h_2h_1)^2=\mathbb{I}_4$ and we have $\bar{H}=D_8/\{\pm\mathbb{I}_4\}\cong D_4$. 
Furthermore, the normalizer $N(H)$ on \eqref{eq:tauZ2Z22}
\begin{align}
    \langle M\begin{psmallmatrix}
        -1\\
        0
    \end{psmallmatrix}M_{(\mathrm{S}^2,\,\mathbb{I}_2)}, \,M_\times M\begin{psmallmatrix}
        0\\
        -2
    \end{psmallmatrix}M_\times
    M
    \begin{psmallmatrix}
        0\\
        1
    \end{psmallmatrix}
    M_{(\mathrm{S}\mathrm{T}^{-1}\mathrm{S}^2,\,\mathrm{S}\mathrm{T}^2\mathrm{S}\mathrm{T})}M\begin{psmallmatrix}
        -1\\
        3
    \end{psmallmatrix},\, 
    M\begin{psmallmatrix}
        -1\\
        0
    \end{psmallmatrix}M_{(\mathrm{S}^2\mathrm{T}^{-1},\,\mathrm{T}^{-1})}\rangle
\end{align}
respectively corresponds to an outer automorphism of $\hat{P}_{\mathrm{Narain}}$, namely the modular group of the symmetric $\mathbb{Z}_2\times\mathbb{Z}_2$ orbifold with discrete Wilson line in the case of supersymmetric heterotic string theory:
\begin{align}
    \langle \hat{M}_1,\, \hat{M}_2, \hat{M}_3\rangle,
\label{eq:modularZ2Z22}
\end{align}
with
\begin{align}
\hat{M}_1&:= \hat{\Theta}_1\hat{\Theta}_2,
\nonumber\\
\hat{M}_2&:= \hat{M}
\hat{W}\begin{psmallmatrix}
    0\\
    -2
\end{psmallmatrix}
\hat{M}
\hat{W}\begin{psmallmatrix}
    0\\
    1
\end{psmallmatrix}
\hat{K}_S\left(\hat{K}_T\right)^{-1}\left(\hat{K}_S\right)^{2}
\hat{C}_S\left(\hat{C}_T\right)^{2}\hat{C}_S\hat{C}_T
\hat{W}\begin{psmallmatrix}
    -1\\
    3
\end{psmallmatrix}
,
\nonumber\\
\hat{M}_3&:= 
\hat{W}\begin{psmallmatrix}
    -1\\
    0
\end{psmallmatrix}
\left(\hat{K}_S\right)^{2}
\left(\hat{K}_T\right)^{-1}
\left(\hat{C}_T\right)^{-1},
\end{align}
obeying
\begin{align}
&\hat{M}_1^2=\hat{M}_2^4=(\hat{M}_2\hat{M}_3)^4=(\hat{M}_3\hat{M}_2)^4=\mathbb{I}_{20},
\nonumber\\
&\hat{M}_1\hat{M}_2(\hat{M}_1)^{-1}=(\hat{M}_2)^{-1},
\quad
[\hat{M}_1, \hat{M}_3]=[(\hat{M}_2)^2, \hat{M}_1]=[(\hat{M}_2)^2, \hat{M}_3]=0,
\end{align}
with $\ell =1,3$. 
It includes $D_4\cong \mathbb{Z}_4^{(\hat{M}_2)}\rtimes \mathbb{Z}_2^{(\hat{M}_1)}$ and the free product $\mathbb{Z}_4\ast \mathbb{Z}_4$ associated with $\hat{M}_2$ and $\hat{M}_2\hat{M}_3$.

In the case of non-supersymmetric heterotic strings, we have to search for $\hat{Z}$ satisfying Eqs. \eqref{condition_T-duality} and \eqref{Z_condition}. 
We list all the possible configurations of $\hat{Z}$ in Appendix \ref{app:Z2Z22}, where the moduli fields 
transform as under the generator of the modular group:
\begin{align}
    T\xrightarrow{\hat{M}_1}T,\qquad 
    T\xrightarrow{\hat{M}_2}-\frac{2T+3}{4T+2},\qquad T\xrightarrow{\hat{M}_3}T-1.
\end{align}
Table \ref{tab:Z2Z22} indicates that the non-supersymmetric string theories have the same modular symmetry as the supersymmetric one for a specific $\hat{Z}$, but the modular symmetries are in general broken down. The remaining symmetries are summarized in Appendix \ref{app:Z2Z22}. 
 
Note that the transformation $\hat{M}_2$ belongs to $\mathrm{SL}(2,\mathbb{R})$ not $\mathrm{SL}(2,\mathbb{Z})$, but it can be embedded to $\mathrm{Sp}(4,\mathbb{Z})$. 
The $\mathcal{CP}$-like transformation, which is an outer automorphism of the Narain space group, 
is defined as
\begin{align}
    \hat{\mathcal{CP}} =
    \hat{W}\begin{psmallmatrix}
    -1\\
    0
\end{psmallmatrix}
    \hat{\Sigma}_\ast,
\end{align}
with $\hat{\mathcal{CP}} \hat{\Theta}_i \hat{\mathcal{CP}}^{-1}=\hat{\Theta}_i^{-1}$ ($i=1,2$).\\

\subsubsection{Asymmetric $S_3$ orbifold with Wilson line}
\label{sec:S3}

As a last example of the moduli space with complex dimension one, we study  an asymmetric $S_3$ orbifold, which is generated by
\begin{align}
    \hat{\Theta}_1 := \hat{M}
    \hat{W}
    \begin{psmallmatrix}
        0\\
        1
    \end{psmallmatrix}\hat{M}\left(\hat{C}_S\right)^2,\qquad \hat{\Theta}_2 := 
    \hat{W}\begin{psmallmatrix}
        0\\
        1
    \end{psmallmatrix}\left(\hat{K}_S\right)^2,
\end{align}
as an element of ${\cal O}_{\hat{\eta}}(18,2,\mathbb{Z})$. 
Note that the Narain twists satisfy $\left(\hat{\Theta}_1\right)^2=\left(\hat{\Theta}_2\right)^2=\left(\hat{\Theta}_1\hat{\Theta}_2\right)^3=\mathbb{I}_{20}$. 
Under this $S_3$ twist, the metric ${\cal H}$ is constrained by \eqref{eq:H_cond}, which leads to the Wilson line modulus $Z=T/2$ and fixed moduli $U=T$ due to $G_{11}=\frac{3}{4}\alpha^\prime$, $G_{12}=B_{12}$, $a_1=1/2$ and $a_2=0$. Then, we arrive at the following configuration of the moduli in the Siegel upper half-plane:
\begin{align}
 \tau=\begin{pmatrix}
        T& T/2\\
        T/2& T\\
    \end{pmatrix}
    .
    \label{eq:tauS3}
\end{align}
In the language of $\mathrm{Sp}(4,\mathbb{Z})$, the above moduli matrix corresponds to the fixed point of $\mathrm{Sp}(4,\mathbb{Z})$ generated by the following stabilizer $\bar{H}\cong S_3$:
\begin{align}
    h_1=\begin{pmatrix}
        -1&1&0&0\\
        0&1&0&0\\
        0&0&-1&0\\
        0&0&1&1\\
    \end{pmatrix},\quad
    h_2=\begin{pmatrix}
        1&0&0&0\\
        1&-1&0&0\\
        0&0&1&1\\
        0&0&0&-1\\
    \end{pmatrix}
\end{align}
satisfying $h_1^2=h_2^2=(h_1h_2)^3=\mathbb{I}_4$. 
Furthermore, the normalizer $N(H)$ on \eqref{eq:tauS3}
\begin{align}
    \langle M_{(\mathrm{S},\,\mathrm{S}^3)}M_\times, \,M_{(\mathrm{S}^3,\,\mathrm{S}^3)} M\begin{psmallmatrix}
        -1\\
        0
    \end{psmallmatrix}M_{(\mathrm{T}^{-2},\,\mathrm{T}^{-2})}M_{(\mathrm{S},\,\mathrm{S})},\, 
    M_\times,\,
    M\begin{psmallmatrix}
        0\\
        1
    \end{psmallmatrix}M_{(\mathrm{S}^2,\,\mathbb{I}_2)}\rangle
\end{align}
respectively correspond to an outer automorphism of $\hat{P}_{\mathrm{Narain}}$, namely the modular group of the asymmetric $S_3$ orbifold with Wilson line in the case of supersymmetric heterotic string theory:
\begin{align}
    \langle \hat{M}_1,\, \hat{M}_2,\,
    \hat{M}_3,\,
    \hat{M}_4\rangle
    \cong \mathrm{PSL}(2,\mathbb{Z})\times D_6,
\label{eq:modularS3}
\end{align}
with
\begin{align}
\hat{M}_1&:= \hat{K}_S\left(\hat{C}_S\right)^{3}\hat{M},
\nonumber\\
\hat{M}_2&:= 
\left(\hat{K}_S\right)^{3}
\left(\hat{C}_S\right)^{3}
\hat{W}\begin{psmallmatrix}
    -1\\
    0
\end{psmallmatrix}
\left(\hat{K}_T\right)^{-2}
\left(\hat{C}_T\right)^{-2}
\hat{K}_S\hat{C}_S,
\nonumber\\
\hat{M}_3&:=\hat{M},
\nonumber\\
\hat{M}_4&:=\hat{W}\begin{psmallmatrix}
        0\\
        1
    \end{psmallmatrix}\left(\hat{K}_S\right)^2,
\end{align}
obeying
\begin{align}
&\hat{M}_1^2=\hat{M}_3^2=\hat{M}_4^2=(\hat{M}_1\hat{M}_2)^6=(\hat{M}_3\hat{M}_4)^6=\mathbb{I}_{20},
\nonumber\\
&[\hat{M}_1, \hat{M}_\ell]=[\hat{M}_2, \hat{M}_\ell]=[(\hat{M}_2\hat{M}_1)^n, \hat{M}_\ell]=[(\hat{M}_3\hat{M}_4)^n, \hat{M}_p]=[(\hat{M}_3\hat{M}_4)^n, (\hat{M}_3\hat{M}_4)^m]=0,
\end{align}
with $n,m\in \mathbb{Z}$, $p=1,2$ and $\ell = 3,4$. 
The $\mathrm{PSL}(2,\mathbb{Z})$ is generated by $\{\hat{M}_1,\hat{M}_2\hat{M}_1\hat{M}_2\}$, and $D_6\cong \mathbb{Z}_6^{(\hat{M}_3\hat{M}_4)}\rtimes \mathbb{Z}_2^{(\hat{M}_4)}$ originates from $\{\hat{M}_3,\hat{M}_4\}$.

In the case of non-supersymmetric heterotic strings, we have to search for $\hat{Z}$ satisfying Eqs. \eqref{condition_T-duality} and \eqref{Z_condition}. 
We list all the possible configurations of $\hat{Z}$ in Appendix \ref{app:S3}, where the moduli fields 
transform as under the generator of the modular group:
\begin{align}
    &T\xrightarrow{\hat{M}_1}-\frac{4}{3T},\qquad T\xrightarrow{\hat{M}_2}\frac{2T}{3T+2},\qquad
    T\xrightarrow{\hat{M}_3}T,\qquad T\xrightarrow{\hat{M}_4}T.
\end{align}
Table \ref{tab:S3} indicates that the non-supersymmetric string theories have the same modular symmetry as the supersymmetric one for a specific $\hat{Z}$, but the modular symmetries are in general broken down. The remaining symmetries are summarized in Appendix \ref{app:S3}:

Note that the transformation of $T$ under $\hat{M}_1$ 
is embedded into $\mathrm{SL}(2,\mathbb{R})$ not on $\mathrm{SL}(2,\mathbb{Z})$, but it can be embedded into $\mathrm{Sp}(4,\mathbb{Z})$.i.e.,
\begin{align}
\begin{pmatrix}
    0 & -\sqrt{\frac{4}{3}}\\
    \sqrt{\frac{3}{4}} & 0
\end{pmatrix}
\in \mathrm{SL}(2,\mathbb{R}),
\end{align}
which is the so-called Scaling duality \cite{Ishiguro:2023flux,Ishiguro:2025pwa}.

Note that the $\mathcal{CP}$-like transformation, which is an outer automorphism of the Narain space group, 
is defined as
\begin{align}
    \hat{\mathcal{CP}} =
    \hat{W}\begin{psmallmatrix}
    -1\\
    0
\end{psmallmatrix}
    \hat{\Sigma}_\ast,
\end{align}
with $\hat{\mathcal{CP}} \hat{\Theta}_i \hat{\mathcal{CP}}^{-1}=\hat{\Theta}_i^{-1}$ ($i=1,2$).

\subsection{$\mathrm{Sp}(4,\mathbb{Z})$ fixed points with dimension 0}
\label{sec:dim0}

In this section, we discuss the $\mathrm{Sp}(4,\mathbb{Z})$ fixed points where the complex dimension of the moduli space is zero. 
There are six non-equivalent isolated fixed points: (i) asymmetric $\mathbb{Z}_5$ orbifold in Sec. \ref{sec:Z5}, (ii) asymmetric $S_4$ orbifold in Sec. \ref{sec:S4}, (iii) asymmetric $(\mathbb{Z}_4\times\mathbb{Z}_2)\rtimes \mathbb{Z}_2$ orbifold in Sec. \ref{sec:Z4Z2Z2}, (iv) asymmetric $S_3\times \mathbb{Z}_6$ orbifold in Sec. \ref{sec:S3Z6}, (v) asymmetric $S_3\times \mathbb{Z}_2$ orbifold in Sec. \ref{sec:S3Z2}, and (vi) asymmetric $\mathbb{Z}_{12}$ orbifold in Sec. \ref{sec:Z12}. 
Note that in these cases, the stabilizer is the same as the normalizer in contrast to previous examples with the moduli space of complex dimensions one and two. 

\subsubsection{Asymmetric $\mathbb{Z}_5$ orbifold with discrete Wilson line}
\label{sec:Z5}

For the moduli space with complex dimension zero, we start with an asymmetric $\mathbb{Z}_5$ orbifold with discrete Wilson line, which is generated by
\begin{align}
\label{eq:ThetaZ5}
    \hat{\Theta} := \hat{C}_T\hat{M}
    \hat{W}
    \begin{psmallmatrix}
        0\\
        1
    \end{psmallmatrix}\left(\hat{K}_S\right)^2\left(\hat{C}_S\right)^3,
\end{align}
as an element of ${\cal O}_{\hat{\eta}}(18,2,\mathbb{Z})$, 
with $\hat{\Theta}^{10}=\mathbb{I}_{20}$.\footnote{Although the generator $\hat{\Theta}$ does not satisfy $\hat{\Theta}^5=\mathbb{I}_{20}$, $\hat{\Theta}^5$ does not change $\mathcal{H}(T,U,Z)$ in \eqref{transformation_moduli}. In this sense, the generator has the $\mathbb{Z}_5$ symmetry.}
Under this $\mathbb{Z}_5$ twist, the metric ${\cal H}$ is constrained by \eqref{eq:H_cond}, which leads to the following configuration of the moduli in the Siegel upper half-plane:
\begin{align}
 \tau=\begin{pmatrix}
        \zeta& \zeta + \zeta^{-2}\\
        \zeta + \zeta^{-2}& -\zeta^{-1}\\
    \end{pmatrix}
    ,
    \label{eq:tauZ5}
\end{align}
due to 
\begin{align}
    G_{11}&=G_{22}=\frac{\alpha^\prime}{2}(-5+3\sqrt{5}),\quad
    G_{12}=\frac{\alpha^\prime}{2}(5-2\sqrt{5}),\quad 
    B_{12}=\frac{\alpha^\prime}{2}(2-\sqrt{5}),
    \nonumber\\
    a_1&=\frac{3-\sqrt{5}}{2},
    \quad
    a_2 =\frac{-1+\sqrt{5}}{2}. 
\end{align}

In the language of $\mathrm{Sp}(4,\mathbb{Z})$, the above moduli matrix corresponds to the fixed point of $\mathrm{Sp}(4,\mathbb{Z})$ generated by the following stabilizer $\bar{H}\cong \mathbb{Z}_5$:
\begin{align}
    h=\begin{pmatrix}
        0&-1&-1&-1\\
        0&0&-1&0\\
        0&0&0&-1\\
        1&0&0&1\\
    \end{pmatrix}
\end{align}
satisfying $h^5=\mathbb{I}_4$. 
Since the normalizer group is the same as the stabilizer group, the modular group in the case of supersymmetric heterotic string theory is the same as the Narain point group, i.e., $\mathbb{Z}_5$ generated by \eqref{eq:ThetaZ5}.

In the case of non-supersymmetric heterotic strings, we have to search for $\hat{Z}$ satisfying Eqs. \eqref{condition_T-duality} and \eqref{Z_condition}. 
We list all the possible configurations of $\hat{Z}$ in Appendix \ref{app:Z5}, indicating that the existence of $\mathbb{Z}_5$ symmetry depends on the choice of $\hat{Z}$ in non-supersymmetric heterotic string theories.    
Note that the $\mathcal{CP}$-like transformation, which is an outer automorphism of the Narain space group, 
is defined as
\begin{align}
    \hat{\mathcal{CP}} =
    \left( \left(\hat{K}_S\hat{C}_S\hat{W}\begin{psmallmatrix}
    -1\\
    0
\end{psmallmatrix}
\right)^3
\hat{W}\begin{psmallmatrix}
    -1\\
    0
\end{psmallmatrix}
\right)^{-1}
    \hat{\Sigma}_\ast,
\end{align}
with $\hat{\mathcal{CP}} \hat{\Theta} \hat{\mathcal{CP}}^{-1}=\hat{\Theta}^{-1}$.

\subsubsection{Asymmetric $S_4$ orbifold with discrete Wilson line}
\label{sec:S4}

We move to an asymmetric $S_4$ orbifold with discrete Wilson line, which is generated by two twists:
\begin{align}
\label{eq:ThetaS4}
    \hat{\Theta}_1 := \hat{M},\qquad 
    \hat{\Theta}_2 :=
    \left(\hat{C}_S\right)^3\hat{M}
    \hat{W}
    \begin{psmallmatrix}
        0\\
        -1
    \end{psmallmatrix}
    \hat{K}_S
    \hat{W}
    \begin{psmallmatrix}
        -1\\
        -1
    \end{psmallmatrix}
    ,
\end{align}
as an element of ${\cal O}_{\hat{\eta}}(18,2,\mathbb{Z})$, 
with $\left(\hat{\Theta}_1\right)^2=\left(\hat{\Theta}_2\right)^4=\left(\hat{\Theta}_1\hat{\Theta}_2\right)^3=\mathbb{I}_{20}$. 
Under this $S_4$ twist, the metric ${\cal H}$ is constrained by \eqref{eq:H_cond}, which leads to the following configuration of the moduli in the Siegel upper half-plane:
\begin{align}
 \tau=\begin{pmatrix}
    \tilde{\eta}&\frac{1}{2}(\tilde{\eta}-1)\\
    \frac{1}{2}(\tilde{\eta}-1)&\tilde{\eta}\\
\end{pmatrix}\quad\mathrm{with}\;\;\ \tilde{\eta}=\frac{1}{3}(1+i2\sqrt{2}),
\label{eq:tauS4}
\end{align}
due to 
\begin{align}
    G_{11}&=G_{22}=\frac{3\alpha^\prime}{4},\quad
    G_{12}=\frac{\alpha^\prime}{4},\quad
    B_{12}=\frac{\alpha^\prime}{2},
    \quad
    a_1=a_2=\frac{1}{2}. 
\end{align}

In the language of $\mathrm{Sp}(4,\mathbb{Z})$, the above moduli matrix corresponds to the fixed point of $\mathrm{Sp}(4,\mathbb{Z})$ generated by the following stabilizer $\bar{H}\cong S_4$:
\begin{align}
    h_1=\begin{pmatrix}
        0&1&0&0\\
        1&0&0&0\\
        0&0&0&1\\
        0&0&1&0\\
    \end{pmatrix},\quad
    h_2=\begin{pmatrix}
        -1&1&1&0\\
        1&0&0&1\\
        -1&0&0&0\\
        1&-1&0&1\\
    \end{pmatrix}
\end{align}
satisfying $h_1^2=h_2^4=(h_1h_2)^3=1$. 
Since the normalizer group is the same as the stabilizer group, the modular group in the case of supersymmetric heterotic string theory is the same as the Narain point group, i.e., $S_4$ generated by \eqref{eq:ThetaS4}.

In the case of non-supersymmetric heterotic strings, we have to search for $\hat{Z}$ satisfying Eqs. \eqref{condition_T-duality} and \eqref{Z_condition}. 
We list all the possible configurations of $\hat{Z}$ in Appendix \ref{app:S4}. 
Table \ref{tab:S4} indicates that the existence of $S_4$ symmetry depends on the choice of $\hat{Z}$ in non-supersymmetric heterotic string theories. 
We list the remaining symmetries in what follows:
\begin{itemize}
    \item $\mathbb{Z}_2$ symmetry if $\hat{\Theta}_1$ remains.
    \item $\mathbb{Z}_4$ symmetry if $\hat{\Theta}_2$ remains.
    \item $S_4$ symmetry if $\hat{\Theta}_1$ and $\hat{\Theta}_2$ remain.
\end{itemize} 
Note that the $\mathcal{CP}$-like transformation, which is an outer automorphism of the Narain space group, 
is defined as \cite{Ding:2021iqp}
\begin{align}
    \hat{\mathcal{CP}} =
\hat{K}_S\hat{C}_S\hat{W}\begin{psmallmatrix}
    -1\\
    0
\end{psmallmatrix}
\left( \hat{C}_T\hat{K}_T\right)^{-1}
\hat{K}_S\hat{C}_S
    \hat{\Sigma}_\ast,
\end{align}
with $\hat{\mathcal{CP}} \hat{\Theta}_i \hat{\mathcal{CP}}^{-1}=\hat{\Theta}_i^{-1}$ for $i=1,2$.

\subsubsection{Asymmetric $(\mathbb{Z}_4\times \mathbb{Z}_2)\rtimes \mathbb{Z}_2$ orbifold without Wilson line}
\label{sec:Z4Z2Z2}

We move to an asymmetric $(\mathbb{Z}_4\times \mathbb{Z}_2)\rtimes \mathbb{Z}_2$ orbifold without Wilson line, which is generated by three twists:
\begin{align}
\label{eq:ThetaZ4Z2Z2}
    \hat{\Theta}_1 := \left(\hat{K}_S\right)^2\hat{C}_S,\qquad 
    \hat{\Theta}_2 :=
    \hat{K}_S\left(\hat{C}_S\right)^3,\qquad 
    \hat{\Theta}_3 :=
    \hat{M},
\end{align}
as an element of ${\cal O}_{\hat{\eta}}(18,2,\mathbb{Z})$, 
with
\begin{align}
\left(\hat{\Theta}_1\right)^4&=\left(\hat{\Theta}_2\right)^2=\left(\hat{\Theta}_3\right)^2=\mathbb{I}_{20},
\quad
[\hat{\Theta}_1,\hat{\Theta}_2]
=
[\hat{\Theta}_2,\hat{\Theta}_3]
=0,
\quad
\hat{\Theta}_3\hat{\Theta}_1\hat{\Theta}_3=\hat{\Theta}_1\hat{\Theta}_2.
\label{eq:propZ4Z2Z2}
\end{align}
Under this $(\mathbb{Z}_4\times \mathbb{Z}_2)\rtimes \mathbb{Z}_2$ twist, the metric ${\cal H}$ is constrained by \eqref{eq:H_cond}, which leads to the following configuration of the moduli in the Siegel upper half-plane:
\begin{align}
 \tau=\begin{pmatrix}
    i&0\\
    0&i
\end{pmatrix}
,
\label{eq:tauZ4Z2Z2}
\end{align}
due to 
\begin{align}
    G_{11}&=G_{22}=\alpha^\prime,\quad
    G_{12}=B_{12}=a_1=a_2=0. 
\end{align}

In the language of $\mathrm{Sp}(4,\mathbb{Z})$, the above moduli matrix corresponds to the fixed point of $\mathrm{Sp}(4,\mathbb{Z})$ generated by the following stabilizer $\bar{H}\cong (\mathbb{Z}_4\times \mathbb{Z}_2)\rtimes \mathbb{Z}_2$:
\begin{align}
    h_1=\begin{pmatrix}
        0&0&1&0\\
        0&-1&0&0\\
        -1&0&0&0\\
        0&0&0&-1\\
    \end{pmatrix},\quad
    h_2=\begin{pmatrix}
        0&0&-1&0\\
        0&0&0&1\\
        1&0&0&0\\
        0&-1&0&0\\
    \end{pmatrix},\quad
    h_3=\begin{pmatrix}
        0&1&0&0\\
        1&0&0&0\\
        0&0&0&1\\
        0&0&1&0\\
    \end{pmatrix}
    ,
\end{align}
satisfying the same properties \eqref{eq:propZ4Z2Z2}. 
Since the normalizer group is the same as the stabilizer group, the modular group in the case of supersymmetric heterotic string theory is the same as the Narain point group, i.e., $(\mathbb{Z}_4\times \mathbb{Z}_2)\rtimes \mathbb{Z}_2$ generated by \eqref{eq:ThetaZ4Z2Z2}.

In the case of non-supersymmetric heterotic strings, we have to search for $\hat{Z}$ satisfying Eqs. \eqref{condition_T-duality} and \eqref{Z_condition}. 
We list all the possible configurations of $\hat{Z}$ in Appendix \ref{app:Z4Z2Z2}. Table \ref{tab:Z4Z2Z2} indicates that the existence of $(\mathbb{Z}_4\times \mathbb{Z}_2)\rtimes \mathbb{Z}_2$ symmetry depends on the choice of $\hat{Z}$ in non-supersymmetric heterotic string theories. 
We list the remaining symmetries in what follows:
\begin{itemize}
    \item $(\mathbb{Z}_4\times \mathbb{Z}_2)\rtimes \mathbb{Z}_2$ symmetry generated by $\{\hat{\Theta}_1,\hat{\Theta}_2, \hat{\Theta}_3\}$.
    \item $\mathbb{Z}_2\times \mathbb{Z}_2$ symmetry generated by $\{\hat{\Theta}_2, \hat{\Theta}_3\}$.
    \item $D_4\cong \mathbb{Z}_4\rtimes \mathbb{Z}_2$ symmetry generated by $\{\hat{\Theta}_1, \hat{\Theta}_3\}$.
    \item $\mathbb{Z}_4\times \mathbb{Z}_2$ symmetry generated by $\{\hat{\Theta}_1,\hat{\Theta}_2\}$.
    \item In the other cases, one can realize the Abelian symmetries such as $\mathbb{Z}_4$ generated by $\hat{\Theta}_1$, $\mathbb{Z}_2$ generated by $\hat{\Theta}_2$, and $\mathbb{Z}_2$ generated by $\hat{M}$.
\end{itemize}
Note that the $\mathcal{CP}$-like transformation, which is an outer automorphism of the Narain space group, 
is defined as \cite{Ding:2021iqp}
\begin{align}
    \hat{\mathcal{CP}} =
    \hat{\Sigma}_\ast,
\end{align}
with $\hat{\mathcal{CP}} \hat{\Theta}_i \hat{\mathcal{CP}}^{-1}=\hat{\Theta}_i^{-1}$ for $i=1,2,3$.

\subsubsection{Asymmetric $S_3 \times \mathbb{Z}_6$ orbifold without Wilson line}
\label{sec:S3Z6}

In this section, we discuss an asymmetric $S_3 \times \mathbb{Z}_6$ orbifold without Wilson line, which is generated by three twists:
\begin{align}
\label{eq:ThetaS3Z6}
    \hat{\Theta}_1 := \left(\hat{K}_T\right)^{-1} \hat{K}_S\left(\hat{C}_S\right)^3\hat{C}_T\hat{M},\qquad 
    \hat{\Theta}_2 := \hat{M},\qquad    
    \hat{\Theta}_3 :=
\left(\hat{K}_S\right)^3\hat{K}_T\hat{C}_S\hat{C}_T,
\end{align}
as an element of ${\cal O}_{\hat{\eta}}(18,2,\mathbb{Z})$, 
with
$\left(\hat{\Theta}_1\right)^2=\left(\hat{\Theta}_2\right)^2=\left(\hat{\Theta}_1\hat{\Theta}_2\right)^3=\mathbb{I}_{20}$. 
Under this $S_3 \times \mathbb{Z}_6$ twist, the metric ${\cal H}$ is constrained by \eqref{eq:H_cond}, which leads to the following configuration of the moduli in the Siegel upper half-plane:
\begin{align}
 \tau=\begin{pmatrix}
    \omega&0\\
    0&\omega
\end{pmatrix}
,
\label{eq:tauS3Z6}
\end{align}
with $\omega=e^{2\pi i/3}$, due to 
\begin{align}
    G_{11}&=G_{22}=\alpha^\prime,\quad
    G_{12}=B_{12}=-\frac{\alpha^\prime}{2},\quad a_1=a_2=0. 
\end{align}

In the language of $\mathrm{Sp}(4,\mathbb{Z})$, the above moduli matrix corresponds to the fixed point of $\mathrm{Sp}(4,\mathbb{Z})$ generated by the following stabilizer $\bar{H}\cong S_3 \times \mathbb{Z}_6$:
\begin{align}
    h_1=\begin{pmatrix}
        0&0&0&-1\\
        1&0&1&0\\
        0&1&0&1\\
        -1&0&0&0\\
    \end{pmatrix},\quad
    h_2=\begin{pmatrix}
        0&1&0&0\\
        1&0&0&0\\
        0&0&0&1\\
        0&0&1&0\\
    \end{pmatrix},\quad
    h_3=\begin{pmatrix}
        0&0&1&0\\
        0&0&0&-1\\
        -1&0&-1&0\\
        0&1&0&1\\
    \end{pmatrix}
\end{align}
satisfying $(h_1)^2 = (h_2)^2 = (h_1h_2)^3 =\mathbb{I}_4$. 
Since the normalizer group is the same as the stabilizer group, the modular group in the case of supersymmetric heterotic string theory is the same as the Narain point group, i.e., $S_3 \times \mathbb{Z}_6$ generated by \eqref{eq:ThetaS3Z6}.

In the case of non-supersymmetric heterotic strings, we have to search for $\hat{Z}$ satisfying Eqs. \eqref{condition_T-duality} and \eqref{Z_condition}. 
We list all the possible configurations of $\hat{Z}$ in Appendix \ref{app:S3Z6}, indicating that the existence of $S_3 \times \mathbb{Z}_6$ symmetry depends on the choice of $\hat{Z}$ in non-supersymmetric heterotic string theories. 
We list the remaining symmetries in what follows:
\begin{itemize}
    \item $S_3\times \mathbb{Z}_6$ symmetry generated by $\{\hat{\Theta}_1, \hat{\Theta}_2, \hat{\Theta}_3\}$.
    \item $S_3$ symmetry generated by $\{\hat{\Theta}_1, \hat{\Theta}_2\}$.
    \item In the other cases, one can realize the Abelian symmetries such as $\mathbb{Z}_2^{(i)}$ generated by $\hat{\Theta}_i$, $\mathbb{Z}_6$ generated by $\hat{\Theta}_3$, and $\mathbb{Z}_2^{(i)}\times \mathbb{Z}_6$ generated by $\{\hat{\Theta}_i,\hat{\Theta}_3\}$ ($i=1,2$).
\end{itemize} 
Note that the $\mathcal{CP}$-like transformation, which is an outer automorphism of the Narain space group, 
is defined as \cite{Ding:2021iqp}
\begin{align}
    \hat{\mathcal{CP}} =
    \left(\hat{C}_T \hat{K}_T\right)^{-1}
    \hat{\Sigma}_\ast,
\end{align}
with $\hat{\mathcal{CP}} \hat{\Theta}_i \hat{\mathcal{CP}}^{-1}\in [\hat{\Theta}_i^{-1}]$ for $i=1,2,3$ indicating an outer automorphism of the Narain group.

\subsubsection{Asymmetric $S_3 \times \mathbb{Z}_2$ orbifold with discrete Wilson line}
\label{sec:S3Z2}

In this section, we discuss an asymmetric $S_3 \times \mathbb{Z}_2$ orbifold with discrete Wilson line, which is generated by three twists:
\begin{align}
\label{eq:ThetaS3Z2}
    \hat{\Theta}_1 := \hat{M}\hat{K}_S \hat{C}_S
    \hat{W}
    \begin{psmallmatrix}
        0\\
        -1
    \end{psmallmatrix},\qquad 
    \hat{\Theta}_2 := \hat{K}_S \hat{C}_S
    \hat{W}
    \begin{psmallmatrix}
        0\\
        -1
    \end{psmallmatrix}\hat{M},\qquad    
    \hat{\Theta}_3 :=
\left(\hat{K}_S\right)^3\hat{C}_S\hat{M},
\end{align}
as an element of ${\cal O}_{\hat{\eta}}(18,2,\mathbb{Z})$, 
with
\begin{align}
\left(\hat{\Theta}_1\right)^2&=\left(\hat{\Theta}_2\right)^2=\left(\hat{\Theta}_1\hat{\Theta}_2\right)^3=\left(\hat{\Theta}_3\right)^2=\mathbb{I}_{20},
\quad
[\hat{\Theta}_3,\hat{\Theta}_1]=[\hat{\Theta}_3,\hat{\Theta}_2]=0.
\end{align}
Under this $S_3 \times \mathbb{Z}_2$ twist, the metric ${\cal H}$ is constrained by \eqref{eq:H_cond}, which leads to the following configuration of the moduli in the Siegel upper half-plane:
\begin{align}
 \tau=\frac{i}{\sqrt{3}}\begin{pmatrix}
    2&1\\
    1&2
\end{pmatrix}
,
\label{eq:tauS3Z2}
\end{align}
due to 
\begin{align}
    G_{11}&=\frac{3\alpha^\prime}{4},\quad G_{22}=\alpha^\prime,\quad
    G_{12}=B_{12}=0,\quad a_1=\frac{1}{2},\quad a_2=0. 
\end{align}

In the language of $\mathrm{Sp}(4,\mathbb{Z})$, the above moduli matrix corresponds to the fixed point of $\mathrm{Sp}(4,\mathbb{Z})$ generated by the following stabilizer $\bar{H}\cong S_3 \times \mathbb{Z}_2$:
\begin{align}
    h_1=\begin{pmatrix}
        0&0&0&1\\
        0&0&1&1\\
        1&-1&0&0\\
        -1&0&0&0\\
    \end{pmatrix},\quad
    h_2=\begin{pmatrix}
        0&0&1&1\\
        0&0&1&0\\
        0&-1&0&0\\
        -1&1&0&0\\
    \end{pmatrix},\quad
    h_3=\begin{pmatrix}
        0&0&0&1\\
        0&0&-1&0\\
        0&-1&0&0\\
        1&0&0&0\\
    \end{pmatrix}
\end{align}
satisfying $(h_1)^2=(h_2)^2=(h_1h_2)^3=(h_3)^2=\mathbb{I}_4$ in $\bar{H}$. 
Since the normalizer group is the same as the stabilizer group, the modular group in the case of supersymmetric heterotic string theory is the same as the Narain point group, i.e., $S_3 \times \mathbb{Z}_2$ generated by \eqref{eq:ThetaS3Z2}.

In the case of non-supersymmetric heterotic strings, we have to search for $\hat{Z}$ satisfying Eqs. \eqref{condition_T-duality} and \eqref{Z_condition}. 
We list all the possible configurations of $\hat{Z}$ in Appendix \ref{app:S3Z2}, indicating that the existence of $S_3 \times \mathbb{Z}_2$ symmetry depends on the choice of $\hat{Z}$ in non-supersymmetric heterotic string theories. 
We list the remaining symmetries in what follows:
\begin{itemize}
    \item $S_3\times \mathbb{Z}_2$ symmetry generated by $\{\hat{\Theta}_1, \hat{\Theta}_2, \hat{\Theta}_3\}$.
    \item $S_3$ symmetry generated by $\{\hat{\Theta}_1, \hat{\Theta}_2\}$.
    \item $\mathbb{Z}_2\times \mathbb{Z}_2$ symmetry generated by $\{\hat{\Theta}_1,\hat{\Theta}_3\}$ or $\{\hat{\Theta}_2,\hat{\Theta}_3\}$.
    \item $\mathbb{Z}_2^i$ symmetry generated by $\hat{\Theta}_i$ ($i=1,2,3$).
\end{itemize}  
Note that the $\mathcal{CP}$-like transformation, which is an outer automorphism of the Narain space group, 
is defined as \cite{Ding:2021iqp}
\begin{align}
    \hat{\mathcal{CP}} =
    \hat{\Sigma}_\ast,
\end{align}
with $\hat{\mathcal{CP}} \hat{\Theta}_i \hat{\mathcal{CP}}^{-1}= \hat{\Theta}_i^{-1}$ for $i=1,2,3$. 

\subsubsection{Asymmetric $\mathbb{Z}_{12}$ orbifold without Wilson line}
\label{sec:Z12}

As a last example for the moduli space of dimension zero, we discuss an asymmetric $\mathbb{Z}_{12}$ orbifold without Wilson line generated by a single element which can be written, in the chosen basis, as the product of three twists:
\begin{align}
\label{eq:ThetaZ12}
    \hat{\Theta} := \hat{K}_S \hat{C}_S
    \hat{C}_T,
\end{align}
as an element of ${\cal O}_{\hat{\eta}}(18,2,\mathbb{Z})$, 
with $\left(\hat{\Theta}\right)^{12}=\mathbb{I}_{20}$. 
Under this $\mathbb{Z}_{12}$ twist, the metric ${\cal H}$ is constrained by \eqref{eq:H_cond}, which leads to the following configuration of the moduli in the Siegel upper half-plane:
\begin{align}
 \tau=\begin{pmatrix}
    \omega&0\\
    0&i
\end{pmatrix}
,
\label{eq:tauZ12}
\end{align}
due to 
\begin{align}
    G_{11}=G_{22}=\frac{2\alpha^\prime}{\sqrt{3}},\quad
    G_{12}=-\frac{\alpha^\prime}{\sqrt{3}},\quad B_{12}=0,\quad a_1=a_2=0. 
\end{align}

In the language of $\mathrm{Sp}(4,\mathbb{Z})$, the above moduli matrix corresponds to the fixed point of $\mathrm{Sp}(4,\mathbb{Z})$ generated by the following stabilizer $\bar{H}\cong \mathbb{Z}_{12}$:
\begin{align}
    h=\begin{pmatrix}
        0&0&1&0\\
        0&0&0&1\\
        -1&0&-1&0\\
        0&-1&0&0\\
    \end{pmatrix}
    ,
\end{align}
satisfying $(h)^{12}=\mathbb{I}_4$ in $\bar{H}$. 
Since the normalizer group is the same as the stabilizer group, the modular group in the case of supersymmetric heterotic string theory is the same as the Narain point group, i.e., $\mathbb{Z}_{12}$ generated by \eqref{eq:ThetaZ12}.

In the case of non-supersymmetric heterotic strings, we have to search for $\hat{Z}$ satisfying Eqs. \eqref{condition_T-duality} and \eqref{Z_condition}. 
We list all the possible configurations of $\hat{Z}$ in Appendix \ref{app:Z12}, indicating that the existence of $\mathbb{Z}_{12}$ symmetry depends on the choice of $\hat{Z}$ in non-supersymmetric heterotic string theories.     
Note that the $\mathcal{CP}$-like transformation, which is an outer automorphism of the Narain space group, 
is defined as \cite{Ding:2021iqp}
\begin{align}
    \hat{\mathcal{CP}} =
    \left(\hat{C}_T\right)^{-1}\hat{\Sigma}_\ast,
\end{align}
with $\hat{\mathcal{CP}} \hat{\Theta} \hat{\mathcal{CP}}^{-1}= \hat{\Theta}^{-1}$.

\section{Conclusion}
\label{sec:conclusion}

We discussed the modular symmetry in non-supersymmetric heterotic string theories on toroidal backgrounds. 
To reveal the modular symmetry, we map the T-duality group ${\cal O}(18, 2, \mathbb{Z})$ to the Siegel modular group $\mathrm{Sp}(4,\mathbb{Z})$ in two-dimensional toroidal compactifications of non-supersymmetric string theories. 
Since the T-duality group in non-supersymmetric string theories is a reduced one in supersymmetric ones due to Scherk-Schwarz compactifications \cite{Itoyama:2021itj}, it is expected to obtain the different modular symmetry from supersymmetric string theories.

In this paper, we focused on $T^2$ and its orbifolds with symmetric and asymmetric twists. 
It is found that the modular symmetries in non-supersymmetric string theories are in general determined by the subgroup of the modular group of supersymmetric string theories for a general configuration of Wilson lines. 
For the $T^2$ compactification, we saw that the gauge symmetry enhancements were determined by the remaining modular symmetries through the examples related to the $SO(16)\times SO(16)$ string. 
Recently, maximal enhancements of gauge groups in the non-supersymmetric heterotic string on $S^1$ were completely classified in \cite{Fraiman:2023cpa}.
It would be interesting to identify maximal enhancements in the $T^2$ case with more general configurations of moduli associated with the part of $O(18,2,\mathbb{Z})$ that cannot be mapped to $\mathrm{Sp}(4,\mathbb{Z})$. 
For the orbifold compactifications, 
a part of moduli fields can be stabilized by the orbifold twists, leading to the non-Abelian discrete symmetries such as $S_3, S_4, D_4, D_6$, as remaining modular symmetries. An existence of $\mathcal{CP}$-like symmetry depends on the configuration $\hat{Z}$ specifying the non-supersymmetric heterotic string theory. 
Since the gauge groups would be fixed by the orbifold twists, it would be interesting to classify the gauge symmetries in orbifold compactifications. We will leave a comprehensive study to future work.

Furthermore, many top-down approaches from the non-supersymmetric strings were proposed in terms of the 
cosmological constant.  We found five tachyon-free vacua corresponding to the maximal enhancements in the non-supersymmetric heterotic strings on $T^2$ directly connected to the $SO(16)\times SO(16)$ string. One of them had a negative cosmological constant and massless scalars, which implies that it sat at an unstable knife edge, while the other four had positive cosmological constants and no massless scalars. Moreover, we found that one of the four maximal enhancements without massless scalars corresponded to a local maximum while the other three saddle points. Local minima could not be found in $T^2$ case, but this is the same situation as in $S^1$ compactification of $SO(16)\times SO(16)$ string \cite{Fraiman:2023cpa}. It may be possible to obtain stable vacua by the orbifold compactifications since moduli can be more stabilized than the toroidal compactifications. The construction of a phenomenologically favored and stable model from both perspectives of the flavor symmetry and the cosmological constant is left for future work.

\acknowledgments

Y.K. and H.O. thank Takuya Hirose for valuable discussions. This work was supported in part by JSPS KAKENHI Grant Numbers JP23H04512 (H.O), JP25H01539 (H.O.), JP26K07087 (H.O.).

\clearpage

\appendix

\section{Modular symmetries on toroidal orbifolds}
\label{app:T2ZN}

In this section, we list the remaining modular transformations for each configuration of $\hat{Z}$ on toroidal orbifolds 
in the context of non-supersymmetric heterotic string theories.

\subsection{Dimension 2}

We present three toroidal orbifolds where the dimension of the moduli space is two.

\subsubsection{Symmetric $\mathbb{Z}_2$ orbifold without Wilson line}
\label{app:Z21}

\begin{center}
\begin{longtable}{|c||c||c|c|}
\caption{Possible configurations of $\left(\hat{q}_{1}\; \mathrm{mod}\;1,\hat{q}_{2}\;\mathrm{mod}\;1;\hat{w}_{1},\hat{w}_{2};\hat{n}_{1},\hat{n}_{2} \right)$ on the symmetric $\mathbb{Z}_2$ orbifold without Wilson line.}
\label{tab:Z21}\\
\hline
    $\#$ & $\left(\hat{q}_{1}\;\mathrm{mod}\;1,\hat{q}_{2}\;\mathrm{mod}\;1;\hat{w}_{1},\hat{w}_{2};\hat{n}_{1},\hat{n}_{2} \right)$ & $G_{\mathrm{modular}}$ & $\hat{\mathcal{CP}}$ \\\hline
    \endfirsthead

    \multicolumn{4}{c}%
    {{\bfseries \tablename \thetable{}--continued from previous page}}\\
    \hline
     $\#$ & $\left(\hat{q}_{1},\hat{q}_{2};\hat{w}_{1},\hat{w}_{2};\hat{n}_{1},\hat{n}_{2} \right)$ & $G_{\mathrm{modular}}$ & $\hat{\mathcal{CP}}$ \\\hline
     \endhead
     \hline \multicolumn{4}{|r|}{{Continued on next page}} \\ \hline
     \endfoot

     \hline
     \endlastfoot
     
    $1$ & $\left(0,0;0,0;0,0\right)$& $\mathrm{SL}(2,\mathbb{Z})\times\mathrm{SL}(2,\mathbb{Z})$ & \checkmark \\
    $2$ & $\left(0,0;0,0;0,1\right)$&$\Gamma_1(2)\times\Gamma_1(2)$ &   \checkmark \\
    $3$ & $\left(0,0;0,0;1,0\right)$& $\Gamma^1(2)\times\Gamma_1(2)$ &  \checkmark \\
    $4$ & $\left(0,0;0,0;1,1\right)$& $\Gamma_\vartheta\times\Gamma_1(2)$ &   \checkmark \\
    $5$ & $\left(0,0;0,1;0,0\right)$& $\Gamma^1(2)\times\Gamma^1(2)$ & \checkmark \\
    $6$ & $\left(0,0;0,1;0,1\right)$& $\Gamma(2)\times\Gamma(2)$ &  \checkmark \\
    $7$ & $\left(0,0;0,1;1,0\right)$& $\Gamma^1(2)\times\Gamma_\vartheta$ & \checkmark \\
    $8$ & $\left(0,0;0,1;1,1\right)$& $\Gamma(2)\times\Gamma(2)$ &   \checkmark \\
    $9$ & $\left(0,0;1,0;0,0\right)$& $\Gamma_1(2)\times\Gamma^1(2)$ &  \checkmark \\
    $10$ & $\left(0,0;1,0;0,1\right)$& $\Gamma_1(2)\times\Gamma_\vartheta$ & \checkmark \\
    $11$ & $\left(0,0;1,0;1,0\right)$& $\Gamma(2)\times\Gamma(2)$ &   \checkmark \\
    $12$ & $\left(0,0;1,0;1,1\right)$& $\Gamma(2)\times\Gamma(2)$ &   \checkmark \\
    $13$ & $\left(0,0;1,1;0,0\right)$& $\Gamma_\vartheta\times\Gamma^1(2)$ &   \checkmark \\
    $14$ & $\left(0,0;1,1;0,1\right)$& $\Gamma(2)\times\Gamma(2)$ &  \checkmark \\
    $15$ & $\left(0,0;1,1;1,0\right)$& $\Gamma(2)\times\Gamma(2)$ &   \checkmark \\
    $16$ & $\left(0,0;1,1;1,1\right)$& $\Gamma_\vartheta\times\Gamma_\vartheta$ &   \checkmark \\
    \hline
\end{longtable}	
\end{center}

\clearpage
\subsubsection{Asymmetric $\mathbb{Z}_2$ orbifold}
\label{app:Z22}

\begin{center}
\begin{longtable}{|c||c||c|c|}
\caption{Possible configurations of $\left(\hat{q}_{1}\;\mathrm{mod}\;1,\hat{q}_{2};\hat{w}_{1},\hat{w}_{2};\hat{n}_{1},\hat{n}_{2} \right)$ on the asymmetric $\mathbb{Z}_2$ orbifold.}
\label{tab:Z22}\\
\hline
     $\#$ & $\left(\hat{q}_{1}\;\mathrm{mod}\;1,\hat{q}_{2};\hat{w}_{1},\hat{w}_{2};\hat{n}_{1},\hat{n}_{2} \right)$ & $G_{\mathrm{modular}}$&   $\hat{\mathcal{CP}}$ \\\hline
    \endfirsthead
     \multicolumn{4}{c}%
    {{\bfseries \tablename \thetable{}--continued from previous page}}\\
    \hline
     $\#$ & $\left(\hat{q}_{1},\hat{q}_{2};\hat{w}_{1},\hat{w}_{2};\hat{n}_{1},\hat{n}_{2} \right)$ & $G_{\mathrm{modular}}$&   $\hat{\mathcal{CP}}$ \\\hline
     \endhead
     \hline \multicolumn{4}{|r|}{{Continued on next page}} \\ \hline
     \endfoot

     \hline
     \endlastfoot
    $1$ & $\left(0,0;0,0;0,0\right)$&$\{\hat{M}_1,\hat{M}_2,\hat{M}_3,\hat{M}_4\}$ &  \checkmark \\
    $2$ & $\left(0,0;0,0;0,1\right)$&$\{(\hat{M}_1\hat{M}_4)^3,\hat{M}_2,\hat{M}_3,\hat{M}_4\}$ & \checkmark \\
    $3$ & $\left(0,0;0,0;1,0\right)$& $\{\hat{M}_3,\hat{M}_4,(\hat{M}_1\hat{M}_4)^2\}$ &  \checkmark \\
    $4$ & $\left(0,0;0,0;1,1\right)$& $\{\hat{M}_3,\hat{M}_4\}$ & \checkmark \\
    $5$ & $\left(0,0;0,1;0,0\right)$& $\{\hat{M}_3,(\hat{M}_1\hat{M}_4)^3\}$ &  \checkmark \\
    $6$ & $\left(0,0;0,1;0,1\right)$& $\{\hat{M}_1,\hat{M}_3,(\hat{M}_1\hat{M}_4)^2\}$ & \checkmark \\
    $7$ & $\left(0,0;0,1;1,0\right)$&$\hat{M}_3$  & \checkmark \\
    $8$ & $\left(0,0;0,1;1,1\right)$& $\{\hat{M}_3,\hat{M}_1\hat{M}_2\}$ & \checkmark \\
    $9$ & $\left(0,0;1,0;0,0\right)$& $\{\hat{M}_3,\hat{M}_4,(\hat{M}_1\hat{M}_4)\}$ & \checkmark \\
    $10$ & $\left(0,0;1,0;0,1\right)$& $\{\hat{M}_3,\hat{M}_4\}$ &   \checkmark \\
    $11$ & $\left(0,0;1,0;1,0\right)$& $\{\hat{M}_1,\hat{M}_2,\hat{M}_3,\hat{M}_4\}$ &  \checkmark \\
    $12$ & $\left(0,0;1,0;1,1\right)$& $\{\hat{M}_3,\hat{M}_4\}$ & \checkmark \\
    $13$ & $\left(0,0;1,1;0,0\right)$& $\hat{M}_3$ &  \checkmark \\
    $14$ & $\left(0,0;1,1;0,1\right)$& $\{\hat{M}_3,\hat{M}_1\hat{M}_2\}$ &   \checkmark \\
    $15$ & $\left(0,0;1,1;1,0\right)$& $\{\hat{M}_3,(\hat{M}_1\hat{M}_4)^3\}$ & \checkmark \\
    $16$ & $\left(0,0;1,1;1,1\right)$& $\{\hat{M}_1,\hat{M}_3,(\hat{M}_1\hat{M}_4)^2\}$ &   \checkmark \\
    $17$ & $\left(0,1;0,0;0,0\right)$& $\{\hat{M}_1,\hat{M}_2,\hat{M}_3\}$ &   \checkmark \\
    $18$ & $\left(0,1;0,0;0,1\right)$& $\{\hat{M}_2,\hat{M}_3\}$ &  \checkmark \\
    $19$ & $\left(0,1;0,0;1,0\right)$& $\hat{M}_3$ &  \checkmark \\
    $20$ & $\left(0,1;0,0;1,1\right)$& $\hat{M}_3$ &  \checkmark \\
    $21$ & $\left(0,1;0,1;0,0\right)$& $\hat{M}_3$ &   \checkmark \\
    $22$ & $\left(0,1;0,1;0,1\right)$& $\{\hat{M}_1,\hat{M}_4\hat{M}_1\hat{M}_4,\hat{M}_3\}$ &   \checkmark \\
    $23$ & $\left(0,1;0,1;1,0\right)$& $\hat{M}_3$ &   \checkmark \\
    $24$ & $\left(0,1;0,1;1,1\right)$&  $\{\hat{M}_3,\hat{M}_1\hat{M}_2, (\hat{M}_1\hat{M}_4)^2\}$ &   \checkmark \\
    $25$ & $\left(0,1;1,0;0,0\right)$& $\hat{M}_3$ &   \checkmark \\
    $26$ & $\left(0,1;1,0;0,1\right)$& $\hat{M}_3$ &   \checkmark \\
    $27$ & $\left(0,1;1,0;1,0\right)$& $\{\hat{M}_1,\hat{M}_2,\hat{M}_3\}$ &   \checkmark \\
    $28$ & $\left(0,1;1,0;1,1\right)$& $\{\hat{M}_2,\hat{M}_3\}$ & \checkmark \\
    $29$ & $\left(0,1;1,1;0,0\right)$& $\hat{M}_3$ &  \checkmark \\
    $30$ & $\left(0,1;1,1;0,1\right)$&$\{\hat{M}_3,\hat{M}_1\hat{M}_2, (\hat{M}_1\hat{M}_4)^2\}$ &   \checkmark \\
    $31$ & $\left(0,1;1,1;1,0\right)$& $\hat{M}_3$ & \checkmark \\
    $32$ & $\left(0,1;1,1;1,1\right)$& $\{\hat{M}_1,\hat{M}_4\hat{M}_1\hat{M}_4,\hat{M}_3\}$ &   \checkmark \\ 
    \hline
\end{longtable}	
\end{center}

Table \ref{tab:Z22} indicates that the non-supersymmetric string theories have the same modular symmetry as the supersymmetric one for a specific $\hat{Z}$, but the modular symmetries are in general broken down. The remaining symmetries are summarized below:
\begin{itemize}
    \item For the configuration of $\hat{Z}$ with $\#1, \#11$,\\
    there exists the same modular symmetry generated by $\{\hat{M}_1,\hat{M}_2,\hat{M}_3,\hat{M}_4\}$ with the supersymmetric heterotic string theory.
    \item For the configuration of $\hat{Z}$ with $\#2, \#12$,\\
    there exists the direct product of $\mathbb{Z}_2^{((\hat{M}_1\hat{M}_4)^3)}$ symmetry, the shift symmetry generated by $\hat{M}_2$ and the infinite dihedral group symmetry $D_\infty \cong \mathbb{Z}\rtimes \mathbb{Z}_2^{(\hat{M}_3)}$ generated by $\{\hat{M}_3,\hat{M}_4\}$.
    \item For the configuration of $\hat{Z}$ with $\#3, \#9$,\\
    there exists the $(\mathbb{Z}\ast \mathbb{Z}_3^{((\hat{M}_1\hat{M}_4)^2)})\rtimes \mathbb{Z}_2^{(\hat{M}_3)}$ generated by $\{\hat{M}_3,\hat{M}_4, (\hat{M}_1\hat{M}_4)^2\}$.
    \item For the configuration of $\hat{Z}$ with $\#4, \#10, \#12$,\\
    there exists the infinite dihedral group $D_\infty \cong \mathbb{Z}\rtimes \mathbb{Z}_2^{(\hat{M}_3)}$ generated by $\{\hat{M}_3,\hat{M}_4\}$.
    \item For the configuration of $\hat{Z}$ with $\#5, \#15$,\\
    there exists the $\mathbb{Z}_2^{(\hat{M}_3)}\times\mathbb{Z}_2^{(\hat{M}_1\hat{M}_4)^3}$ symmetry generated by $\{\hat{M}_3,(\hat{M}_1\hat{M}_4)^3\}$.
    \item For the configuration of $\hat{Z}$ with $\#6, \#16$,\\
    there exists the $\mathbb{Z}_2^{(\hat{M}_1\hat{M}_4)^3}\times \mathbb{Z}_2^{(\hat{M}_1)}\times \mathbb{Z}_2^{(\hat{M}_3)}$ symmetry generated by $\{\hat{M}_1,\hat{M}_3, (\hat{M}_1\hat{M}_4)^3\}$.
    \item For the configuration of $\hat{Z}$ with $\#7, \#13, \#19, \#20, \#21, \#23, \#25, \#26, \#29, \#31$,\\
    there exists the $\mathbb{Z}_2^{(\hat{M}_3)}$ symmetry generated by $\hat{M}_3$.
    \item For the configuration of $\hat{Z}$ with $\#8, \#14$,\\
    there exists the $\mathbb{Z}_2^{(\hat{M}_3)}\times\mathbb{Z}_3^{(\hat{M}_1\hat{M}_2)}$ symmetry generated by $\{\hat{M}_3,\hat{M}_1\hat{M}_2\}$.
    \item For the configuration of $\hat{Z}$ with $\#17$, $\#27$,\\
    there exists the $\mathrm{PSL}(2,\mathbb{Z})\times \mathbb{Z}_2^{(\hat{M}_3)}$ symmetry generated by $\{\hat{M}_1, \hat{M}_2, \hat{M}_3\}$.
    \item For the configuration of $\hat{Z}$ with $\#18$, $\#28$,\\
    there exists a direct product of the shift symmetry and $\mathbb{Z}_2^{(\hat{M}_3)}$ symmetry generated by $\{\hat{M}_2, \hat{M}_3\}$.
    \item For the configuration of $\hat{Z}$ with $\#22$, $\#32$,\\
    there exists the $\mathrm{PGL}(2,\mathbb{Z})$ symmetry:
    \begin{align*}
        \mathrm{PGL}(2,\mathbb{Z})=\langle S, T, U\,|\, S^2=(ST)^3=U^2=\mathbb{I}, \quad UTU^{-1}=T^{-1}\rangle, 
    \end{align*}
    with $S\equiv \hat{M}_1$, $T\equiv \hat{M}_4\hat{M}_1\hat{M}_4$, and $U=\hat{M}_3$. 
    \item For the configuration of $\hat{Z}$ with $\#24$, $\#30$,\\
    there exists the $\mathbb{Z}_3^{((\hat{M}_1\hat{M}_4)^2)}\ast(\mathbb{Z}_3^{(\hat{M}_1\hat{M}_2)}\times \mathbb{Z}_2^{(\hat{M}_3)})$ symmetry generated by $\{\hat{M}_3,\hat{M}_1\hat{M}_2, (\hat{M}_1\hat{M}_4)^2\}$.
\end{itemize}
\clearpage

\subsubsection{Symmetric $\mathbb{Z}_2$ orbifold with discrete Wilson line}
\label{app:Z23}

We can obtain the same modular symmetries in Table \ref{tab:Z22}.

\subsection{Dimension 1}

We present five toroidal orbifolds where the dimension of the moduli space is one.

\subsubsection{Symmetric $\mathbb{Z}_4$ orbifold without Wilson line}
\label{app:Z4}

\begin{center}
\begin{longtable}{|c||c||c|c|}
\caption{Possible configurations of $\left(\hat{q}_{1}\;\mathrm{mod}\;1,\hat{q}_{2}\;\mathrm{mod}\;1;\hat{w}_{1},\hat{w}_{2};\hat{n}_{1},\hat{n}_{2} \right)$ on the symmetric $\mathbb{Z}_4$ orbifold without Wilson line.}
\label{tab:Z4}\\
\hline
   $\#$ & $\left(\hat{q}_{1}\;\mathrm{mod}\;1,\hat{q}_{2}\;\mathrm{mod}\;1;\hat{w}_{1},\hat{w}_{2};\hat{n}_{1},\hat{n}_{2} \right)$ & $G_{\mathrm{modular}}$ & $\hat{\mathcal{CP}}$ \\\hline
    \endfirsthead
     \multicolumn{4}{c}%
    {{\bfseries \tablename \thetable{}--continued from previous page}}\\
    \hline
    $\#$ & $\left(\hat{q}_{1},\hat{q}_{2};\hat{w}_{1},\hat{w}_{2};\hat{n}_{1},\hat{n}_{2} \right)$ &  $G_{\mathrm{modular}}$&  $\hat{\mathcal{CP}}$ \\\hline
     \endhead
     \hline \multicolumn{4}{|r|}{{Continued on next page}} \\ \hline
     \endfoot

     \hline
     \endlastfoot
    $1$ & $\left(0,0;0,0;0,0\right)$& $(\mathrm{SL}(2,\mathbb{Z})\times\mathbb{Z}_4)/\mathbb{Z}_2$ &   \checkmark \\
    $2$ & $\left(0,0;0,0;0,1\right)$&$\Gamma_1(2)\times\mathbb{Z}_2$ &   \checkmark \\
    $3$ & $\left(0,0;0,0;1,0\right)$& $\Gamma_1(2)\times\mathbb{Z}_2$ & \checkmark \\
    $4$ & $\left(0,0;0,0;1,1\right)$& $\Gamma_1(2)\times\mathbb{Z}_4$ &   \checkmark \\
    $5$ & $\left(0,0;0,1;0,0\right)$& $\Gamma^1(2)\times\mathbb{Z}_2$ &    \checkmark \\
    $6$ & $\left(0,0;0,1;0,1\right)$& $\Gamma(2)\times\mathbb{Z}_2$ &   \checkmark \\
    $7$ & $\left(0,0;0,1;1,0\right)$& $\Gamma_\vartheta\times\mathbb{Z}_2$ &\checkmark \\
    $8$ & $\left(0,0;0,1;1,1\right)$& $\Gamma(2)\times\mathbb{Z}_2$ &    \checkmark \\
    $9$ & $\left(0,0;1,0;0,0\right)$& $\Gamma^1(2)\times\mathbb{Z}_2$ &  \checkmark \\
    $10$ & $\left(0,0;1,0;0,1\right)$& $\Gamma_\vartheta\times\mathbb{Z}_2$ &  \checkmark \\
    $11$ & $\left(0,0;1,0;1,0\right)$& $\Gamma(2)\times\mathbb{Z}_2$ &  \checkmark \\
    $12$ & $\left(0,0;1,0;1,1\right)$& $\Gamma(2)\times\mathbb{Z}_2$ &  \checkmark \\
    $13$ & $\left(0,0;1,1;0,0\right)$& $\Gamma^1(2)\times\mathbb{Z}_4$ &    \checkmark \\
    $14$ & $\left(0,0;1,1;0,1\right)$& $\Gamma(2)\times\mathbb{Z}_2$ &   \checkmark \\
    $15$ & $\left(0,0;1,1;1,0\right)$& $\Gamma(2)\times\mathbb{Z}_2$ &   \checkmark \\
    $16$ & $\left(0,0;1,1;1,1\right)$&$\Gamma_\vartheta\times\mathbb{Z}_4$ &  \checkmark \\
    \hline
\end{longtable}	
\end{center}

\clearpage
\subsubsection{Symmetric $\mathbb{Z}_3$ and $\mathbb{Z}_6$ orbifolds without Wilson line}
\label{app:Z6}

\begin{center}
\begin{longtable}{|c||c||c|c|}
\caption{Possible configurations of $\left(\hat{q}_{1}\;\mathrm{mod}\;1,\hat{q}_{2}\;\mathrm{mod}\;1;\hat{w}_{1},\hat{w}_{2};\hat{n}_{1},\hat{n}_{2} \right)$ on the $\mathbb{Z}_3$ and $\mathbb{Z}_6$ orbifolds without Wilson line.}
\label{tab:Z6}\\
\hline
   $\#$ & $\left(\hat{q}_{1}\;\mathrm{mod}\;1,\hat{q}_{2}\;\mathrm{mod}\;1;\hat{w}_{1},\hat{w}_{2};\hat{n}_{1},\hat{n}_{2} \right)$ & $G_{\mathrm{modular}}$&  $\hat{\mathcal{CP}}$ \\\hline
    \endfirsthead
     \multicolumn{4}{c}%
    {{\bfseries \tablename \thetable{}--continued from previous page}}\\
    \hline
    $\#$ & $\left(\hat{q}_{1},\hat{q}_{2};\hat{w}_{1},\hat{w}_{2};\hat{n}_{1},\hat{n}_{2} \right)$ &  $G_{\mathrm{modular}}$ &   $\hat{\mathcal{CP}}$ \\\hline
     \endhead
     \hline \multicolumn{4}{|r|}{{Continued on next page}} \\ \hline
     \endfoot

     \hline
     \endlastfoot
    $1$ & $\left(0,0;0,0;0,0\right)$& $(\mathrm{SL}(2,\mathbb{Z})\times\mathbb{Z}_6)/\mathbb{Z}_2$ &   \checkmark \\
    $2$ & $\left(0,0;0,0;0,1\right)$&$\Gamma_1(2)\times\mathbb{Z}_2$ &  \checkmark \\
    $3$ & $\left(0,0;0,0;1,0\right)$& $\Gamma_1(2)\times\mathbb{Z}_2$ &    --- \\
    $4$ & $\left(0,0;0,0;1,1\right)$& $\Gamma_1(2)\times\mathbb{Z}_2$ &  --- \\
    $5$ & $\left(0,0;0,1;0,0\right)$& $\Gamma^1(2)\times\mathbb{Z}_2$ &   --- \\
    $6$ & $\left(0,0;0,1;0,1\right)$& $\Gamma(2)\times\mathbb{Z}_2$ &    --- \\
    $7$ & $\left(0,0;0,1;1,0\right)$& $\Gamma_\vartheta\times\mathbb{Z}_2$ &  --- \\
    $8$ & $\left(0,0;0,1;1,1\right)$& $\Gamma(2)\times\mathbb{Z}_2$ &   --- \\
    $9$ & $\left(0,0;1,0;0,0\right)$& $\Gamma^1(2)\times\mathbb{Z}_2$ &   \checkmark \\
    $10$ & $\left(0,0;1,0;0,1\right)$& $\Gamma_\vartheta\times\mathbb{Z}_2$ & \checkmark \\
    $11$ & $\left(0,0;1,0;1,0\right)$& $\Gamma(2)\times\mathbb{Z}_2$ & --- \\
    $12$ & $\left(0,0;1,0;1,1\right)$& $\Gamma(2)\times\mathbb{Z}_2$ & --- \\
    $13$ & $\left(0,0;1,1;0,0\right)$& $\Gamma^1(2)\times\mathbb{Z}_2$ &   --- \\
    $14$ & $\left(0,0;1,1;0,1\right)$& $\Gamma(2)\times\mathbb{Z}_2$ &  --- \\
    $15$ & $\left(0,0;1,1;1,0\right)$& $\Gamma(2)\times\mathbb{Z}_2$ & --- \\
    $16$ & $\left(0,0;1,1;1,1\right)$& $\Gamma_\vartheta\times\mathbb{Z}_2$  &   --- \\
    \hline
\end{longtable}	
\end{center}

\clearpage

\subsubsection{Asymmetric $\mathbb{Z}_2\times \mathbb{Z}_2$ orbifold without Wilson line}
\label{app:Z2Z21}

\begin{center}
\begin{longtable}{|c||c||c|c|}
\caption{Possible configurations of $\left(\hat{q}_{1}\;\mathrm{mod}\;1,\hat{q}_{2}\;\mathrm{mod}\;1;\hat{w}_{1},\hat{w}_{2};\hat{n}_{1},\hat{n}_{2} \right)$ on the asymmetric $\mathbb{Z}_2\times \mathbb{Z}_2$ orbifold without Wilson line.}
\label{tab:Z2Z21}\\
\hline
  $\#$ & $\left(\hat{q}_{1}\;\mathrm{mod}\;1,\hat{q}_{2}\;\mathrm{mod}\;1;\hat{w}_{1},\hat{w}_{2};\hat{n}_{1},\hat{n}_{2} \right)$ & $G_{\mathrm{modular}}$ &  $\hat{\mathcal{CP}}$\\\hline
    \endfirsthead
     \multicolumn{4}{c}%
    {{\bfseries \tablename \thetable{}--continued from previous page}}\\
    \hline
   $\#$ & $\left(\hat{q}_{1},\hat{q}_{2};\hat{w}_{1},\hat{w}_{2};\hat{n}_{1},\hat{n}_{2} \right)$ & $G_{\mathrm{modular}}$ &  $\hat{\mathcal{CP}}$ \\\hline
     \endhead
     \hline \multicolumn{4}{|r|}{{Continued on next page}} \\ \hline
     \endfoot

     \hline
     \endlastfoot
    $1$ & $\left(0,0;0,0;0,0\right)$& $\mathrm{PSL}(2,\mathbb{Z})\times\mathbb{Z}_2\times\mathbb{Z}^{\hat{M}}_2$ &   \checkmark \\
    $2$ & $\left(0,0;0,0;0,1\right)$&$\Gamma_1(2)\times\mathbb{Z}_2\times\mathbb{Z}^{\hat{M}}_2$ &\checkmark \\
    $3$ & $\left(0,0;0,0;1,0\right)$& $\Gamma(2)\times\mathbb{Z}_2\times\mathbb{Z}^{\hat{M}}_2$ & \checkmark \\
    $4$ & $\left(0,0;0,0;1,1\right)$&$\Gamma(2)\times\mathbb{Z}_2\times\mathbb{Z}^{\hat{M}}_2$ &  \checkmark \\
    $5$ & $\left(0,0;0,1;0,0\right)$& $\Gamma^1(2)\times\mathbb{Z}_2\times\mathbb{Z}^{\hat{M}}_2$ &  \checkmark \\
    $6$ & $\left(0,0;0,1;0,1\right)$& $\Gamma_\vartheta\times\mathbb{Z}_2\times\mathbb{Z}^{\hat{M}}_2$ &  \checkmark \\
    $7$ & $\left(0,0;0,1;1,0\right)$&$\Gamma(2)\times\mathbb{Z}_2\times\mathbb{Z}^{\hat{M}}_2$ & \checkmark \\
    $8$ & $\left(0,0;0,1;1,1\right)$& $\mathbb{Z}_3$ &   \checkmark \\
    $9$ & $\left(0,0;1,0;0,0\right)$& $\Gamma(2)\times\mathbb{Z}_2\times\mathbb{Z}^{\hat{M}}_2$ &  \checkmark \\
    $10$ & $\left(0,0;1,0;0,1\right)$& $\Gamma(2)\times\mathbb{Z}_2\times\mathbb{Z}^{\hat{M}}_2$ &  \checkmark \\
    $11$ & $\left(0,0;1,0;1,0\right)$& $\mathrm{PSL}(2,\mathbb{Z})\times\mathbb{Z}_2\times\mathbb{Z}^{\hat{M}}_2$ &  \checkmark \\
    $12$ & $\left(0,0;1,0;1,1\right)$& $\Gamma_1(2)\times\mathbb{Z}_2\times\mathbb{Z}^{\hat{M}}_2$ &  \checkmark \\
    $13$ & $\left(0,0;1,1;0,0\right)$& $\Gamma(2)\times\mathbb{Z}_2\times\mathbb{Z}^{\hat{M}}_2$ & \checkmark \\
    $14$ & $\left(0,0;1,1;0,1\right)$& $\mathrm{Residue\; class\; of}\;\Gamma(2)$ & \checkmark \\
    $15$ & $\left(0,0;1,1;1,0\right)$& $\Gamma^1(2)\times\mathbb{Z}_2\times\mathbb{Z}^{\hat{M}}_2$ &   \checkmark \\
    $16$ & $\left(0,0;1,1;1,1\right)$& $\Gamma_\vartheta\times\mathbb{Z}_2\times\mathbb{Z}^{\hat{M}}_2$ &   \checkmark \\
    \hline
\end{longtable}	
\end{center}

\clearpage

\subsubsection{Asymmetric $\mathbb{Z}_2\times \mathbb{Z}_2$ orbifold with discrete Wilson line}
\label{app:Z2Z22}

\begin{center}
\begin{longtable}{|c||c||c|c|}
\caption{Possible configurations of $\left(\hat{q}_{1}\;\mathrm{mod}\;1,\hat{q}_{2};\hat{w}_{1},\hat{w}_{2};\hat{n}_{1},\hat{n}_{2} \right)$ on the asymmetric $\mathbb{Z}_2\times \mathbb{Z}_2$ orbifold with discrete Wilson line.}
\label{tab:Z2Z22}\\
\hline
  $\#$ & $\left(\hat{q}_{1}\;\mathrm{mod}\;1,\hat{q}_{2};\hat{w}_{1},\hat{w}_{2};\hat{n}_{1},\hat{n}_{2} \right)$& $G_{\mathrm{modular}}$ &  $\hat{\mathcal{CP}}$ \\\hline
    \endfirsthead
     \multicolumn{4}{c}%
    {{\bfseries \tablename \thetable{}--continued from previous page}}\\
    \hline
   $\#$ & $\left(\hat{q}_{1},\hat{q}_{2};\hat{w}_{1},\hat{w}_{2};\hat{n}_{1},\hat{n}_{2} \right)$& $G_{\mathrm{modular}}$&   $\hat{\mathcal{CP}}$ \\\hline
     \endhead
     \hline \multicolumn{4}{|r|}{{Continued on next page}} \\ \hline
     \endfoot

     \hline
     \endlastfoot
    $1$ & $\left(0,0;0,0;0,0\right)$&$\{\hat{M}_1,\hat{M}_2,\hat{M}_3\}$ &    \checkmark \\
    $2$ & $\left(0,0;0,0;0,1\right)$& $\{\hat{M}_1,(\hat{M}_2)^2\hat{M}_3\}$ &  \checkmark \\
    $3$ & $\left(0,0;0,0;1,0\right)$& $\{\hat{M}_1,\hat{M}_3\}$ &  \checkmark \\
    $4$ & $\left(0,0;0,0;1,1\right)$& $\{\hat{M}_1,\hat{M}_3\}$ & \checkmark \\
    $5$ & $\left(0,0;0,1;0,0\right)$& --- &  --- \\
    $6$ & $\left(0,0;0,1;0,1\right)$& --- &  --- \\
    $7$ & $\left(0,0;0,1;1,0\right)$& ---- &  --- \\
    $8$ & $\left(0,0;0,1;1,1\right)$& --- & --- \\
    $9$ & $\left(0,0;1,0;0,0\right)$& $\{\hat{M}_1,\hat{M}_3\}$ & \checkmark \\
    $10$ & $\left(0,0;1,0;0,1\right)$& $\{\hat{M}_1,\hat{M}_3\}$ &  \checkmark \\
    $11$ & $\left(0,0;1,0;1,0\right)$& $\{\hat{M}_1,(\hat{M}_2)^2\hat{M}_3\}$ & \checkmark \\
    $12$ & $\left(0,0;1,0;1,1\right)$& $\{\hat{M}_1,\hat{M}_2,\hat{M}_3\}$ &  \checkmark \\
    $13$ & $\left(0,0;1,1;0,0\right)$& --- &   --- \\
    $14$ & $\left(0,0;1,1;0,1\right)$& --- &   --- \\
    $15$ & $\left(0,0;1,1;1,0\right)$& --- & --- \\
    $16$ & $\left(0,0;1,1;1,1\right)$& --- &  --- \\
    $17$ & $\left(0,1;0,0;0,0\right)$& --- &    --- \\
    $18$ & $\left(0,1;0,0;0,1\right)$& --- &    --- \\
    $19$ & $\left(0,1;0,0;1,0\right)$& $\hat{M}_3$ &  --- \\
    $20$ & $\left(0,1;0,0;1,1\right)$& $\hat{M}_3$ &   --- \\
    $21$ & $\left(0,1;0,1;0,0\right)$& $\{(\hat{M}_2)^2,(\hat{M}_2\hat{M}_3)^2\}$ &  --- \\
    $22$ & $\left(0,1;0,1;0,1\right)$& $\{(\hat{M}_2)^2,(\hat{M}_2\hat{M}_3)^2\}$ & --- \\
    $23$ & $\left(0,1;0,1;1,0\right)$& --- &  --- \\
    $24$ & $\left(0,1;0,1;1,1\right)$& --- &  --- \\
    $25$ & $\left(0,1;1,0;0,0\right)$& $\hat{M}_3$ &   --- \\
    $26$ & $\left(0,1;1,0;0,1\right)$& $\hat{M}_3$ &  --- \\
    $27$ & $\left(0,1;1,0;1,0\right)$& --- &   --- \\
    $28$ & $\left(0,1;1,0;1,1\right)$& --- &   --- \\
    $29$ & $\left(0,1;1,1;0,0\right)$& --- &  --- \\
    $30$ & $\left(0,1;1,1;0,1\right)$& --- &   --- \\
    $31$ & $\left(0,1;1,1;1,0\right)$& $\{(\hat{M}_2)^2,(\hat{M}_2\hat{M}_3)^2\}$ &   --- \\
    $32$ & $\left(0,1;1,1;1,1\right)$& $\{(\hat{M}_2)^2,(\hat{M}_2\hat{M}_3)^2\}$ & --- \\
    \hline
\end{longtable}	
\end{center}

Table \ref{tab:Z2Z22} indicates that the non-supersymmetric string theories have the same modular symmetry as the supersymmetric one for a specific $\hat{Z}$, but the modular symmetries are in general broken down. The remaining symmetries are summarized below:
\begin{itemize}
    \item For the configuration of $\hat{Z}$ with $\#1, \#12$,\\
    there exists the same modular symmetry generated by $\{\hat{M}_1,\hat{M}_2,\hat{M}_3\}$ with the supersymmetric heterotic string theory.
    \item For the configuration of $\hat{Z}$ with $\#2, \#11$,\\
    there exists the direct product of $\mathbb{Z}_2^{(\hat{M}_1)}\times \mathbb{Z}_2^{((\hat{M}_2)^2)}$ symmetry and the shift symmetry generated by $\hat{M}_3$.
    \item For the configuration of $\hat{Z}$ with $\#3, \#4, \#9, \#10$,\\
    there exists the direct product of $\mathbb{Z}_2^{(\hat{M}_1)}$ symmetry and the shift symmetry generated by $\hat{M}_3$.
    \item For the configuration of $\hat{Z}$ with $\#19, \#20, \#25, \#26$,\\
    there exists the shift symmetry generated by $\hat{M}_3$.
    \item For the configuration of $\hat{Z}$ with $\#21, \#22, \#31, \#32$,\\
    there exists the $\mathbb{Z}_2^{((\hat{M}_2)^2)}\times\mathbb{Z}_2^{((\hat{M}_2\hat{M}_3)^2)}$ symmetry generated by $\{(\hat{M}_2)^2,(\hat{M}_2\hat{M}_3)^2\}$.
    \item For the other configurations of $\hat{Z}$, there is no remaining symmetry.
\end{itemize}

\clearpage

\subsubsection{Asymmetric $S_3$ orbifold with Wilson line}
\label{app:S3}

\begin{center}
\begin{longtable}{|c||c||c|c|}
\caption{Possible configurations of $\left(\hat{q}_{1}\;\mathrm{mod}\;1,\hat{q}_{2};\hat{w}_{1},\hat{w}_{2};\hat{n}_{1},\hat{n}_{2} \right)$ on the asymmetric $S_3$ orbifold with Wilson line.}
\label{tab:S3}\\
\hline
   $\#$ & $\left(\hat{q}_{1}\;\mathrm{mod}\;1,\hat{q}_{2};\hat{w}_{1},\hat{w}_{2};\hat{n}_{1},\hat{n}_{2} \right)$& $G_{\mathrm{modular}}$&   $\hat{\mathcal{CP}}$ \\\hline
    \endfirsthead
     \multicolumn{4}{c}%
    {{\bfseries \tablename \thetable{}--continued from previous page}}\\
    \hline
   $\#$ & $\left(\hat{q}_{1},\hat{q}_{2};\hat{w}_{1},\hat{w}_{2};\hat{n}_{1},\hat{n}_{2} \right)$& $G_{\mathrm{modular}}$&   $\hat{\mathcal{CP}}$ \\\hline
     \endhead
     \hline \multicolumn{4}{|r|}{{Continued on next page}} \\ \hline
     \endfoot

     \hline
     \endlastfoot
    $1$ & $\left(0,0;0,0;0,0\right)$& $\{\hat{M}_1,\hat{M}_2,\hat{M}_3,\hat{M}_4\}$ & \checkmark \\
    $2$ & $\left(0,0;0,0;0,1\right)$& $\{(\hat{M}_1\hat{M}_2)^3,\hat{M}_3,\hat{M}_4\}$ &  \checkmark \\
    $3$ & $\left(0,0;0,0;1,0\right)$&$\{\hat{M}_1,\hat{M}_2\hat{M}_1\hat{M}_2,\hat{M}_4,\hat{M}_3\hat{M}_4\hat{M}_3\hat{M}_4\hat{M}_3\}$ & \checkmark \\
    $4$ & $\left(0,0;0,0;1,1\right)$& $\{\hat{M}_4,\hat{M}_3\hat{M}_4\hat{M}_3\hat{M}_4\hat{M}_3\}$ &\checkmark \\
    $5$ & $\left(0,0;0,1;0,0\right)$& $\{\hat{M}_2\hat{M}_1\hat{M}_2,\hat{M}_3,\hat{M}_4\}$ &  --- \\
    $6$ & $\left(0,0;0,1;0,1\right)$& $\{\hat{M}_3,\hat{M}_4,\hat{M}_1,\hat{M}_2\hat{M}_1\hat{M}_2\hat{M}_1\hat{M}_2\}$ & --- \\
    $7$ & $\left(0,0;0,1;1,0\right)$& $\{\hat{M}_4,\hat{M}_2\hat{M}_1\hat{M}_2,\hat{M}_3\hat{M}_4\hat{M}_3\hat{M}_4\hat{M}_3\}$ &--- \\
    $8$ & $\left(0,0;0,1;1,1\right)$& $\{\hat{M}_1,\hat{M}_4,\hat{M}_2\hat{M}_1\hat{M}_2\hat{M}_1\hat{M}_2,\hat{M}_3\hat{M}_4\hat{M}_3\hat{M}_4\hat{M}_3\}$ & --- \\
    $9$ & $\left(0,0;1,0;0,0\right)$& $\{\hat{M}_1,\hat{M}_2,(\hat{M}_3\hat{M}_4)^2\}$ & \checkmark \\
    $10$ & $\left(0,0;1,0;0,1\right)$& $\{(\hat{M}_3\hat{M}_4)^3\}$ & \checkmark \\
    $11$ & $\left(0,0;1,0;1,0\right)$& $\{\hat{M}_1,\hat{M}_3,\hat{M}_2\hat{M}_1\hat{M}_2,\hat{M}_4\hat{M}_3\hat{M}_4\hat{M}_3\hat{M}_4\}$ & \checkmark \\
    $12$ & $\left(0,0;1,0;1,1\right)$& $\{\hat{M}_3,\hat{M}_4\hat{M}_3\hat{M}_4\hat{M}_3\hat{M}_4\}$ &  \checkmark \\
    $13$ & $\left(0,0;1,1;0,0\right)$& $\hat{M}_2\hat{M}_1\hat{M}_2$ &  --- \\
    $14$ & $\left(0,0;1,1;0,1\right)$& $\{\hat{M}_1,\hat{M}_2\hat{M}_1\hat{M}_2\hat{M}_1\hat{M}_2\}$ &  --- \\
    $15$ & $\left(0,0;1,1;1,0\right)$& $\{\hat{M}_3,\hat{M}_2\hat{M}_1\hat{M}_2,\hat{M}_4\hat{M}_3\hat{M}_4\hat{M}_3\hat{M}_4\}$ & --- \\
    $16$ & $\left(0,0;1,1;1,1\right)$& $\{\hat{M}_1,\hat{M}_3,\hat{M}_2\hat{M}_1\hat{M}_2\hat{M}_1\hat{M}_2,\hat{M}_3\hat{M}_4\hat{M}_3\hat{M}_4\hat{M}_3\}$ & --- \\
    $17$ & $\left(0,1;0,0;0,0\right)$& $\{\hat{M}_1,\hat{M}_3\}$ &  --- \\
    $18$ & $\left(0,1;0,0;0,1\right)$& $\hat{M}_3$ & --- \\
    $19$ & $\left(0,1;0,0;1,0\right)$& $\hat{M}_1$ & --- \\
    $20$ & $\left(0,1;0,0;1,1\right)$& --- &  --- \\
    $21$ & $\left(0,1;0,1;0,0\right)$& $\hat{M}_3$ &--- \\
    $22$ & $\left(0,1;0,1;0,1\right)$& $\{\hat{M}_1,\hat{M}_3,\hat{M}_2\hat{M}_1\hat{M}_2\}$ &--- \\
    $23$ & $\left(0,1;0,1;1,0\right)$& --- &  --- \\
    $24$ & $\left(0,1;0,1;1,1\right)$& $\{\hat{M}_1,\hat{M}_2\hat{M}_1\hat{M}_2\}$ &  --- \\
    $25$ & $\left(0,1;1,0;0,0\right)$& $\hat{M}_1$ & --- \\
    $26$ & $\left(0,1;1,0;0,1\right)$& --- &--- \\
    $27$ & $\left(0,1;1,0;1,0\right)$& $\{\hat{M}_1,\hat{M}_3,(\hat{M}_3\hat{M}_4)^2\}$ &--- \\
    $28$ & $\left(0,1;1,0;1,1\right)$& $\{\hat{M}_3,(\hat{M}_3\hat{M}_4)^2\}$ & --- \\
    $29$ & $\left(0,1;1,1;0,0\right)$& --- & --- \\
    $30$ & $\left(0,1;1,1;0,1\right)$& $\{\hat{M}_1,\hat{M}_2\hat{M}_1\hat{M}_2\}$ & --- \\
    $31$ & $\left(0,1;1,1;1,0\right)$& $\{\hat{M}_3,(\hat{M}_3\hat{M}_4)^2\}$ &  --- \\
    $32$ & $\left(0,1;1,1;1,1\right)$& $\{\hat{M}_1,\hat{M}_3,\hat{M}_2\hat{M}_1\hat{M}_2,(\hat{M}_3\hat{M}_4)^2\}$ & --- \\
    \hline
\end{longtable}	
\end{center}

Table \ref{tab:S3} indicates that the non-supersymmetric string theories have the same modular symmetry as the supersymmetric one for a specific $\hat{Z}$, but the modular symmetries are in general broken down. The remaining symmetries are summarized below:
\begin{itemize}
    \item For the configuration of $\hat{Z}$ with $\#1$,\\
    there exists the same $\mathrm{PSL}(2,\mathbb{Z})\times D_6$ modular symmetry generated by $\{\hat{M}_1,\hat{M}_2,\hat{M}_3,\hat{M}_4\}$ with the supersymmetric heterotic string theory.
    \item For the configuration of $\hat{Z}$ with $\#2$,\\
    there exists the $\mathbb{Z}_2^{((\hat{M}_1\hat{M}_2)^3)}\times D_6$ symmetry generated by $\{(\hat{M}_1\hat{M}_2)^3, \hat{M}_3, \hat{M}_4\}$.
    \item For the configuration of $\hat{Z}$ with $\#3$,\\
    there exists the $\mathrm{PSL}(2,\mathbb{Z})\times \mathbb{Z}_2^{(\hat{M}_4)} \times \mathbb{Z}_2^{(\hat{M}_3\hat{M}_4\hat{M}_3\hat{M}_4\hat{M}_3)}$ symmetry, where we use $(\hat{M}_4)^2=1$ and $[\hat{M}_4, \hat{M}_3\hat{M}_4\hat{M}_3\hat{M}_4\hat{M}_3]=0$.
    \item For the configuration of $\hat{Z}$ with $\#4$,\\
     there exists the $\mathbb{Z}_2^{(\hat{M}_4)} \times \mathbb{Z}_2^{(\hat{M}_3\hat{M}_4\hat{M}_3\hat{M}_4\hat{M}_3)}$ symmetry generated by $\{\hat{M}_4, \hat{M}_3\hat{M}_4\hat{M}_3\hat{M}_4\hat{M}_3\}$ with $(\hat{M}_4)^2=1$ and $[\hat{M}_4, \hat{M}_3\hat{M}_4\hat{M}_3\hat{M}_4\hat{M}_3]=0$.
    \item For the configuration of $\hat{Z}$ with $\#5$,\\
    there exists a direct product of the shift symmetry generated by $\hat{M}_2\hat{M}_1\hat{M}_2$ and $D_6$ symmetry generated by $\{\hat{M}_3, \hat{M}_4\}$.
    \item For the configuration of $\hat{Z}$ with $\#6$,\\
    there exists a direct product of $D_\infty \cong \mathbb{Z}\rtimes \mathbb{Z}_2^{(\hat{M}_1)}$ generated by $\{\hat{M}_1, \hat{M}_2\hat{M}_1\hat{M}_2\hat{M}_1\hat{M}_2\}$ and $D_6$ symmetry generated by $\{\hat{M}_3, \hat{M}_4\}$.
    \item For the configuration of $\hat{Z}$ with $\#7$,\\
    there exists a direct product of the shift symmetry generated by $\hat{M}_2\hat{M}_1\hat{M}_2$ and $\mathbb{Z}_2^{(\hat{M}_4)} \times \mathbb{Z}_2^{(\hat{M}_3\hat{M}_4\hat{M}_3\hat{M}_4\hat{M}_3)}$ symmetry, where we use $(\hat{M}_4)^2=1$ and $[\hat{M}_4, \hat{M}_3\hat{M}_4\hat{M}_3\hat{M}_4\hat{M}_3]=0$.    
    \item For the configuration of $\hat{Z}$ with $\#8$,\\
    there exists a direct product of $D_\infty \cong \mathbb{Z}\rtimes \mathbb{Z}_2^{(\hat{M}_1)}$ generated by $\{\hat{M}_1, \hat{M}_2\hat{M}_1\hat{M}_2\hat{M}_1\hat{M}_2\}$ and $\mathbb{Z}_2^{(\hat{M}_4)} \times \mathbb{Z}_2^{(\hat{M}_3\hat{M}_4\hat{M}_3\hat{M}_4\hat{M}_3)}$ symmetry.    
    \item For the configuration of $\hat{Z}$ with $\#9$,\\
    there exists the $\mathrm{PSL}(2,\mathbb{Z})\times \mathbb{Z}_2^{((\hat{M}_3\hat{M}_4)^3)}$ modular symmetry generated by $\{\hat{M}_1,\hat{M}_2,(\hat{M}_3\hat{M}_4)^3\}$.
    \item For the configuration of $\hat{Z}$ with $\#10$,\\
    there exists the $\mathbb{Z}_2^{((\hat{M}_3\hat{M}_4)^3)}$ symmetry generated by $(\hat{M}_3\hat{M}_4)^3$.
    \item For the configuration of $\hat{Z}$ with $\#11$,\\
    there exists the $\mathrm{PSL}(2,\mathbb{Z})\times \mathbb{Z}_2^{(\hat{M}_3)} \times \mathbb{Z}_2^{(\hat{M}_4\hat{M}_3\hat{M}_4\hat{M}_3\hat{M}_4)}$ symmetry, where we use $(\hat{M}_3)^2=1$ and $[\hat{M}_3, \hat{M}_4\hat{M}_3\hat{M}_4\hat{M}_3\hat{M}_4]=0$.
    \item For the configuration of $\hat{Z}$ with $\#12$,\\
    there exists the $\mathbb{Z}_2^{(\hat{M}_3)} \times \mathbb{Z}_2^{(\hat{M}_4\hat{M}_3\hat{M}_4\hat{M}_3\hat{M}_4)}$ symmetry, where we use $(\hat{M}_3)^2=1$ and $[\hat{M}_3, \hat{M}_4\hat{M}_3\hat{M}_4\hat{M}_3\hat{M}_4]=0$.
    \item For the configuration of $\hat{Z}$ with $\#13$, there exists the shift symmetry generated by $\hat{M}_2\hat{M}_1\hat{M}_2$.    
    \item For the configuration of $\hat{Z}$ with $\#14$,\\
    there exists the $D_\infty \cong \mathbb{Z}\rtimes \mathbb{Z}_2^{(\hat{M}_1)}$ symmetry generated by $\{\hat{M}_1, \hat{M}_2\hat{M}_1\hat{M}_2\hat{M}_1\hat{M}_2\}$.
    \item For the configuration of $\hat{Z}$ with $\#15$,\\
    there exists a direct product of the shift symmetry generated by $\hat{M}_2\hat{M}_1\hat{M}_2$ and $\mathbb{Z}_2^{(\hat{M}_3)} \times \mathbb{Z}_2^{(\hat{M}_4\hat{M}_3\hat{M}_4\hat{M}_3\hat{M}_4)}$ symmetry, where we use $(\hat{M}_3)^2=1$ and $[\hat{M}_3, \hat{M}_4\hat{M}_3\hat{M}_4\hat{M}_3\hat{M}_4]=0$.    
    \item For the configuration of $\hat{Z}$ with $\#16$,\\
    there exists a direct product of $D_\infty \cong \mathbb{Z}\rtimes \mathbb{Z}_2^{(\hat{M}_1)}$ generated by $\{\hat{M}_1, \hat{M}_2\hat{M}_1\hat{M}_2\hat{M}_1\hat{M}_2\}$ and $\mathbb{Z}_2^{(\hat{M}_3)} \times \mathbb{Z}_2^{(\hat{M}_4\hat{M}_3\hat{M}_4\hat{M}_3\hat{M}_4)}$ symmetry, where we use $(\hat{M}_3)^2=1$ and $[\hat{M}_3, \hat{M}_4\hat{M}_3\hat{M}_4\hat{M}_3\hat{M}_4]=0$.    
    \item For the configuration of $\hat{Z}$ with $\#17$, there exists the $\mathbb{Z}_2^{(\hat{M}_1)} \times \mathbb{Z}_2^{(\hat{M}_3)}$ symmetry.    
    \item For the configuration of $\hat{Z}$ with $\#18, \#21$, there exists the $\mathbb{Z}_2^{(\hat{M}_3)}$ symmetry.  
    \item For the configuration of $\hat{Z}$ with $\#19, \#25$, there exists the $\mathbb{Z}_2^{(\hat{M}_1)}$ symmetry.  
    \item For the configuration of $\hat{Z}$ with $\#22$, there exists the $\mathrm{PSL}(2,\mathbb{Z})\times \mathbb{Z}_2^{(\hat{M}_3)}$ symmetry.  
    \item For the configuration of $\hat{Z}$ with $\#24, \#30$, there exists the $\mathrm{PSL}(2,\mathbb{Z})$ symmetry.  
    \item For the configuration of $\hat{Z}$ with $\#27$, there exists the $\mathbb{Z}_2^{(\hat{M}_1)}\times S_3$ symmetry due to
    \begin{align*}
        (\hat{M}_1)^2=(A)^2=(B)^3=(AB)^2=\mathbb{I},
    \end{align*}
    with $A=\hat{M}_3$ and $B=(\hat{M}_3\hat{M}_4)^2$.
    \item For the configuration of $\hat{Z}$ with $\#28, \#31$, there exists the $S_3$ symmetry generated by $\{\hat{M}_3, (\hat{M}_3\hat{M}_4)^2\}$.  
    \item For the configuration of $\hat{Z}$ with $\#32$, there exists the $\mathrm{PSL}(2,\mathbb{Z})\times S_3$ symmetry.  
    \item For the other configurations of $\hat{Z}$ such as $\#20, \#23, \#26, \#29$, there is no remaining symmetry.
\end{itemize}
\clearpage

\subsection{Dimension 0}

We present six toroidal orbifolds where the dimension of the moduli space is zero.

\subsubsection{Asymmetric $\mathbb{Z}_5$ orbifold with discrete Wilson line}
\label{app:Z5}

\begin{center}
\begin{longtable}{|c||c||c|c|}
\caption{Possible configurations of $\left(\hat{q}_{1}\;\mathrm{mod}\;1,\hat{q}_{2};\hat{w}_{1},\hat{w}_{2};\hat{n}_{1},\hat{n}_{2} \right)$ on the asymmetric $\mathbb{Z}_5$ orbifold with discrete Wilson line.}
\label{tab:Z5}\\
\hline
   $\#$ & $\left(\hat{q}_{1}\;\mathrm{mod}\;1,\hat{q}_{2};\hat{w}_{1},\hat{w}_{2};\hat{n}_{1},\hat{n}_{2} \right)$ & $G_{\mathrm{modular}}$ &  $\hat{\mathcal{CP}}$ \\\hline
    \endfirsthead
    \multicolumn{4}{c}%
    {{\bfseries \tablename \thetable{}--continued from previous page}}\\
    \hline
    $\#$ & $\left(\hat{q}_{1},\hat{q}_{2};\hat{w}_{1},\hat{w}_{2};\hat{n}_{1},\hat{n}_{2} \right)$ & $G_{\mathrm{modular}}$ &  $\hat{\mathcal{CP}}$ \\\hline
     \endhead
     \hline \multicolumn{4}{|r|}{{Continued on next page}} \\ \hline
     \endfoot

     \hline
     \endlastfoot
    $1$ & $\left(0,0;0,0;0,0\right)$& $\hat{\Theta}$ &    \checkmark \\
    $2$ & $\left(0,0;0,0;0,1\right)$&--- &    \checkmark \\
    $3$ & $\left(0,0;0,0;1,0\right)$& --- &   --- \\
    $4$ & $\left(0,0;0,0;1,1\right)$& --- &   --- \\
    $5$ & $\left(0,0;0,1;0,0\right)$&--- &   --- \\
    $6$ & $\left(0,0;0,1;0,1\right)$& --- &   --- \\
    $7$ & $\left(0,0;0,1;1,0\right)$& --- &   --- \\
    $8$ & $\left(0,0;0,1;1,1\right)$& --- &    --- \\
    $9$ & $\left(0,0;1,0;0,0\right)$& --- &   --- \\
    $10$ & $\left(0,0;1,0;0,1\right)$& --- &   --- \\
    $11$ & $\left(0,0;1,0;1,0\right)$& --- &    \checkmark \\
    $12$ & $\left(0,0;1,0;1,1\right)$& --- &   \checkmark \\
    $13$ & $\left(0,0;1,1;0,0\right)$& --- &    --- \\
    $14$ & $\left(0,0;1,1;0,1\right)$& --- &    --- \\
    $15$ & $\left(0,0;1,1;1,0\right)$& --- &   --- \\
    $16$ & $\left(0,0;1,1;1,1\right)$& --- &     --- \\
    $17$ & $\left(0,1;0,0;0,0\right)$& --- &   --- \\
    $18$ & $\left(0,1;0,0;0,1\right)$& --- &     --- \\
    $19$ & $\left(0,1;0,0;1,0\right)$& --- &  --- \\
    $20$ & $\left(0,1;0,0;1,1\right)$& --- &    --- \\
    $21$ & $\left(0,1;0,1;0,0\right)$& --- &     \checkmark \\
    $22$ & $\left(0,1;0,1;0,1\right)$& --- &   \checkmark \\
    $23$ & $\left(0,1;0,1;1,0\right)$& --- &     --- \\
    $24$ & $\left(0,1;0,1;1,1\right)$& --- &   --- \\
    $25$ & $\left(0,1;1,0;0,0\right)$& --- &    --- \\
    $26$ & $\left(0,1;1,0;0,1\right)$& --- &   --- \\
    $27$ & $\left(0,1;1,0;1,0\right)$& --- &    --- \\
    $28$ & $\left(0,1;1,0;1,1\right)$& --- &     --- \\
    $29$ & $\left(0,1;1,1;0,0\right)$& --- &    --- \\
    $30$ & $\left(0,1;1,1;0,1\right)$& --- &   --- \\
    $31$ & $\left(0,1;1,1;1,0\right)$& --- &   \checkmark \\
    $32$ & $\left(0,1;1,1;1,1\right)$& --- &   \checkmark \\
    \hline

\end{longtable}	
\end{center}

\clearpage

\subsubsection{Asymmetric $S_4$ orbifold with discrete Wilson line}
\label{app:S4}

\begin{center}
\begin{longtable}{|c||c||c|c|}
\caption{Possible configurations of $\left(\hat{q}_{1}\;\mathrm{mod}\;1,\hat{q}_{2};\hat{w}_{1},\hat{w}_{2};\hat{n}_{1},\hat{n}_{2} \right)$ on the asymmetric $S_4$ orbifold with discrete Wilson line.}
\label{tab:S4}\\
\hline
    $\#$ & $\left(\hat{q}_{1}\;\mathrm{mod}\;1,\hat{q}_{2};\hat{w}_{1},\hat{w}_{2};\hat{n}_{1},\hat{n}_{2} \right)$ &$G_{\mathrm{modular}}$ &  $\hat{\mathcal{CP}}$ \\\hline
    \endfirsthead
    \multicolumn{4}{c}%
    {{\bfseries \tablename \thetable{}--continued from previous page}}\\
    \hline
    $\#$ & $\left(\hat{q}_{1},\hat{q}_{2};\hat{w}_{1},\hat{w}_{2};\hat{n}_{1},\hat{n}_{2} \right)$ & $G_{\mathrm{modular}}$&  $\hat{\mathcal{CP}}$ \\\hline
     \endhead
     \hline \multicolumn{4}{|r|}{{Continued on next page}} \\ \hline
     \endfoot

     \hline
     \endlastfoot
    $1$ & $\left(0,0;0,0;0,0\right)$& $\{\hat{\Theta}_1,\hat{\Theta}_2\}$ &   \checkmark \\
    $2$ & $\left(0,0;0,0;0,1\right)$& $\hat{\Theta}_1$ & --- \\
    $3$ & $\left(0,0;0,0;1,0\right)$& $\hat{\Theta}_1\hat{\Theta}_2$ &--- \\
    $4$ & $\left(0,0;0,0;1,1\right)$&  $(\hat{\Theta}_2)^2$ &  --- \\
    $5$ & $\left(0,0;0,1;0,0\right)$& $\hat{\Theta}_1$ & \checkmark \\
    $6$ & $\left(0,0;0,1;0,1\right)$& $\hat{\Theta}_1$ & --- \\
    $7$ & $\left(0,0;0,1;1,0\right)$& --- &  --- \\
    $8$ & $\left(0,0;0,1;1,1\right)$& --- & --- \\
    $9$ & $\left(0,0;1,0;0,0\right)$& --- &  --- \\
    $10$ & $\left(0,0;1,0;0,1\right)$& --- &  --- \\
    $11$ & $\left(0,0;1,0;1,0\right)$& $\hat{\Theta}_1$ & \checkmark \\
    $12$ & $\left(0,0;1,0;1,1\right)$& $\hat{\Theta}_1$ &  --- \\
    $13$ & $\left(0,0;1,1;0,0\right)$& $\hat{\Theta}_1\hat{\Theta}_2$ & --- \\
    $14$ & $\left(0,0;1,1;0,1\right)$& $(\hat{\Theta}_2)^2$ &  --- \\
    $15$ & $\left(0,0;1,1;1,0\right)$& $\{\hat{\Theta}_1,\hat{\Theta}_2\}$ & \checkmark \\
    $16$ & $\left(0,0;1,1;1,1\right)$&  $\hat{\Theta}_1$ &    --- \\
    $17$ & $\left(0,1;0,0;0,0\right)$& $\hat{\Theta}_1$ &  --- \\
    $18$ & $\left(0,1;0,0;0,1\right)$& $\hat{\Theta}_1$ &  --- \\
    $19$ & $\left(0,1;0,0;1,0\right)$& --- &  \checkmark \\
    $20$ & $\left(0,1;0,0;1,1\right)$&  --- &  --- \\
    $21$ & $\left(0,1;0,1;0,0\right)$& $\hat{\Theta}_1$ & --- \\
    $22$ & $\left(0,1;0,1;0,1\right)$& $\{\hat{\Theta}_1,\hat{\Theta}_2\}$ &   --- \\
    $23$ & $\left(0,1;0,1;1,0\right)$& $(\hat{\Theta}_2)^2$ &   \checkmark \\
    $24$ & $\left(0,1;0,1;1,1\right)$& $\hat{\Theta}_1\hat{\Theta}_2$ &   --- \\
    $25$ & $\left(0,1;1,0;0,0\right)$& $(\hat{\Theta}_2)^2$ &  \checkmark \\
    $26$ & $\left(0,1;1,0;0,1\right)$& $\hat{\Theta}_1\hat{\Theta}_2$ &  --- \\
    $27$ & $\left(0,1;1,0;1,0\right)$& $\hat{\Theta}_1$ &  --- \\
    $28$ & $\left(0,1;1,0;1,1\right)$& $\hat{\Theta}_1\hat{\Theta}_2$ & --- \\
    $29$ & $\left(0,1;1,1;0,0\right)$& --- &   \checkmark \\
    $30$ & $\left(0,1;1,1;0,1\right)$& --- &  --- \\
    $31$ & $\left(0,1;1,1;1,0\right)$& $\hat{\Theta}_1$ &  --- \\
    $32$ & $\left(0,1;1,1;1,1\right)$&  $\hat{\Theta}_1$ &--- \\
    \hline
\end{longtable}	
\end{center}

\clearpage

\subsubsection{Asymmetric $(\mathbb{Z}_4\times \mathbb{Z}_2)\rtimes \mathbb{Z}_2$ orbifold without Wilson line}
\label{app:Z4Z2Z2}

\begin{center}
\begin{longtable}{|c||c||c|c|}
\caption{Possible configurations of $\left(\hat{q}_{1}\;\mathrm{mod}\;1,\hat{q}_{2}\;\mathrm{mod}\;1;\hat{w}_{1},\hat{w}_{2};\hat{n}_{1},\hat{n}_{2} \right)$ on the asymmetric $(\mathbb{Z}_4\times \mathbb{Z}_2)\rtimes \mathbb{Z}_2$ orbifold without Wilson line.}
\label{tab:Z4Z2Z2}\\
\hline
   $\#$ & $\left(\hat{q}_{1}\;\mathrm{mod}\;1,\hat{q}_{2}\;\mathrm{mod}\;1;\hat{w}_{1},\hat{w}_{2};\hat{n}_{1},\hat{n}_{2} \right)$ & $G_{\mathrm{modular}}$ &   $\hat{\mathcal{CP}}$ \\\hline
    \endfirsthead
    \multicolumn{4}{c}%
    {{\bfseries \tablename \thetable{}--continued from previous page}}\\
    \hline
    $\#$ & $\left(\hat{q}_{1},\hat{q}_{2};\hat{w}_{1},\hat{w}_{2};\hat{n}_{1},\hat{n}_{2} \right)$ & $G_{\mathrm{modular}}$ &   $\hat{\mathcal{CP}}$ \\\hline
     \endhead
     \hline \multicolumn{4}{|r|}{{Continued on next page}} \\ \hline
     \endfoot

     \hline
     \endlastfoot
    $1$ & $\left(0,0;0,0;0,0\right)$& $\{\hat{\Theta}_1,\hat{\Theta}_2,\hat{\Theta}_3\}$ &    \checkmark \\
    $2$ & $\left(0,0;0,0;0,1\right)$& $\{(\hat{\Theta}_1)^2,\hat{\Theta}_3\}$ &  \checkmark \\
    $3$ & $\left(0,0;0,0;1,0\right)$& $(\hat{\Theta}_1)^2$ &   \checkmark \\
    $4$ & $\left(0,0;0,0;1,1\right)$& $\hat{\Theta}_1$ &    \checkmark \\
    $5$ & $\left(0,0;0,1;0,0\right)$& $\{(\hat{\Theta}_1)^2,\hat{\Theta}_3\}$ &  \checkmark \\
    $6$ & $\left(0,0;0,1;0,1\right)$& $\{(\hat{\Theta}_1)^2,\hat{\Theta}_2,\hat{\Theta}_3\}$ &    \checkmark \\
    $7$ & $\left(0,0;0,1;1,0\right)$& $(\hat{\Theta}_1)^2$ &  \checkmark \\
    $8$ & $\left(0,0;0,1;1,1\right)$& $(\hat{\Theta}_1)^2$ &  \checkmark \\
    $9$ & $\left(0,0;1,0;0,0\right)$& $(\hat{\Theta}_1)^2$ &  \checkmark \\
    $10$ & $\left(0,0;1,0;0,1\right)$& $(\hat{\Theta}_1)^2$ &  \checkmark \\
    $11$ & $\left(0,0;1,0;1,0\right)$& $\{(\hat{\Theta}_1)^2,\hat{\Theta}_2,\hat{\Theta}_3\}$ &  \checkmark \\
    $12$ & $\left(0,0;1,0;1,1\right)$& $\{(\hat{\Theta}_1)^2,\hat{\Theta}_3\}$ &  \checkmark \\
    $13$ & $\left(0,0;1,1;0,0\right)$& $\hat{\Theta}_1$ &  \checkmark \\
    $14$ & $\left(0,0;1,1;0,1\right)$& $(\hat{\Theta}_1)^2$ &  \checkmark \\
    $15$ & $\left(0,0;1,1;1,0\right)$& $\{(\hat{\Theta}_1)^2,\hat{\Theta}_3\}$ &  \checkmark \\
    $16$ & $\left(0,0;1,1;1,1\right)$& $\{\hat{\Theta}_1,\hat{\Theta}_2,\hat{\Theta}_3\}$ & \checkmark \\
    \hline

\end{longtable}
\end{center}

\clearpage

\subsubsection{Asymmetric $S_3 \times \mathbb{Z}_6$ orbifold without Wilson line}
\label{app:S3Z6}

\begin{center}
\begin{longtable}{|c||c||c|c|}
\caption{Possible configurations of $\left(\hat{q}_{1}\;\mathrm{mod}\;1,\hat{q}_{2}\;\mathrm{mod}\;1;\hat{w}_{1},\hat{w}_{2};\hat{n}_{1},\hat{n}_{2} \right)$ on the asymmetric $S_3 \times \mathbb{Z}_6$ orbifold without Wilson line.}
\label{tab:S3Z6}\\
\hline
  $\#$ & $\left(\hat{q}_{1}\;\mathrm{mod}\;1,\hat{q}_{2}\;\mathrm{mod}\;1;\hat{w}_{1},\hat{w}_{2};\hat{n}_{1},\hat{n}_{2} \right)$ & $G_{\mathrm{modular}}$ &  $\hat{\mathcal{CP}}$ \\\hline
    \endfirsthead
    \multicolumn{4}{c}%
    {{\bfseries \tablename \thetable{}--continued from previous page}}\\
    \hline
    $\#$ & $\left(\hat{q}_{1},\hat{q}_{2};\hat{w}_{1},\hat{w}_{2};\hat{n}_{1},\hat{n}_{2} \right)$ & $G_{\mathrm{modular}}$ &  $\hat{\mathcal{CP}}$ \\\hline
     \endhead
     \hline \multicolumn{4}{|r|}{{Continued on next page}} \\ \hline
     \endfoot

     \hline
     \endlastfoot
    $1$ & $\left(0,0;0,0;0,0\right)$& $\{\hat\Theta_1,\hat\Theta_2,\hat\Theta_3\}$ &   \checkmark \\
    $2$ & $\left(0,0;0,0;0,1\right)$& $\hat\Theta_2$ & \checkmark \\
    $3$ & $\left(0,0;0,0;1,0\right)$& $\hat\Theta_1$ &   --- \\
    $4$ & $\left(0,0;0,0;1,1\right)$& --- &  --- \\
    $5$ & $\left(0,0;0,1;0,0\right)$& $\hat\Theta_2$ &  --- \\
    $6$ & $\left(0,0;0,1;0,1\right)$& $\{\hat\Theta_1,\hat\Theta_2\}$ &  --- \\
    $7$ & $\left(0,0;0,1;1,0\right)$& --- & --- \\
    $8$ & $\left(0,0;0,1;1,1\right)$& $\{\hat\Theta_1,\hat\Theta_3\}$ & --- \\
    $9$ & $\left(0,0;1,0;0,0\right)$& --- & --- \\
    $10$ & $\left(0,0;1,0;0,1\right)$& $\hat\Theta_1$ &  --- \\
    $11$ & $\left(0,0;1,0;1,0\right)$& $\{\hat\Theta_2,\hat\Theta_3\}$ &  \checkmark \\
    $
    12$ & $\left(0,0;1,0;1,1\right)$& $\{\hat\Theta_1,\hat\Theta_2\}$ &  \checkmark \\
    $13$ & $\left(0,0;1,1;0,0\right)$& $\hat\Theta_1$ &  --- \\
    $14$ & $\left(0,0;1,1;0,1\right)$& $\hat\Theta_3$ &   --- \\
    $15$ & $\left(0,0;1,1;1,0\right)$& $\{\hat\Theta_1,\hat\Theta_2\}$ &  --- \\
    $16$ & $\left(0,0;1,1;1,1\right)$& $\hat\Theta_2$ & --- \\
    \hline

\end{longtable}
\end{center}

\clearpage

\subsubsection{Asymmetric $S_3 \times \mathbb{Z}_2$ orbifold with discrete Wilson line}
\label{app:S3Z2}

\begin{center}
\begin{longtable}{|c||c||c|c|}
\caption{Possible configurations of $\left(\hat{q}_{1}\;\mathrm{mod}\;1,\hat{q}_{2};\hat{w}_{1},\hat{w}_{2};\hat{n}_{1},\hat{n}_{2} \right)$ on the asymmetric $S_3 \times \mathbb{Z}_2$ orbifold with discrete Wilson line.}
\label{tab:S3Z2}\\
\hline
  $\#$ & $\left(\hat{q}_{1}\;\mathrm{mod}\;1,\hat{q}_{2};\hat{w}_{1},\hat{w}_{2};\hat{n}_{1},\hat{n}_{2} \right)$ & $G_{\mathrm{modular}}$ &  $\hat{\mathcal{CP}}$ \\\hline
    \endfirsthead
      \multicolumn{4}{c}%
    {{\bfseries \tablename \thetable{}--continued from previous page}}\\
    \hline
     $\#$ & $\left(\hat{q}_{1},\hat{q}_{2};\hat{w}_{1},\hat{w}_{2};\hat{n}_{1},\hat{n}_{2} \right)$ & $G_{\mathrm{modular}}$ & $\hat{\mathcal{CP}}$ \\\hline
     \endhead
     \hline \multicolumn{4}{|r|}{{Continued on next page}} \\ \hline
     \endfoot

     \hline
     \endlastfoot
    $1$ & $\left(0,0;0,0;0,0\right)$& $\{\hat\Theta_1,\hat\Theta_2,\hat\Theta_3\}$ &  \checkmark \\
    $2$ & $\left(0,0;0,0;0,1\right)$& --- &  \checkmark \\
    $3$ & $\left(0,0;0,0;1,0\right)$& $\{\hat\Theta_1,\hat\Theta_3\}$ &  \checkmark \\
    $4$ & $\left(0,0;0,0;1,1\right)$& --- &  \checkmark \\
    $5$ & $\left(0,0;0,1;0,0\right)$& --- & \checkmark \\
    $6$ & $\left(0,0;0,1;0,1\right)$& $\{\hat\Theta_1,\hat\Theta_2,\hat\Theta_3\}$ &  \checkmark \\
    $7$ & $\left(0,0;0,1;1,0\right)$& --- & \checkmark \\
    $8$ & $\left(0,0;0,1;1,1\right)$& $\{\hat\Theta_1,\hat\Theta_3\}$ &  \checkmark \\
    $9$ & $\left(0,0;1,0;0,0\right)$& $\{\hat\Theta_2,\hat\Theta_3\}$ & \checkmark \\
    $10$ & $\left(0,0;1,0;0,1\right)$& --- &  \checkmark \\
    $11$ & $\left(0,0;1,0;1,0\right)$& $\hat\Theta_3$ & \checkmark \\
    $12$ & $\left(0,0;1,0;1,1\right)$& --- &  \checkmark \\
    $13$ & $\left(0,0;1,1;0,0\right)$& --- &  \checkmark \\
    $14$ & $\left(0,0;1,1;0,1\right)$& $\{\hat\Theta_2,\hat\Theta_3\}$ &   \checkmark \\
    $15$ & $\left(0,0;1,1;1,0\right)$& --- &  \checkmark \\
    $16$ & $\left(0,0;1,1;1,1\right)$& $\hat\Theta_3$ &   \checkmark \\
    $17$ & $\left(0,1;0,0;0,0\right)$& $\hat\Theta_3$ &   \checkmark \\
    $18$ & $\left(0,1;0,0;0,1\right)$& --- &  \checkmark \\
    $19$ & $\left(0,1;0,0;1,0\right)$& $\hat\Theta_3$ &   \checkmark \\
    $20$ & $\left(0,1;0,0;1,1\right)$& --- &   \checkmark \\
    $21$ & $\left(0,1;0,1;0,0\right)$& --- &  \checkmark \\
    $22$ & $\left(0,1;0,1;0,1\right)$& $\hat\Theta_3$ &   \checkmark \\
    $23$ & $\left(0,1;0,1;1,0\right)$& --- &  \checkmark \\
    $24$ & $\left(0,1;0,1;1,1\right)$& $\hat\Theta_3$ &   \checkmark \\
    $25$ & $\left(0,1;1,0;0,0\right)$& $\hat\Theta_3$ &   \checkmark \\
    $26$ & $\left(0,1;1,0;0,1\right)$& --- &   \checkmark \\
    $27$ & $\left(0,1;1,0;1,0\right)$& $\hat\Theta_3$ &  \checkmark \\
    $28$ & $\left(0,1;1,0;1,1\right)$& --- &  \checkmark \\
    $29$ & $\left(0,1;1,1;0,0\right)$& --- & \checkmark \\
    $30$ & $\left(0,1;1,1;0,1\right)$& $\hat\Theta_3$ & \checkmark \\
    $31$ & $\left(0,1;1,1;1,0\right)$& --- &  \checkmark \\
    $32$ & $\left(0,1;1,1;1,1\right)$& $\hat\Theta_3$ &  \checkmark \\
    \hline

\end{longtable}
\end{center}

\clearpage

\subsubsection{Asymmetric $\mathbb{Z}_{12}$ orbifold without Wilson line}
\label{app:Z12}

\begin{center}
\begin{longtable}{|c||c||c|c|}
\caption{Possible configurations of $\left(\hat{q}_{1}\;\mathrm{mod}\;1,\hat{q}_{2}\;\mathrm{mod}\;1;\hat{w}_{1},\hat{w}_{2};\hat{n}_{1},\hat{n}_{2} \right)$ on the asymmetric $\mathbb{Z}_{12}$ orbifold without Wilson line.}
\label{tab:Z12}\\
\hline
 $\#$ & $\left(\hat{q}_{1}\;\mathrm{mod}\;1,\hat{q}_{2}\;\mathrm{mod}\;1;\hat{w}_{1},\hat{w}_{2};\hat{n}_{1},\hat{n}_{2} \right)$& $G_{\mathrm{modular}}$&   $\hat{\mathcal{CP}}$ \\\hline
    \endfirsthead
     \multicolumn{4}{c}%
    {{\bfseries \tablename \thetable{}--continued from previous page}}\\
    \hline
     $\#$ & $\left(\hat{q}_{1},\hat{q}_{2};\hat{w}_{1},\hat{w}_{2};\hat{n}_{1},\hat{n}_{2} \right)$& $G_{\mathrm{modular}}$ &  $\hat{\mathcal{CP}}$ \\\hline
     \endhead
     \hline \multicolumn{4}{|r|}{{Continued on next page}} \\ \hline
     \endfoot

     \hline
     \endlastfoot
    $1$ & $\left(0,0;0,0;0,0\right)$& $\{\hat\Theta,(\hat\Theta)^2,(\hat\Theta)^3,(\hat\Theta)^4,(\hat\Theta)^6\}$& \checkmark \\
    $2$ & $\left(0,0;0,0;0,1\right)$&$(\hat\Theta)^6$  &\checkmark \\
    $3$ & $\left(0,0;0,0;1,0\right)$& $(\hat\Theta)^6$  &   --- \\
    $4$ & $\left(0,0;0,0;1,1\right)$& $(\hat\Theta)^6$ &  --- \\
    $5$ & $\left(0,0;0,1;0,0\right)$&$(\hat\Theta)^6$  & --- \\
    $6$ & $\left(0,0;0,1;0,1\right)$& $(\hat\Theta)^6$  &--- \\
    $7$ & $\left(0,0;0,1;1,0\right)$& $\{(\hat\Theta)^3,(\hat\Theta)^6\}$  & --- \\
    $8$ & $\left(0,0;0,1;1,1\right)$& $(\hat\Theta)^6$ & --- \\
    $9$ & $\left(0,0;1,0;0,0\right)$& $(\hat\Theta)^6$ &\checkmark \\
    $10$ & $\left(0,0;1,0;0,1\right)$& $\{(\hat\Theta)^3,(\hat\Theta)^6\}$ &\checkmark \\
    $11$ & $\left(0,0;1,0;1,0\right)$& $(\hat\Theta)^6$ & --- \\
    $12$ & $\left(0,0;1,0;1,1\right)$& $(\hat\Theta)^6$ &  --- \\
    $13$ & $\left(0,0;1,1;0,0\right)$& $(\hat\Theta)^6$ & --- \\
    $14$ & $\left(0,0;1,1;0,1\right)$& $(\hat\Theta)^6$ &  --- \\
    $15$ & $\left(0,0;1,1;1,0\right)$& $(\hat\Theta)^6$ & --- \\
    $16$ & $\left(0,0;1,1;1,1\right)$& $\{(\hat\Theta)^3,(\hat\Theta)^6\}$ &  --- \\
    \hline

\end{longtable}	
\end{center}



\bibliography{references}{}
\bibliographystyle{JHEP}

\end{document}